\RequirePackage{fix-cm}
\documentclass[twocolumn,epjc3]{svjour3}

\RequirePackage[T1]{fontenc}

\RequirePackage{amssymb}
\RequirePackage{bm}
\RequirePackage{dcolumn}
\RequirePackage{amssymb}
\RequirePackage{amsmath}
\RequirePackage{graphicx,subfig}
\RequirePackage{booktabs}
\RequirePackage{multirow}
\RequirePackage{setspace}
\RequirePackage{array}
\RequirePackage{threeparttable}
\RequirePackage{txfonts}
\RequirePackage{float}
\RequirePackage{tabularx}
\newcommand{\PreserveBackslash}[1]{\let\temp=\\#1\let\\=\temp}
\newcolumntype{C}[1]{>{\PreserveBackslash\centering}p{#1}}
\newcolumntype{R}[1]{>{\PreserveBackslash\raggedleft}p{#1}}
\newcolumntype{L}[1]{>{\PreserveBackslash\raggedright}p{#1}}

\smartqed  % flush right qed marks, e.g. at end of proof

\RequirePackage{graphicx}
\RequirePackage{mathptmx}      % use Times fonts if available on your TeX system
\RequirePackage{flushend}
\RequirePackage[numbers,sort&compress]{natbib}
\RequirePackage[colorlinks,citecolor=blue,urlcolor=blue,linkcolor=blue]{hyperref}

\journalname{Eur. Phys. J. C}

\begin{document}

\title{Light speed variation from gamma ray bursts: criteria for low energy photons}

%\subtitle{criteria for low energy photons}

\author{Yue Liu\thanksref{addr1}
        \and
        Bo-Qiang Ma\thanksref{e2,addr1,addr2,addr3,addr4} %etc.
}

%%\thankstext[$\star$]{t1}{Thanks to the title}
\thankstext{e2}{e-mail: mabq@pku.edu.cn}

\institute{School of Physics and State Key Laboratory of Nuclear Physics and Technology, Peking University, Beijing 100871, China\label{addr1}
          \and
          Collaborative Innovation Center of Quantum Matter, Beijing, China\label{addr2}
          \and
          Center for High Energy Physics, Peking University, Beijing 100871, China\label{addr3}
          \and
          Center for History and Philosophy of Science, Peking University, Beijing 100871,
          China\label{addr4}
}

\date{Received: date / Accepted: date}

\maketitle

\begin{abstract}
We examine a method to detect the light speed variation from gamma ray burst data observed by the Fermi Gamma-ray Space Telescope (FGST).
We suggest new criteria to determine the characteristic time for low energy photons by the energy curve and the average energy curve respectively, and obtain similar results compared with those from the light curve. We offer a new criterion with both the light curve and the average energy curve to determine the characteristic time for low energy photons.
We then apply the new criteria to the GBM NaI data, the GBM BGO data, and the LAT LLE data, and obtain consistent results for three different sets of low energy photons from different FERMI detectors.

\end{abstract}

\section{Introduction}

According to Einstein's relativity, the speed of light is a constant $c$ in free space. However, it is speculated that the effect of quantum gravity may bring a tiny correction to the light speed of the order $E/E_{\rm Pl}$, where $E$ is the photon energy and $E_{\rm Pl} = \sqrt{\hbar c^5 /G}\approx 1.22\times 10^{19} \rm GeV$ is the Planck energy. The matter effect of the universe may also cause a modification to the light speed in the cosmological space. It is very difficult to measure the light speed variation by ordinary experiments on Earth because such a variation of $c$ is extremely small. One approach to solve this problem is to focus on photons from far away astrophysical objects. Amelino-Camelia \textit{et al.} first suggested
detecting light speed variation due to the Lorentz invariance
violation (LV) from gamma-ray bursts~(GRBs)~\cite{intro-1,intro-2}. Gamma-ray bursts are extremely energetic and rather quick processes in the universe. During a GRB, photons with different energies are emitted from the source and these photons travel through the cosmological space to reach the detectors. Some of the GRBs are so energetic that they emit high energy photons with energy $E_{\rm high, src}\gtrsim30$~GeV at the source~(src). Meanwhile, a GRB also emits numerous low energy photons: $E_{\rm low, src}\lesssim 1$~MeV. Although $E_{\rm high, src}$ and $E_{\rm low, src}$ are still very small compared to the Planck energy, the little difference in the light speed can be accumulated during the very long distance of travel. The tiny light speed variation leads to the different time of travel, so it contributes to the difference between the arrival times of high energy photons and low energy photons.

Ellis~{\it et al.} first analyzed the GRB data of photons aiming to
detect quantum gravity induced light speed variance~\cite{ellis}, and
they also developed a robust method to collect high energy photon data from different observations and
analyzed them collectively~\cite{jrellis}. In later work,
refs.~\cite{slj,zhangshu} combined high energy photons from different GRBs of
the Fermi telescope, and analyzed them collectively according to the method prescribed in ref.~\cite{jrellis}.
Reference~\cite{xhw} used the first main peak in the low energy light
curve rather than the trigger time as the low energy characteristic
time. Some progress has been made in References~\cite{slj,zhangshu,xhw,xhw160509,camelia,Xu:2018ien} on detecting the light speed variation from analysis of energetic photon events detected by the Fermi Gamma-ray Space Telescope (FGST)~\cite{fermi-1,fermi-2}.
By analyzing the time lags between energetic photon events and the corresponding low energy photon signals for several GRBs with known redshifts, a regularity was found for the time lags between photons of different energies. Such a regularity suggests a tiny light speed variation of the form $v(E)=c(1-E/E_{\rm LV})$, where $E_{\rm LV} \simeq 3.6 \times 10^{17}$~GeV. In this work we check this method carefully, mainly focusing on the determination of a remarkable low energy photon signal for each GRB in the analysis.
Besides light curves, we apply more criteria on choosing a low energy characteristic time from the photon energy released per unit time and average energy of photons.
We thus offer a new criterion to include the average energy curve in addition to the light curve to determine the characteristic time for low energy photons. Such a criterion not only includes the widely used light curve but also takes into account the changes of GRB energy distribution.
We then apply the new criteria to the GBM NaI data, the GBM BGO data, and the LAT LLE data, and obtain similar results for three different sets of low energy photons from different FERMI detectors.
With the new analysis,
we arrive at the results that are consistent with each other. %However, from the results, no light speed variation can be seen clearly.

In sect.~\ref{tpeaksec}, we review the basic method to detect the light speed variation from GRBs.
The usual criterion is based on the light curve of a GRB~(sect.~\ref{tpeaksec}).
%By examining the results under different conditions, we find the results not sensitive to different choices.
In sect.~\ref{options}, we provide more options on the criterion for the characteristic time of low energy photons. We use more criteria considering different aspects of intrinsic nature of the source. In sect.~\ref{c1}, we focus on the energy released. In sect.~\ref{c2}, we use the average energy per photon of an energy band. In sect.~\ref{c3}, we combine the criterion of the light curve in sect.~\ref{tpeaksec} and the criterion of the average energy in sect.~\ref{c2} and give our recommendation of the characteristic times.  In sect.~\ref{instruments}, we apply the new criteria to the GBM NaI data, the GBM BGO data and the LAT LLE data, and obtain consistent results for three different sets of low energy photons. In this section we also provide the estimation on uncertainties. Section~\ref{conclusionsec} serves as a summary and conclusion.

\section{Detecting the light speed variation}\label{tpeaksec}
\subsection{The method}\label{method}

We now explain the method on detecting the light speed variation from analyzing GRB data.

Consider that an astrophysical object has an energetic process such as a gamma ray burst, emitting a large number of gamma-rays, so that it can be detected by an observer on Earth~(or a detector in space). This object emits both high energy photons and low energy photons. Consider two photons with different energies, $E_{\rm high, src}$ for the high energy photon and $E_{\rm low, src}$ for the low energy photon, in the source frame.
These two photons need not to be emitted at the same time, so we use $\Delta t_{\rm in}$ to represent the intrinsic time lag between the high energy photon and the low energy photon when they are emitted, i.e., $\Delta t_{\rm in}=t_{\rm high, src}-t_{\rm low, src}$, where $t_{\rm high(low), src}$ is the emitting time of the high (low) energy photon in the source reference frame.
For each photon, it travels through the cosmological space and is captured by a detector, and its observed energy can be written as
\begin{equation}\label{energytransform}
E_{\rm obs} = E_{\rm src}/(1+z),
\end{equation}
where $z$ is the redshift of the GRB, and $E_{\rm src}$ is the energy of photon when it is emitted from the source.
So the energies of the two photons can be written as
\begin{equation}
E_{\rm high, obs} = E_{\rm high, src}/(1+z),
\end{equation}
and
\begin{equation}\label{eq:e-z}
E_{\rm low, obs} = E_{\rm low, src}/(1+z).
\end{equation}
The observed time lag between the high energy photon and the low energy photon is caused by two factors: the intrinsic time lag~($\Delta t_{\rm in}$) at the source, and the time lag caused by the difference of light speed~(e.g., the time lag caused by the LV effect or by matter effect of the universe), which is represented by $\Delta t_{\rm LV}$ here. Thus we have
\begin{equation}\label{eq:t-k}
\Delta t_{\rm obs} = \Delta t_{\rm LV} + \Delta t_{\rm in} (1+z).
\end{equation}
Now we consider the general form of dispersion relation of a photon. For a photon with energy $E$, if $E\ll E_{\rm Pl}$, the LV effect leads to a modified form
of dispersion relation
\begin{equation}
E^2 = p^2 c^2 \left[1-s_n \left(\frac{pc}{E_{\rm LV,n}}\right)^n\right],
\end{equation}
where $s_n = \pm 1$ indicates whether the high energy photon travels faster ($s_n = -1$) or slower ($s_n = +1$) than the low energy photon, and $E_{\rm LV,n}$ denotes the $n$th-order Lorentz invariance violation scale to be determined.

We assume $n=1$ in this article. Thus, we can derive the speed of light by $v=\partial E / \partial p$, and obtain
\begin{equation}\label{eq:speedoflight}
v(E)=c\left[1-s_1 \left( \frac{E}{E_{\rm LV}}\right) \right],
\end{equation}
where $E_{\rm LV}$ represents $E_{\rm LV, 1}$.

Once we obtain the relationship between speed and energy, we can apply it to the two photons mentioned above.
To calculate the time lag, $\Delta t_{\rm LV}$, we use the $\rm \Lambda$CDM Universe model that the universe consists of matter and dark energy~(cosmological constant).
It can be proved that $\Delta t_{\rm LV}$~(in the observer reference frame) can be written as~\cite{eq-1,eq-2}
\begin{equation}
\Delta t_{\rm LV} = (1+z)\frac{K}{E_{\rm LV}},
\label{eq-k-t}
\end{equation}
with
\begin{equation}\label{eq:K}
K = s_1 \frac{E_{\rm high,obs} - E_{\rm low,obs}}{H_0} \frac{1}{(1+z)} \int_{0}^{z} \frac{(1+z')\mathrm{d}z'}{\sqrt{\Omega_m (1+z')^3 + \Omega_{\Lambda}}},
\end{equation}
where $H_0 = 67.3 \pm 1.2 {\rm kms^{-1} Mpc^{-1}}$ is the present day Hubble constant, and $\Omega_m = 0.315^{+0.016}_{-0.017}$ and $\Omega_{\Lambda}=0.685^{+0.017}_{-0.016}$ are the matter density and dark energy density~\cite{lcdm}.

In the derivation above, $\Delta t_{\rm in}$ and $E_{\rm LV}$ are unknown but other parameters: $z$, $E_{\rm low,obs}$, $E_{\rm high,obs}$ and $\Delta t_{\rm obs}$ are known in principle.
What we should do next is to obtain $E_{\rm LV}$ based on these parameters and some reasonable assumptions.

In refs.~\cite{slj,zhangshu}, a method to look for the light speed variation from GRBs was introduced.
According to eq.~\ref{eq-k-t}, the time lag caused by the light speed variation~($\Delta t_{\rm LV}$) can be extracted by $K$, a factor that depends on the low and high photon energies and the redshift of the GRB.
What we can obtain from data is the observed time lag between high and low energy photons
\begin{equation}\label{eq:t}
\Delta t_{\rm obs} = t_{\rm high,obs} - t_{\rm low,obs},
\end{equation}
which is the sum of $\Delta t_{\rm LV}$ and the intrinsic emission time lag~($\Delta t_{\rm in}$) in eq.~\ref{eq:t-k}.
Then we get
\begin{equation}\label{eq:diag}
\frac{\Delta t_{\rm obs}}{1+z} = \frac{K}{E_{\rm LV}} + \Delta t_{\rm in}.
\end{equation}
If LV really exists, we can expect that there is a relationship between $K$ and $\Delta t_{\rm obs}$ due to the light speed variation, which infers that $E_{\rm LV}$ is of certain value, so our method is to plot the events in a $\Delta t_{\rm obs}/(1+z)$ versus $K$ plot and find out whether there is a correlation between $K$ and $\Delta t_{\rm obs}/(1+z)$.

\begin{table*}
	\centering
	\caption{The data of high energy ($>30$~GeV) photon events from GRBs with known redshifts, read from Pass-8 data~\cite{pass8hp} of the FERMI telescope. $t_{\rm high,obs}$ refers to the observed arrival time since trigger.}
	\label{tab:phinfo}
	\begin{tabular*}{1\textwidth}{@{\extracolsep{\fill}}cccccc}
		\hline
		\hline
		GRB   &$z$ &$t_{\rm high,obs}$ (s) & $E_{\rm high,obs}$ (GeV)& $E_{\rm high,src}$ (GeV)& $K \rm (10^{18} s\cdot GeV)$\\
		\hline
		080916C(1) & 4.35$\pm 0.15$  &40.509  & 27.4  & 146.7 & 9.87 \\
		080916C(2) & 4.35$\pm 0.15$ &16.545  & 12.4  & 66.5  & 4.47 \\
		080916C(3) & 4.35$\pm 0.15$  &43.999  & 5.71  & 30.5  & 2.05 \\
		080916C(4) & 4.35$\pm 0.15$  &28.210  & 6.72  & 36.0  & 2.42 \\
		\hline
		090510 & 0.903$\pm 0.003$ &0.828  & 29.9  & 56.9  & 7.21 \\
		\hline
		090902B(1) & 1.822 &81.746  & 39.9  & 112.5 & 12.9 \\
		090902B(2) & 1.822 &26.168  & 18.1  & 51.1  & 5.85 \\
		090902B(3) & 1.822 &45.608  & 15.4  & 43.5  & 4.97 \\
		090902B(4) & 1.822 &14.167  & 14.2  & 40.1  & 4.59 \\
		090902B(5) & 1.822 &42.374  & 12.7 & 35.7  & 4.10\\
		090902B(6) & 1.822&11.671  & 11.9  & 33.5  & 3.84\\
		\hline
		090926A & 2.1071$\pm 0.0001$ &24.838  & 19.5  & 60.5  & 6.52 \\
		\hline
		100414A & 1.368 &33.368  & 29.8  & 70.6  & 8.73 \\
		\hline
		130427A(1) & 0.3399$\pm 0.0002$ &18.644  & 77.1  & 103.3 & 9.58 \\
		130427A(2) & 0.3399$\pm 0.0002$ &47.588  & 28.4  & 38.1 & 3.53 \\
		130427A(3) & 0.3399$\pm 0.0002$ &84.749  & 26.9  & 36.0 & 3.34 \\
		130427A(4) & 0.3399$\pm 0.0002$ &78.397  & 38.7  & 51.8  & 4.80 \\
		\hline
		140619B & 2.67$\pm 0.37$  &0.613  & 22.7  & 83.5  & 7.96 \\
		\hline
		160509A & 1.17  &76.506  & 51.9  & 112.6 & 14.23 \\
		
		\hline
		\hline
	\end{tabular*}%
\end{table*}%

What we do next is based on the data from the FERMI telescope.
The FERMI telescope consists of the Fermi Large Area Telescope ~(LAT)~\cite{fermi-1,fermi-3,fermi-4} and the Gamma-Ray Burst Monitor (GBM)~\cite{fermi-2,gbminfo} detectors.
LAT is mainly used to record high energy events while GBM is used to record low energy events. GBM consists of 12 Sodium Iodide (NaI) detectors and 2 Bismuth Germanate (BGO) detectors. The energy range of NaI detectors is about $8-1000$ keV. The BGO detectors provide energy coverage from about 150 keV to 40 MeV~\cite{gbminfo,gbminfo2}.
The time accuracy of GBM and LAT is smaller than 10 $\mu$s relative to spacecraft time~\cite{gbmoverview,latoverview}. Therefore, we can treat these two instruments as well-synchronized.
We choose the photons whose intrinsic energies are higher than 30~GeV at source as high energy events and the photons detected by the GBM NaI detectors as the low energy events.
From the LAT telescope, the energies and the observed arrival time of the high energy events can be read directly.
Although the LAT data trace the direction of the photon events, the GBM detectors do not record the directions.
Thus, the background contamination can be big. Fortunately, we focus on the characteristic time when the light curve deviation from the background is significant. Since the background is expected to be stable over time, it merely adds a constant pedestal to the light curve without inducing any time-varying features. We therefore do not perform the background subtraction below because the background actually does not influence the determination of $t_{\rm low}$, as shown later in sect.~\ref{uncertainty}.

We search the high energy photons in all the GRBs detected by the FERMI telescope before 2016.12.31 and list them in table~\ref{tab:phinfo}.
The latest Pass 8 data~\cite{pass8hp,latgrbweb} are used to read the energies and arrival times of photons.
Here we choose those photons whose energies are higher than 30~GeV when they are emitted at the source~(after they are emitted, the redshift effect reduces their observed energies).
In former works, the selection rule of high energy photons is over 10~GeV in the observer frame in refs.~\cite{slj,zhangshu,xhw,xhw160509}
and it is changed to over 40~GeV in the source frame in ref.~\cite{camelia}. We here choose photons with energies over 30~GeV at source as an optional choice with some subsidiary information.

As the number of photons with such a high energy from a single GRB is quite small, we need to combine different GRBs together.
As the sample GRBs we choose must have known redshifts and photons with enough high energies, the GRBs that meet these restrictions are: GRB 080916C, GRB 090510, GRB 090902B, GRB 090926A, GRB 100414A, GRB 130427A, GRB 140619B and GRB 160509A (the redshift of GRB 140619B is obtained from ref.~\cite{140619B} and other redshifts are obtained from ref.~\cite{gcnredshift}, also see refs.~\cite{080916C,090902B,090926A,100414A,130427A,fermiredshift}).
For high energy events, each of GRB 080919C and GRB 130427A has 4 events, and GRB 090902B has 6 events. Each of the other GRBs has only one event.
Therefore, there are 19 high energy events that meet the requirements above.

The energies of GBM (NaI) photons are less than $2$~MeV, and thus are negligible when calculating $K$, because $E_{\rm low} \ll E_{\rm high}$. Therefore, $K$ can be written approximately as
\begin{equation}\label{eq:appro-K}
K = s_1 \frac{E_{\rm high,obs}}{H_0} \frac{1}{(1+z)} \int_{0}^{z} \frac{(1+z')\mathrm{d}z'}{\sqrt{\Omega_m (1+z')^3 + \Omega_{\Lambda}}}.
\end{equation}
The discussion above adopts a situation consisting of a high energy photon and a low energy photon.
For each GRB, it is reasonable to use the arrival time of a single high energy photon to mark a high energy process, but it is more complicated to mark the time of low energy process because there are plenty of low energy photons emitted during the whole GRB process.
We need to form a low energy event for each GRB, considering all the low energy photons at all time, to fit the situation above.
The approximate expression of $K$ tells us that the energy of a low energy event does not matter, so we do not need to focus on the exact value of energy of low energy photons.
What remains unclear is the time of a low energy event~($t_{\rm low,obs}$), which does not have a clear definition because while $t_{\rm high,obs}$ is a property of one single high energy photon, $t_{\rm low,obs}$ can only be obtained from \textbf{a set of} low energy photons with different energies and arrival times.

It is reasonable to assume that the low energy process has effect on the number of photons emitted per unit of time, so the characteristic time of a low energy event in the observer reference frame, i.e., $t_{\rm low,obs}$, must correlate with a characteristic point in the light curve.
In refs.~\cite{slj,zhangshu}, $t_{\rm low,obs}$ is chosen as the trigger time of GBM detector, while in refs.~\cite{xhw,xhw160509}, $t_{\rm low,obs}$ is the time of the first main pulse in the light curve, i.e., $t_{\rm peak,obs}$.
As the trigger time is strongly affected by the performance of GBM detectors and the distance of the source, we choose the time of the first main peak in the light curve as the signal time of low energy photons.
As the peaks of a light curve mark the moments with the largest densities of photons, the first main peak can serve as a significant benchmark that represents the intrinsic property of the GRB objectively.
We choose the highest point of this peak as $t_{\rm low,obs}$, and the details are discussed in the next subsection.

As this method depends on the criterion for choosing low energy events, we mainly focus on the determination of low energy characteristic times from different viewpoints and different data in the following sections. We try to provide a more comprehensive criterion and test it with more data.

\subsection{Determination of low energy characteristic time from light curves}\label{sss-obs}

From discussion in the previous section, the arrival time of a high energy photon can be read directly from the data.
We now need to find a proper way to determine $t_{\rm low,obs}$. Following refs.~\cite{xhw,xhw160509}, we choose the first peak in the light curve of low energy photons as $t_{\rm low,obs}$ to mark an intrinsic low energy process of GRB. It is considered to be a better choice than the trigger time~\cite{zhangshu,xhw}, because the latter is also sensitive to the performance of detectors besides the properties of the GRB itself.
In ref.~\cite{xhw}, photons ranging from 8-260~keV are used to plot the low energy light curve, but here we choose another energy band.
The number of recorded photons is determined not only by the numbers of emitted photons, but also by the efficiency of the detector.
Due to the materials in front of NaI detectors in GBM and other factors, the effective area of the detector depends on energy (see fig.~11 in ref.~\cite{fermi-2}).
NaI detectors can record photons ranging from 8-1000~keV, but the effective area of the detector drops quickly when the energy is lower than 20~keV and higher than 200~keV.
In the band of $20$-$200$~keV, the effective area is nearly the same.
If we choose low energy photons regardless of the dependence of effective area on energy, we actually assume implicitly that the light curves for each low energy band are of the same shape.
However, the light curves for different energy bands are not necessarily the same. %%see figures!
In ref.~\cite{xhw}, the low energy band is set to be 8-260~keV, but in that work, the variance of effective area is not considered. We use another low energy band here, 20-200~keV, in which the effective area is nearly constant.
Thus, we take all the photons in 20-200~keV band from GBM data into account, and in this band, the dependence of effective area on energy is much smaller than that of 8-260~keV (we call the 20-200~keV band as Band-Obs later).

Following former works~\cite{slj,zhangshu,xhw,xhw160509}, we choose the time of the first peak in the light curve in Band-Obs as the low energy photon arrival time $t_{\rm low,obs}$ and put it into eq.~\ref{eq:t} to calculate $\Delta t_{\rm obs}$ for each corresponding high energy event.

\begin{figure*}
	\centering
	\begin{minipage}[t]{0.48\linewidth}
		\centering
		\includegraphics[width=0.90\linewidth]{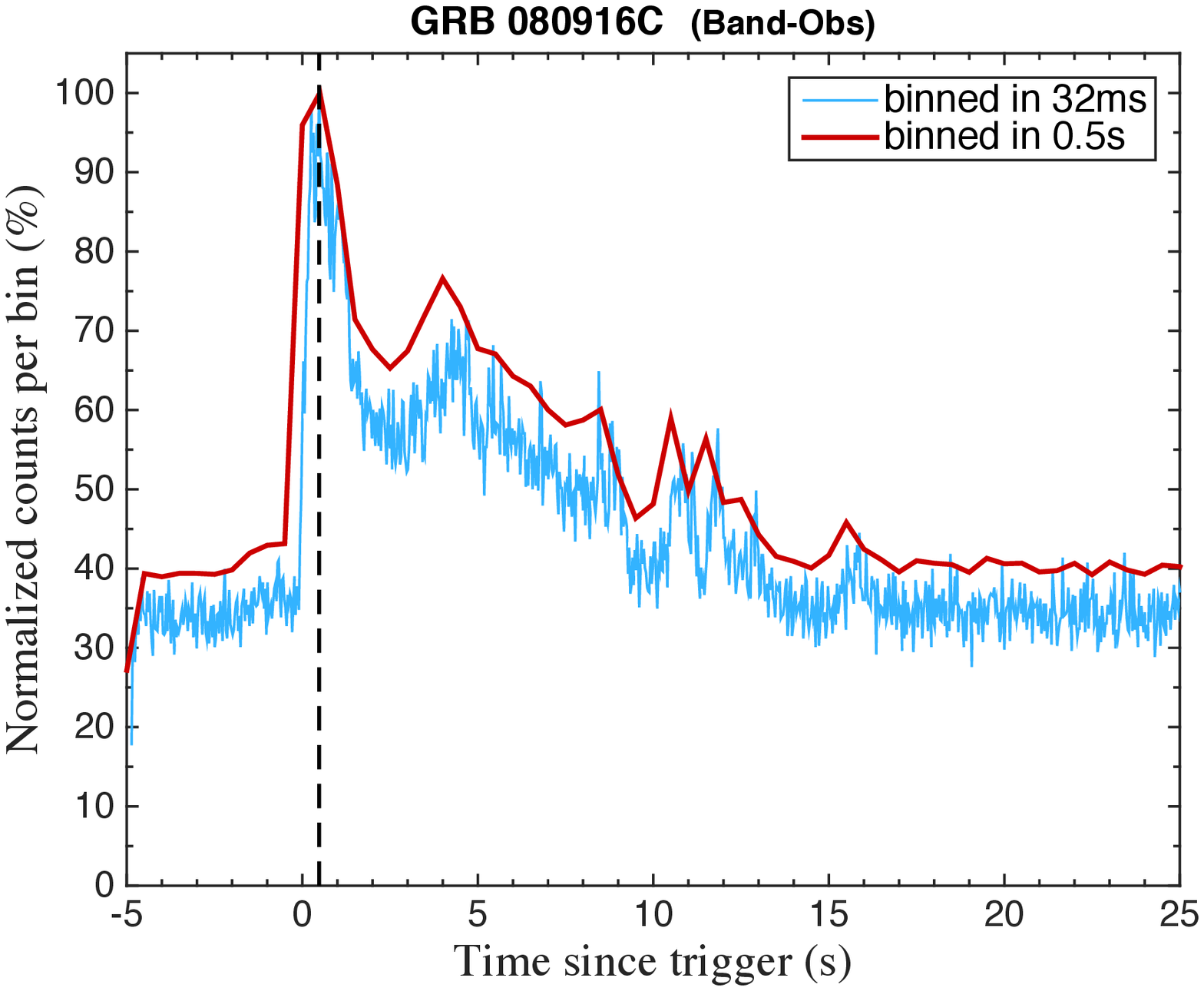}
		\includegraphics[width=0.90\linewidth]{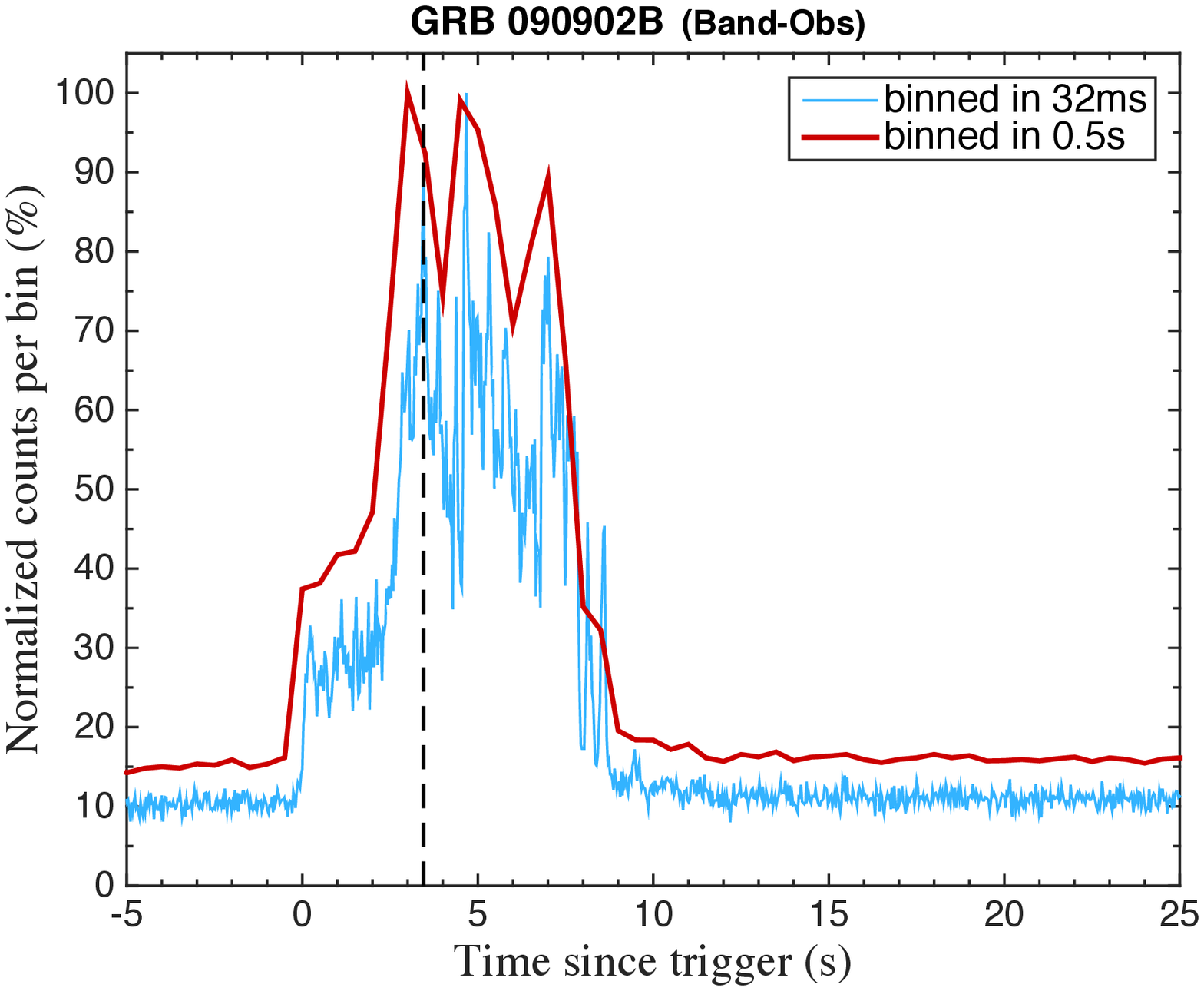}
		\includegraphics[width=0.90\linewidth]{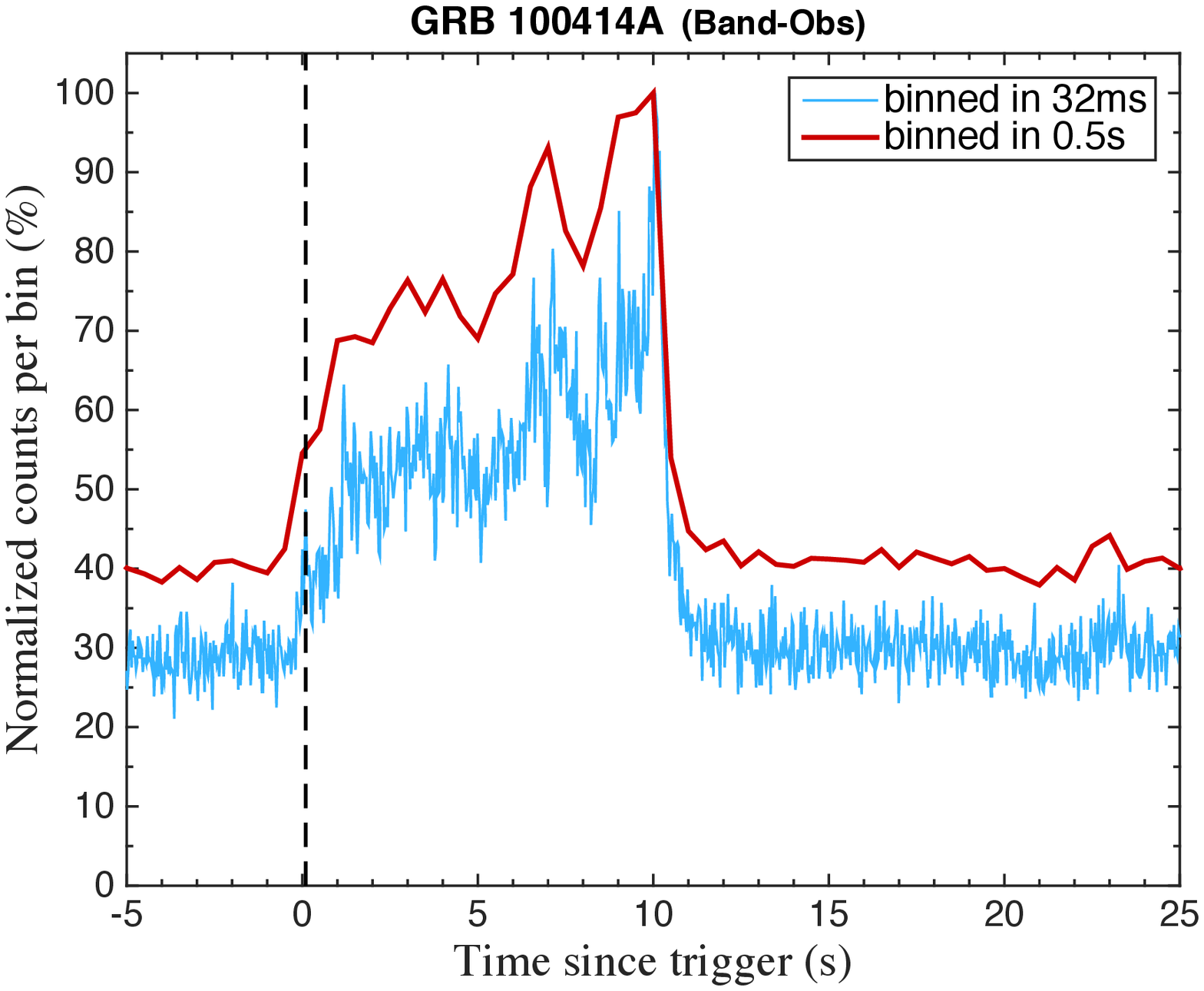}
		\includegraphics[width=0.90\linewidth]{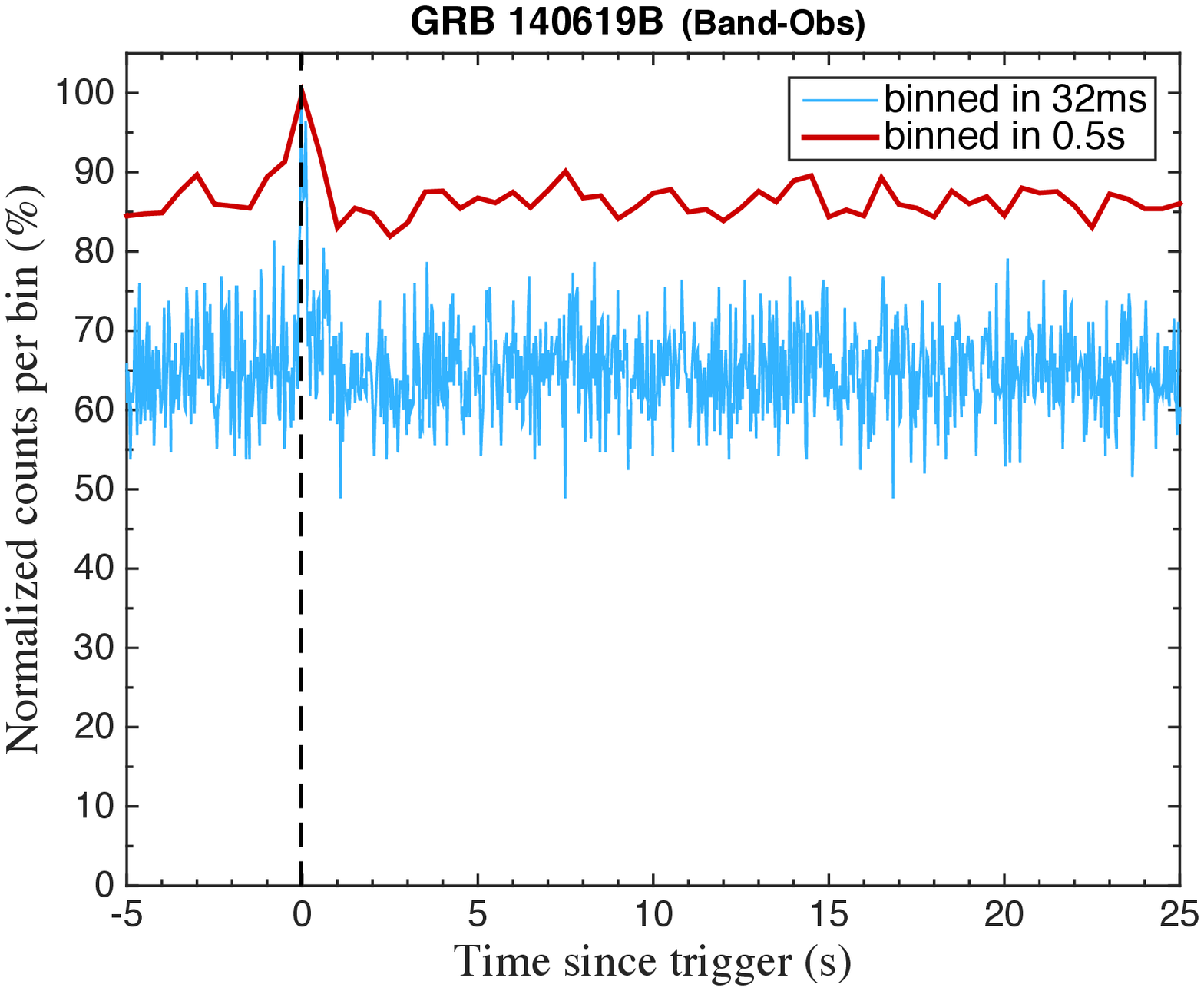}		
	\end{minipage}
	\begin{minipage}[t]{0.48\linewidth}
		\centering
		\includegraphics[width=0.90\linewidth]{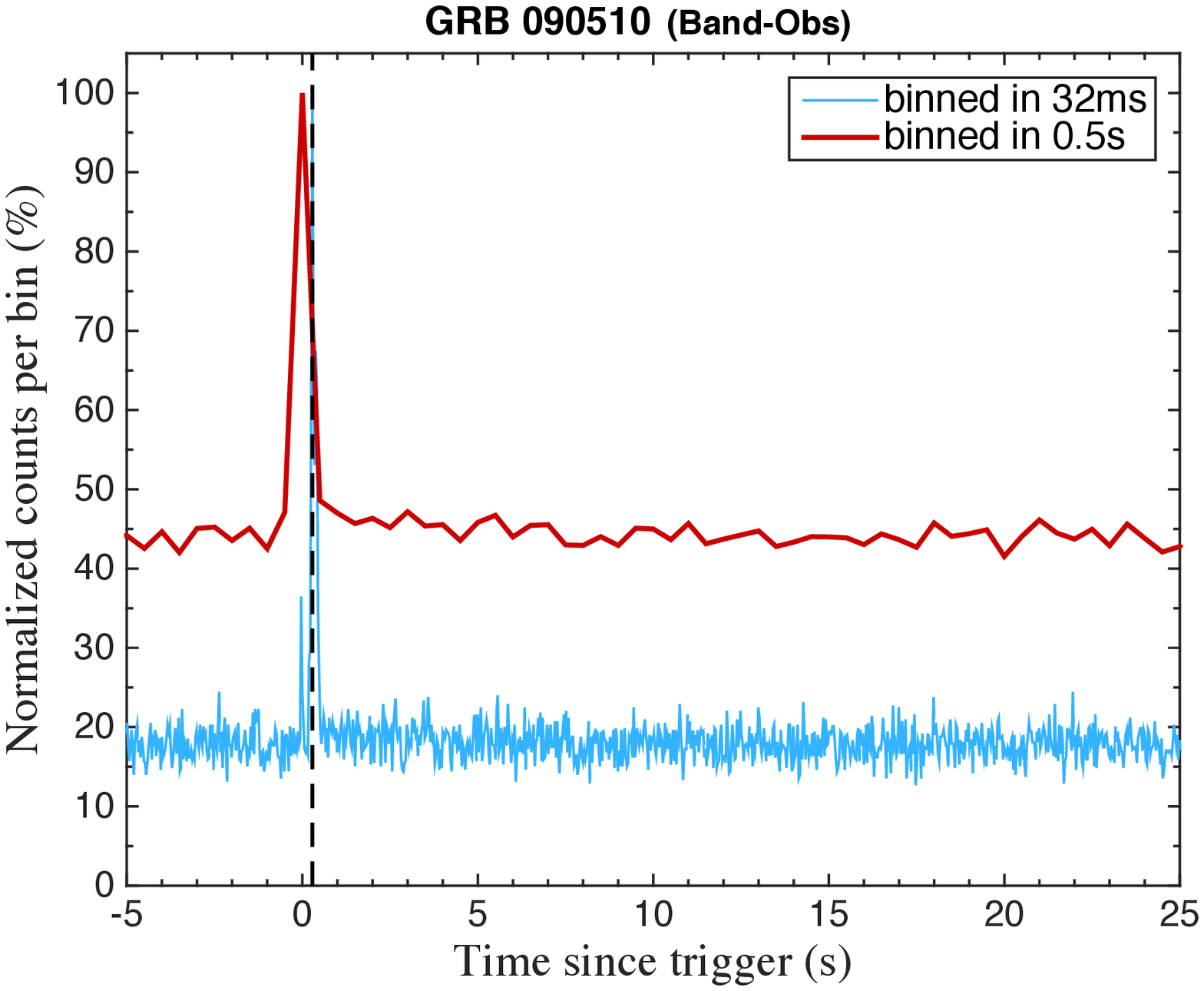}
		\includegraphics[width=0.90\linewidth]{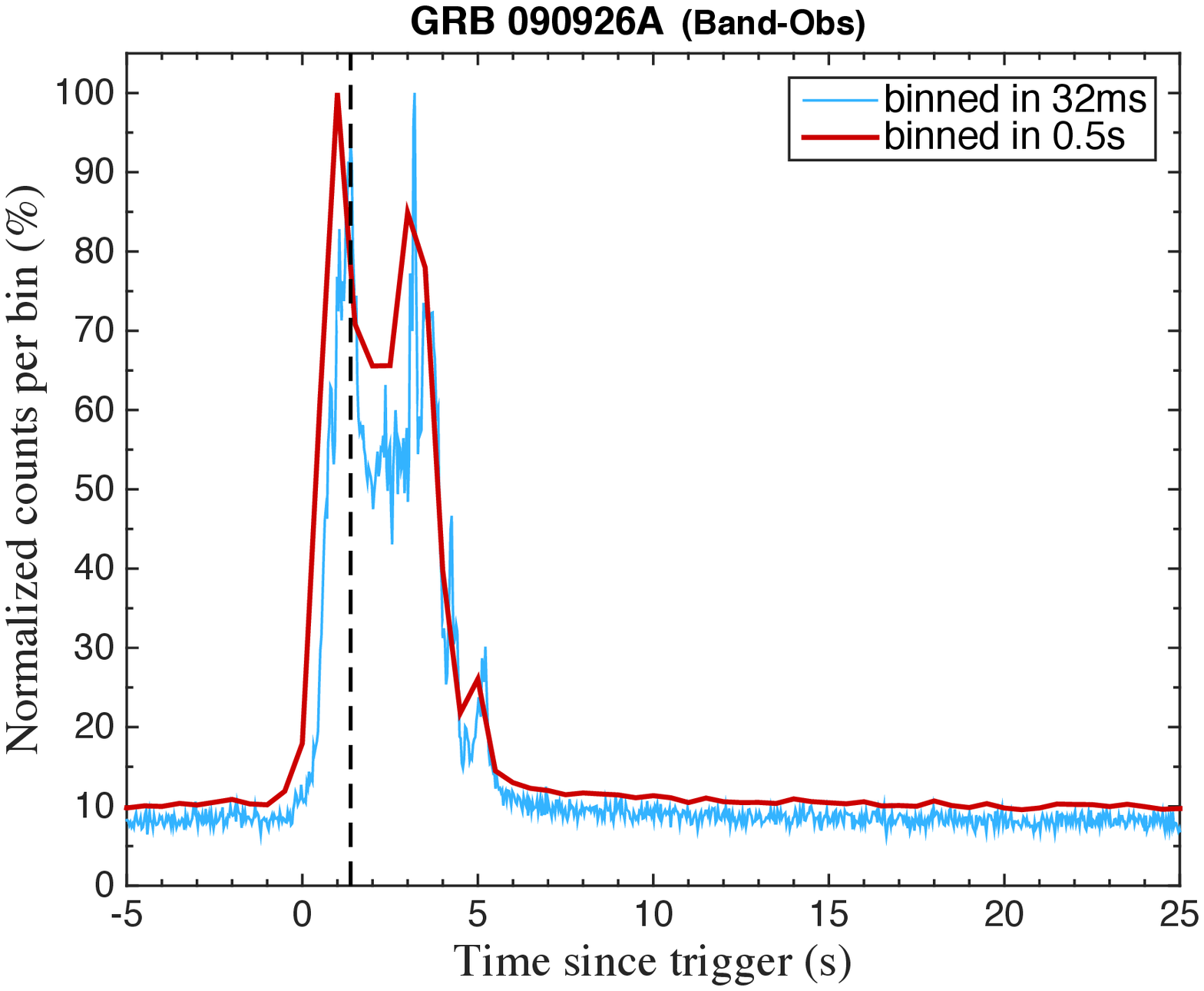}
		\includegraphics[width=0.90\linewidth]{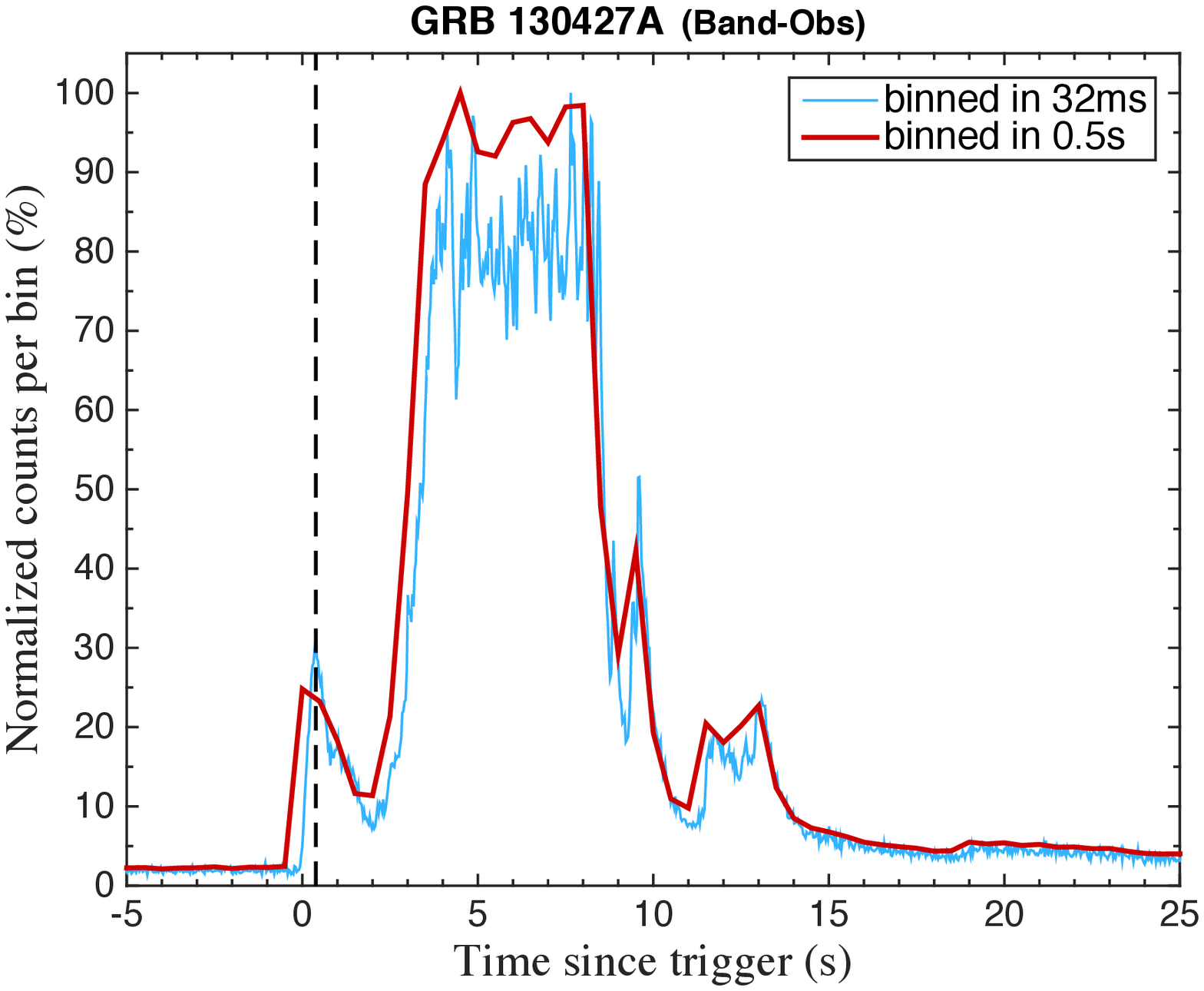}
		\includegraphics[width=0.90\linewidth]{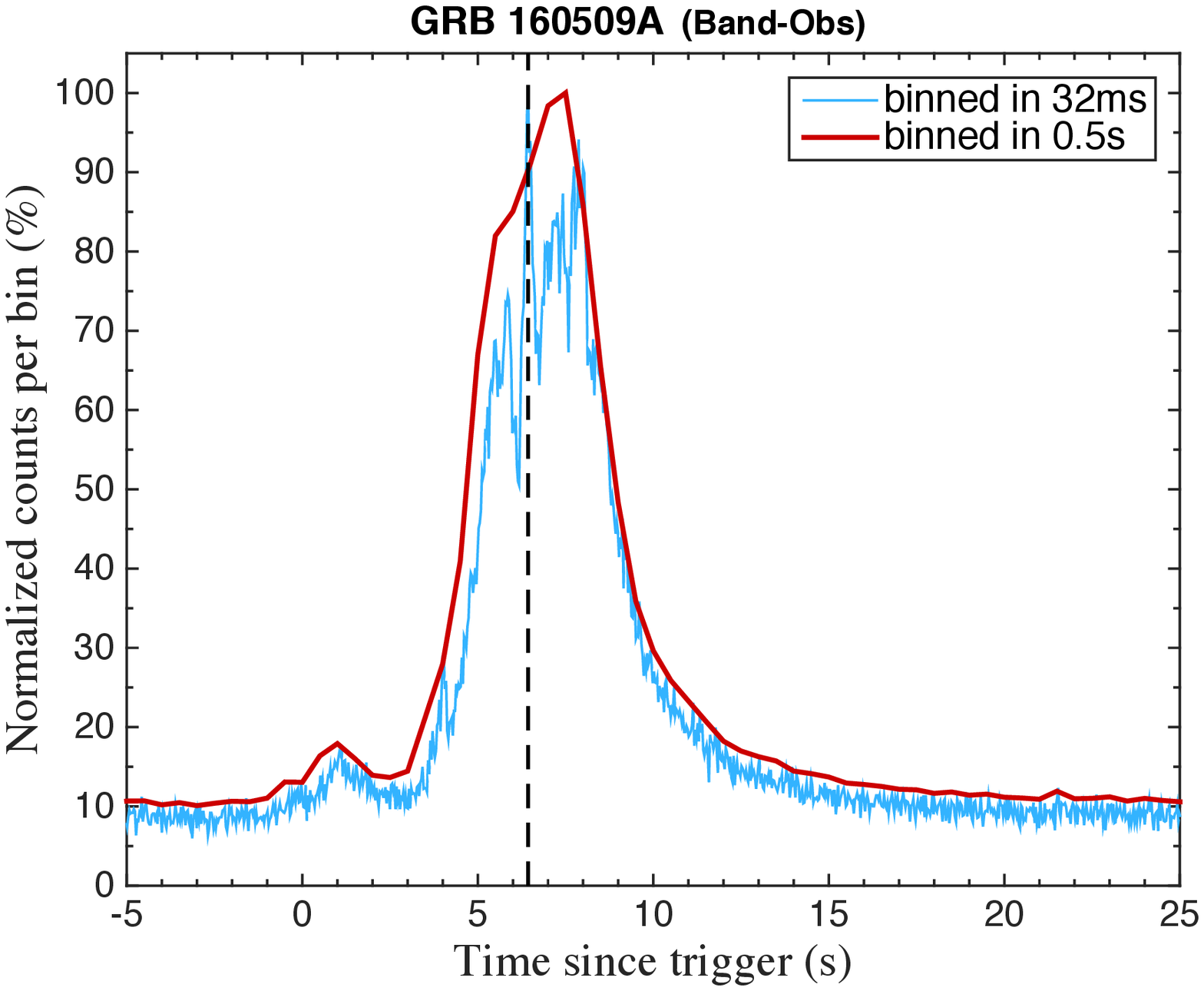}
		
	\end{minipage}
	\caption{The light curves in Band-Obs of 8 GRBs. The $x$-axis refers to the time $t_{\rm obs}/(1+z)$. The thin (blue) curves are the light curves binned in 0.032~s and the thick (red) curves are those binned in 0.5~s. The $y$-axis is the normalized counts, i.e., the counts per bin divided by the maximum counts per bin of the corresponding light curve. The vertical dashed lines (black) refer to the peaks we choose for each GRB.}\label{Band-obs--Time-obs}
\end{figure*}

In order to combine different GRBs, a more plausible way to analyze the data is to set the time axis to $t_{\rm low,obs}/(1+z)$, because that refers to the time at source.
Thus, to be specific, we use the data from two triggered NaI detectors with most detected photons and bin their events in 0.5 second to find the first main peak and choose the highest point in the light curve binned in 32~ms around this peak as the characteristic time for low energy photons%~(i.e., $t_{\rm low,obs}$)
\footnote{Remark: In this subsection, we use the peaks chosen in refs.~\cite{xhw,xhw160509}. More options will be discussed later.}.
In this way, with both wide bins (0.5~s) and narrow bins (0.032~s), we can avoid stochastic fluctuations of the light curve when the GRB is not very bright (see fig.~\ref{Band-obs--Time-obs} and table~\ref{tab:tlow-bandobs}).

\begin{table}[htbp]
	\centering
	\caption{$t_{\rm low,obs}/(1+z)$ chosen from light curves in Band-Obs.}
	\label{tab:tlow-bandobs}%
	\begin{tabular*}{0.48\textwidth}{@{\extracolsep{\fill}}ccc}
		\hline
		\hline
		GRB   & $z$ &   $t_{\rm low,obs}/(1+z)$~(s) \\
		\hline
		080916C & 4.35 & 0.480  \\
		090510 & 0.903   & 0.288  \\
		090902B & 1.822   & 3.456  \\
		090926A & 2.1071  & 1.376  \\
		100414A & 1.368  & 0.096  \\
		130427A & 0.3399   & 0.384  \\
		140619B & 2.67    & -0.032  \\
		160509A & 1.17    & 6.432  \\
		\hline
	\end{tabular*}%
\end{table}%

\begin{figure}[htbp]
	\centering
	\includegraphics[width=1\linewidth]{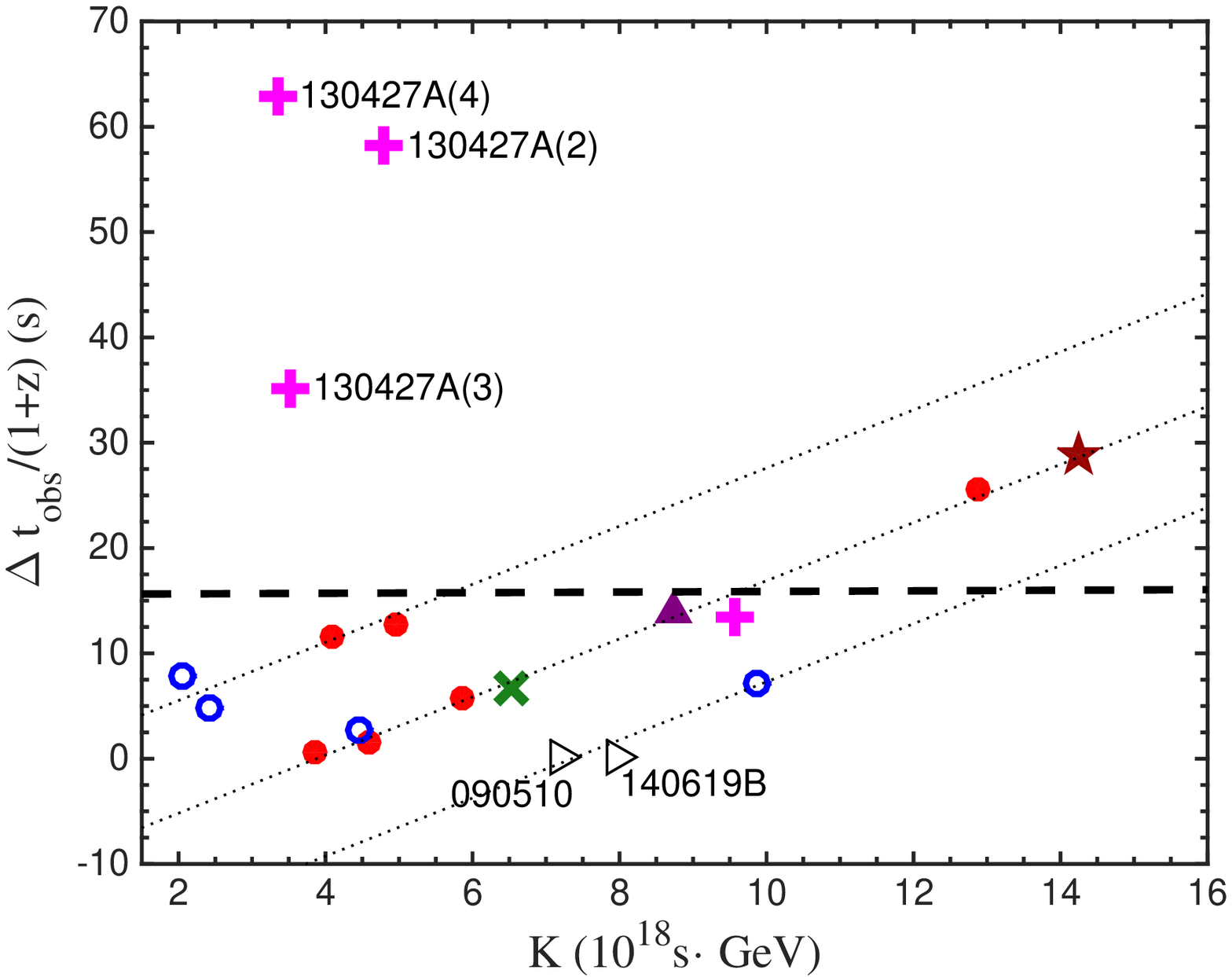}
	\caption{The $\Delta t_{\rm obs}/(1+z)$-$K$ plot for all the events in table~\ref{tab:phinfo}. Two hollow triangles refer to two events from short bursts, while the others come from long bursts. We obtain this plot by analyzing the light curves in Band-Obs. The hollow circles refer to events of GRB 080916C. The solid circles refer to events of 090902B. The plus signs refer to events of GRB 130427A. The cross refers to the event 090926A. The solid triangle refers to 100414A and the star refers to 160509A. The three parallel dotted lines are suggested in refs.~\cite{xhw,xhw160509}. The dashed line is the fit of all points.}\label{allphotonsdiag}
\end{figure}

Combining $t_{\rm low,obs}$ in table~\ref{tab:tlow-bandobs} with $t_{\rm high,obs}$ and $K$ in table~\ref{tab:phinfo}, we can obtain $K$ and $\Delta t_{\rm obs}/(1+z)$ using eqs.~\ref{eq:t} and \ref{eq:appro-K} for each energetic photon event, and then draw the $\Delta t_{\rm obs}/(1+z)$-$K$ plot as shown in fig.~\ref{allphotonsdiag}. From this figure, we notice that if we ignore 130427(2), 130427(3) and 130427(4), the rest events roughly fall on three parallel lines~(dashed lines in fig.~\ref{allphotonsdiag}). 9 events~(080916C(2), 090902B(1), 090902B(2), 090902B(4), 090902B(6), 090926A, 100414A, 130427A(1) and 160509A) fall surprisingly on the middle line, which is called the ``mainline" in refs.~\cite{xhw,xhw160509}.
Besides, 4 events~(080916C(3), 080916C(4), 090902(3) and 090902(5)) fall on the upper line and 3 events~(080916C(1), 090510, and 140619B) fall on the lower line.
It is noticed in ref.~\cite{xhw160509} that the events on the lower line have relatively higher energies (in the source frame) while the events on the upper line roughly have lower energies.
An assumption may be made that photons with different intrinsic energies have different intrinsic time lag statistically, and therefore we introduce three lines to fit higher, medium, and lower energy photons.
That is to say, for each line, it is assumed that the events on this line may share a same intrinsic time lag, and the slope of the line represents $1/E_{\rm LV}$ according to sect.~\ref{method}. Different parallel lines imply a same $E_{\rm LV}$ with different intrinsic time lags.

However, in our analysis, three events, 130427(2), 130427(3) and 130427(4), which were not included in former analyses in refs.~\cite{xhw,xhw160509}, fall far from the three lines suggested by those works. Thus, the 3 parallel lines seem less plausible, though all the long bursts have events (including 130427(1)) on the mainline. As there is no convincing reason to ignore the three events off three lines, we fit all points with a single line and obtain the slope $(0\pm3)\times 10^{-18} \rm GeV^{-1}$, from which we obtain a lower bound on $E_{\rm LV}$ with $|E_{\rm LV}|\ge 3\times 10^{17}$~GeV.

\section{Other criteria}\label{options}

One remaining question is how to select the characteristic time for low energy photons from more convincing viewpoints.
In the discussion above, we select the first main peak of the light curve.
That is to say, we choose the time when the density of photons is sharp.
We need to offer more options on the selection of $t_{\rm low}$ to explore whether $t_{\rm low}$ determined above represents a characteristic mechanism. If the mechanism at $t_{\rm low}$ is significant, we should expect that it not only results in notable changes in photon numbers but also changes in other aspects such as the energy distribution. If the characteristic times determined from different viewpoints are consistent with each other, we may assume that there is an important process at around $t_{\rm low}$ which results in significant phenomena in different aspects.

In this section, we offer other criteria to determine the characteristic time for low energy photons. As before, our data is obtained from two GBM NaI detectors with most detected photons.

In consideration of the fact that the light curve shows the number of photons but not the energy released from source, our next method is to plot the whole energy received per bin within a low energy band versus time, and then choose the first main peak of this ``energy curve" as the characteristic time.

Another method is to calculate the average energy per photon in the low energy band and draw the average energy versus time plot. This method is based on a speculation that the energy distribution may change severely during a characteristic process so that the average energy per photon also changes a lot.
In this way, the first significant peak (or dip) of the average energy curve~(the plot of average energy per photon versus time) can represent the change point of the energy distribution and thus a characteristic time of the GRB.

At last, we offer a new criterion based on both light curve and average energy curve in subsection~\ref{c3}.

\subsection{Criterion 1: Energy received in a certain band}\label{c1}
In this subsection, we focus on the total energy received per bin within the low energy band.
We choose the band, Band-Obs, and select all the photons in this energy band. Then the total energy of selected photons in each time bin is calculated.
The energy resolution of NaI detectors varies with the energy. Within Band-Obs, the energy resolution is $\sim 1~\rm keV$. We choose the mean value of the energy bin in which the photon falls as the energy of a photon.
The energy received per bin is used to draw an energy curve and the characteristic time of low energy photons is chosen as the first main peak of the energy curve.
%Considering the detector performance, %and the fact that it is more reasonable to discuss in the source frame
%we analyze the energy curves for Band-Obs, Band-I and Band-II separately and determine their characteristic times, as shown in table~\ref{tab:energycurve}, figs.~\ref{energyobs}, \ref{energyI} and \ref{energyII}.
For the sake of uniformity, we bin the energy curve in 0.5~s and 0.032~s, and the time axis refers to $t_{\rm low,obs}/(1+z)$.

\begin{table}[htbp]
	\centering
	\caption{The $t_{\rm low,obs}/(1+z)$ determined by the energy criterion (Criterion 1).}
	\label{tab:energycurve}%
	\begin{tabular*}{0.48\textwidth}{@{\extracolsep{\fill}}ccc}
		\hline
		\hline
		
		GRB &   z    &$t_{\rm low,obs}/(1+z)$ \\
		\hline
		080916C & 4.35  & 0.256  \\
		090510 & 0.903 & 0.288  \\
		090902B & 1.822 & 3.456 \\
		090926A & 2.1071 & 1.376  \\
		100414A & 1.368 & 0.096 \\
		130427A & 0.3399 & 0.384   \\
		140619B & 2.67  & 0.096 \\
		160509A & 1.17  & 6.432  \\
		\hline
	\end{tabular*}%
\end{table}%

\begin{figure*}[htbp]
	\centering
	\begin{minipage}[t]{0.48\linewidth}
		\centering
		\includegraphics[width=0.90\linewidth]{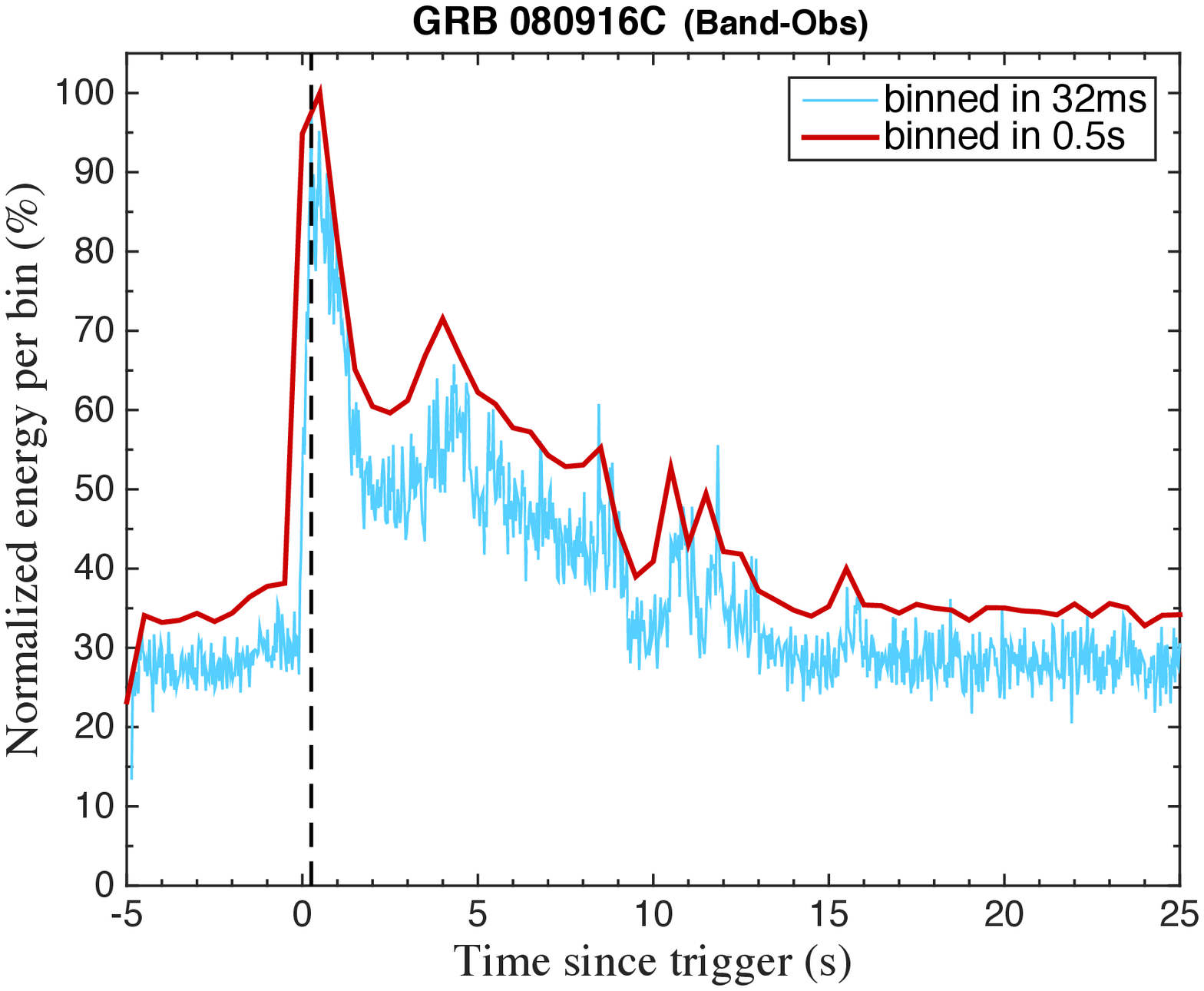}
		\includegraphics[width=0.90\linewidth]{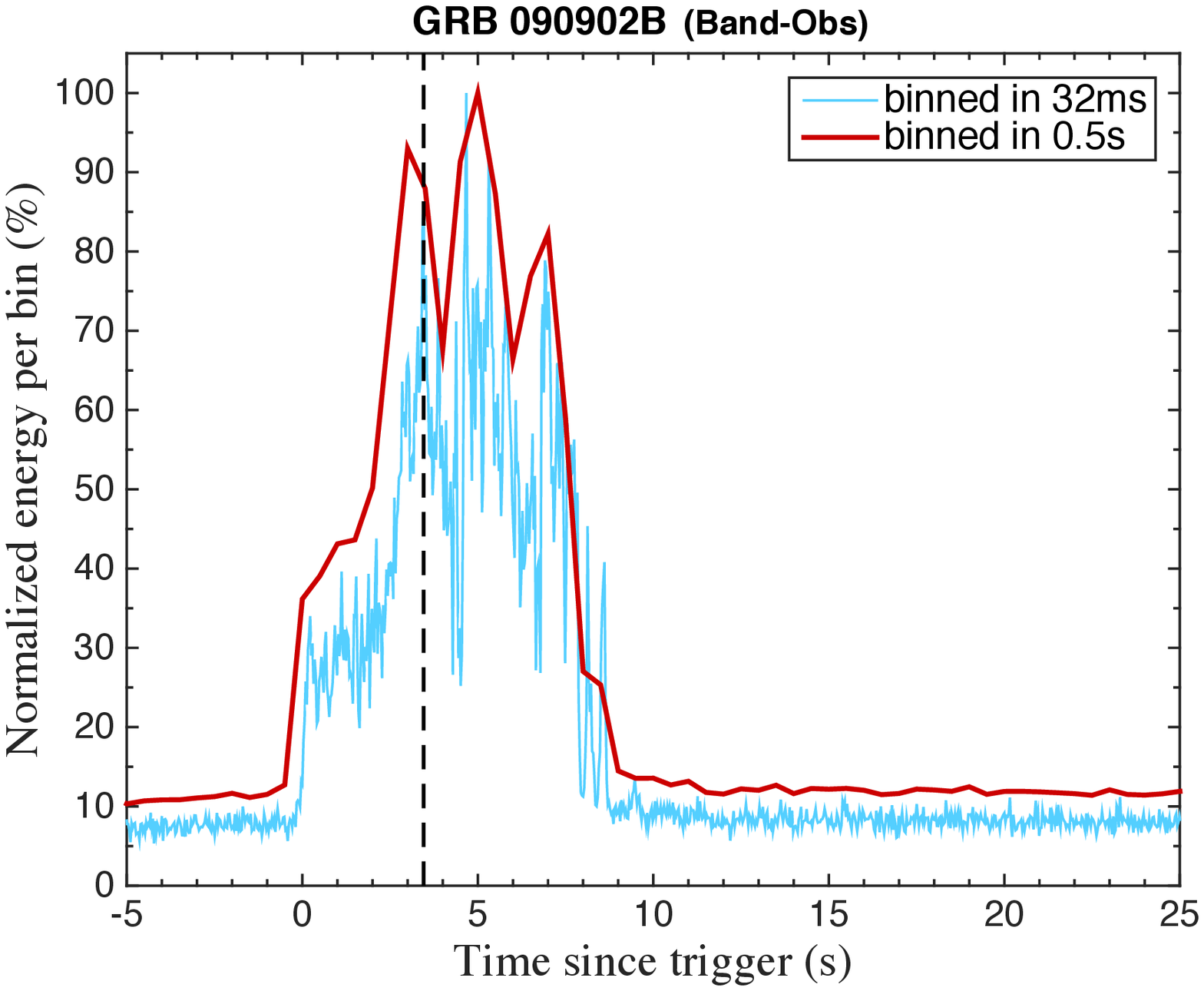}
		\includegraphics[width=0.90\linewidth]{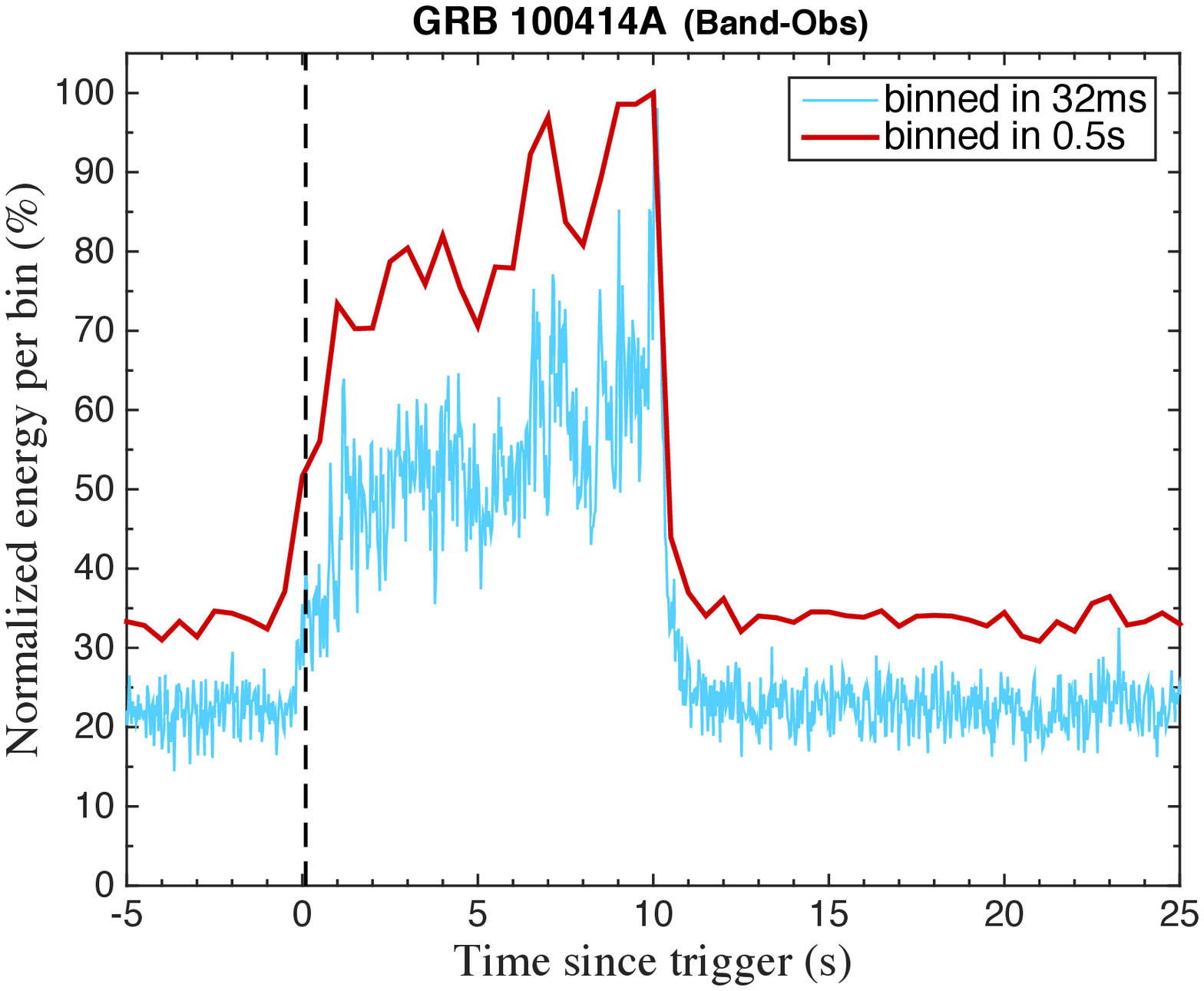}
		\includegraphics[width=0.90\linewidth]{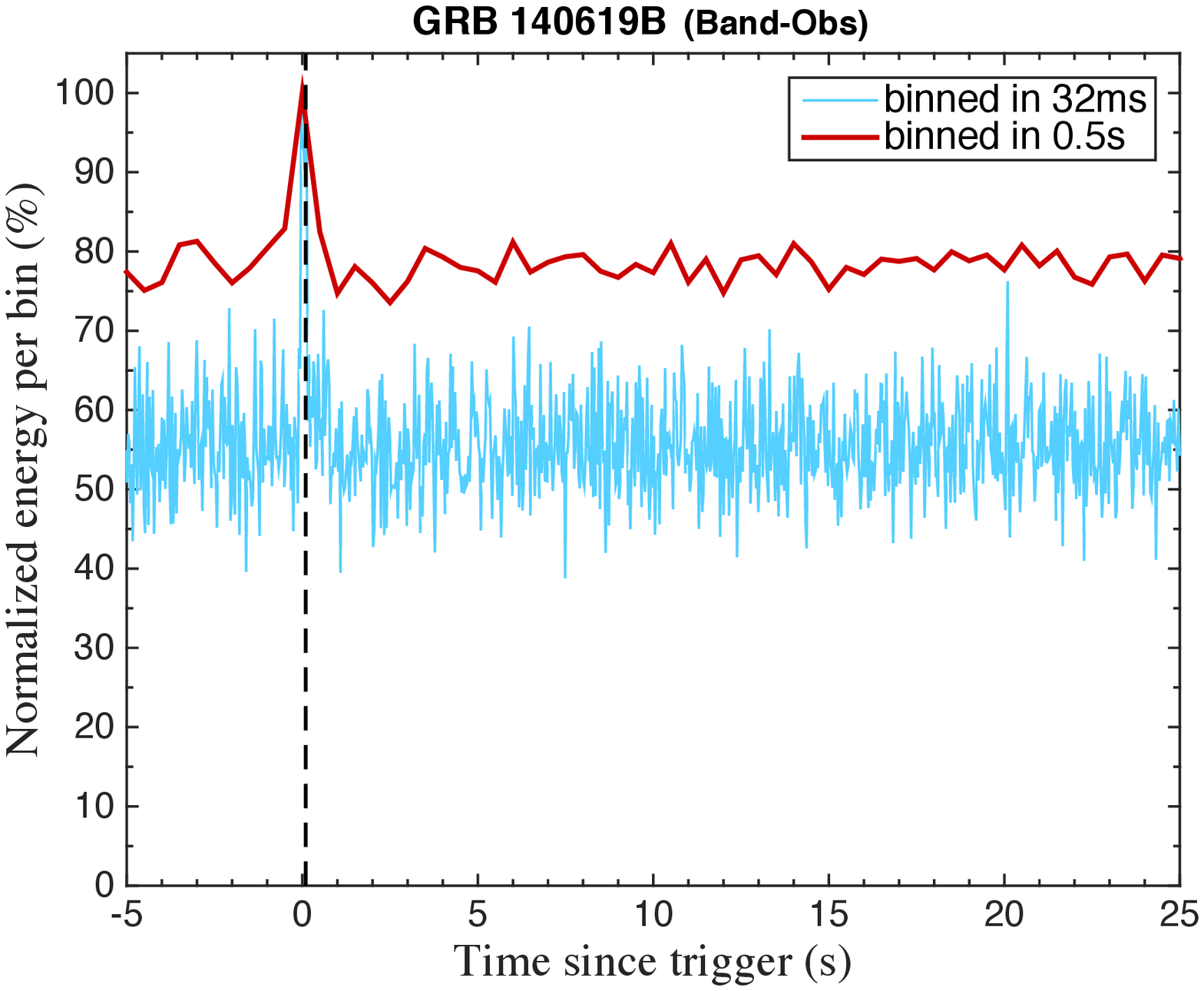}		
	\end{minipage}
	\begin{minipage}[t]{0.48\linewidth}
		\centering
		\includegraphics[width=0.90\linewidth]{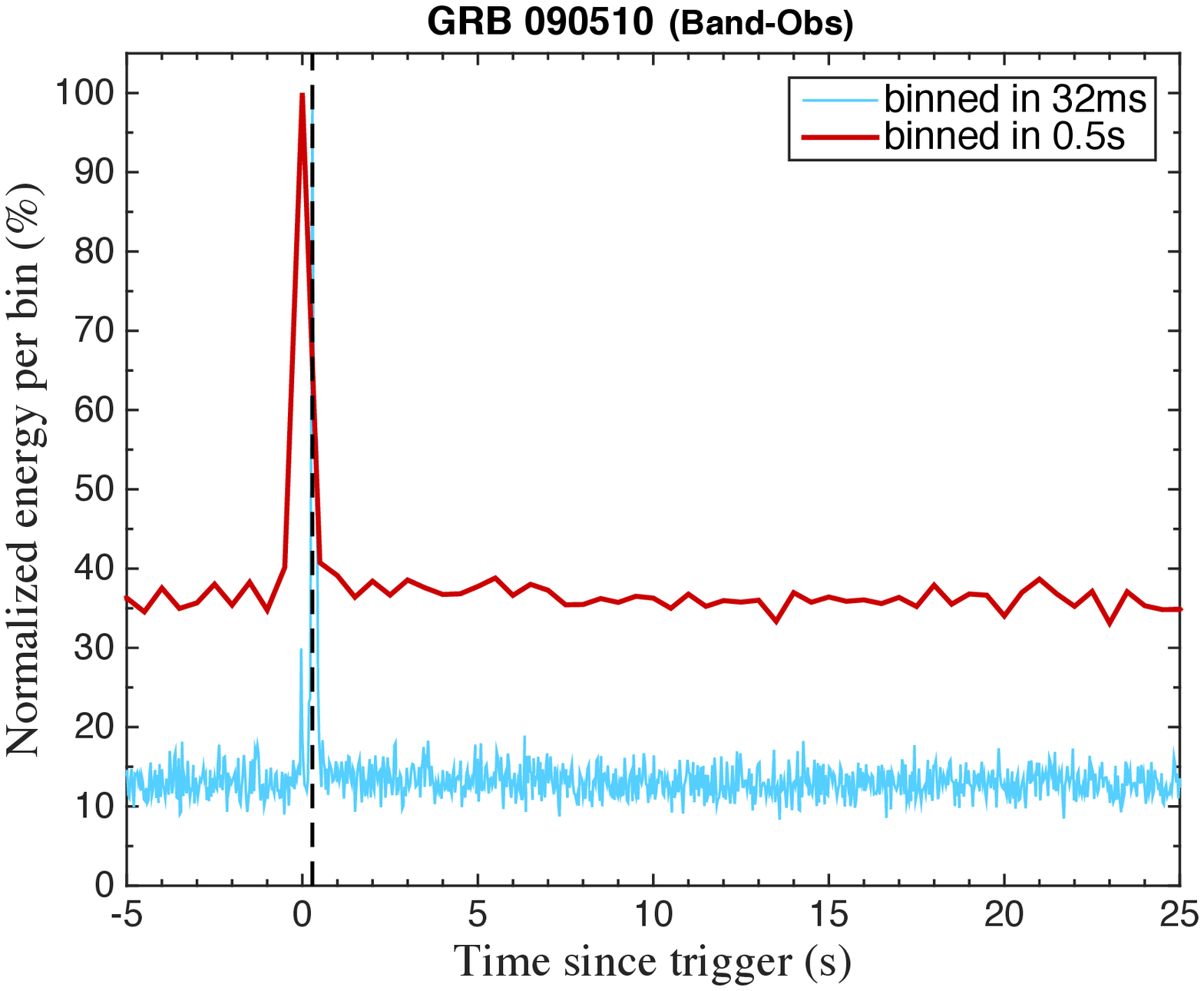}
		\includegraphics[width=0.90\linewidth]{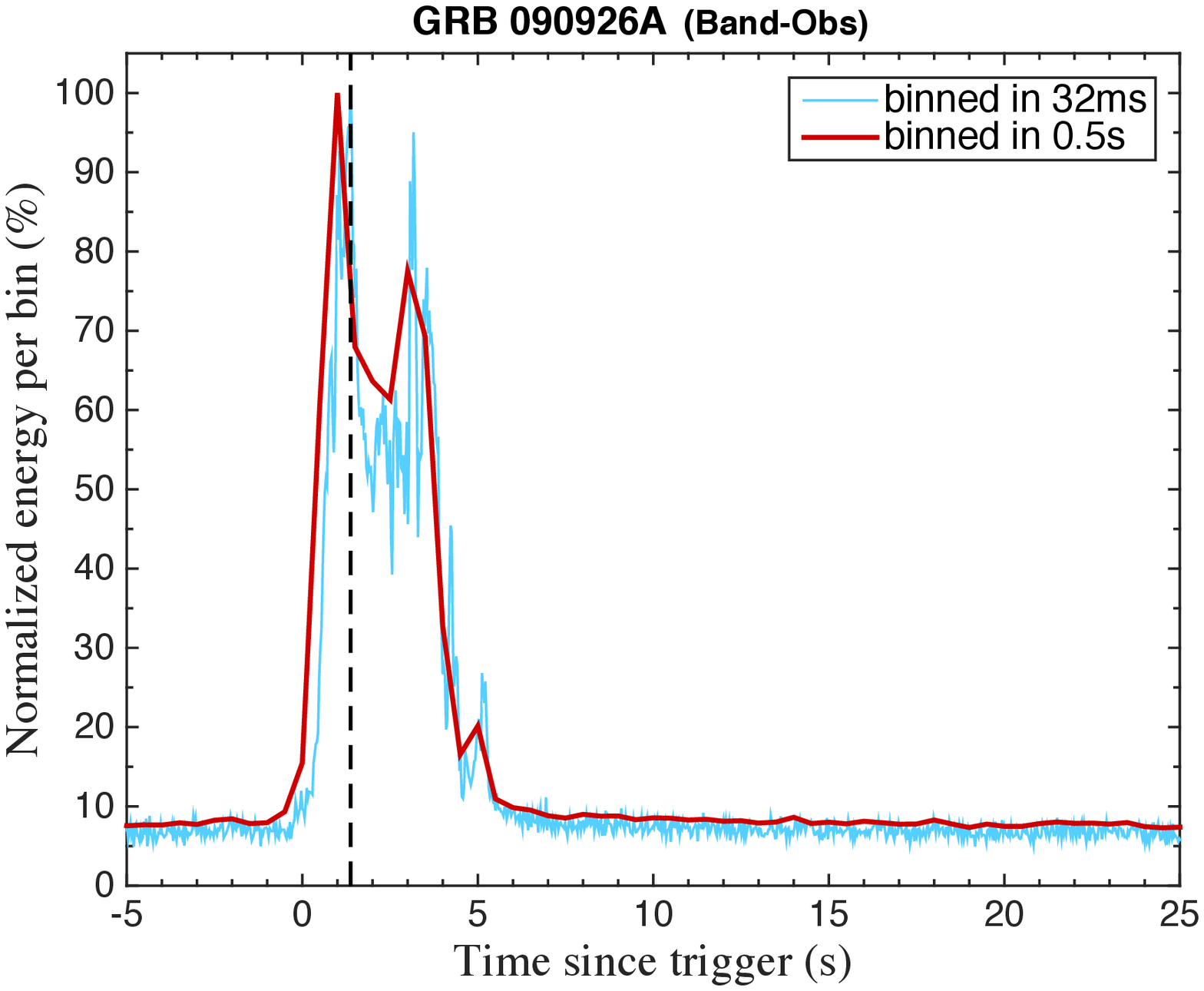}
		\includegraphics[width=0.90\linewidth]{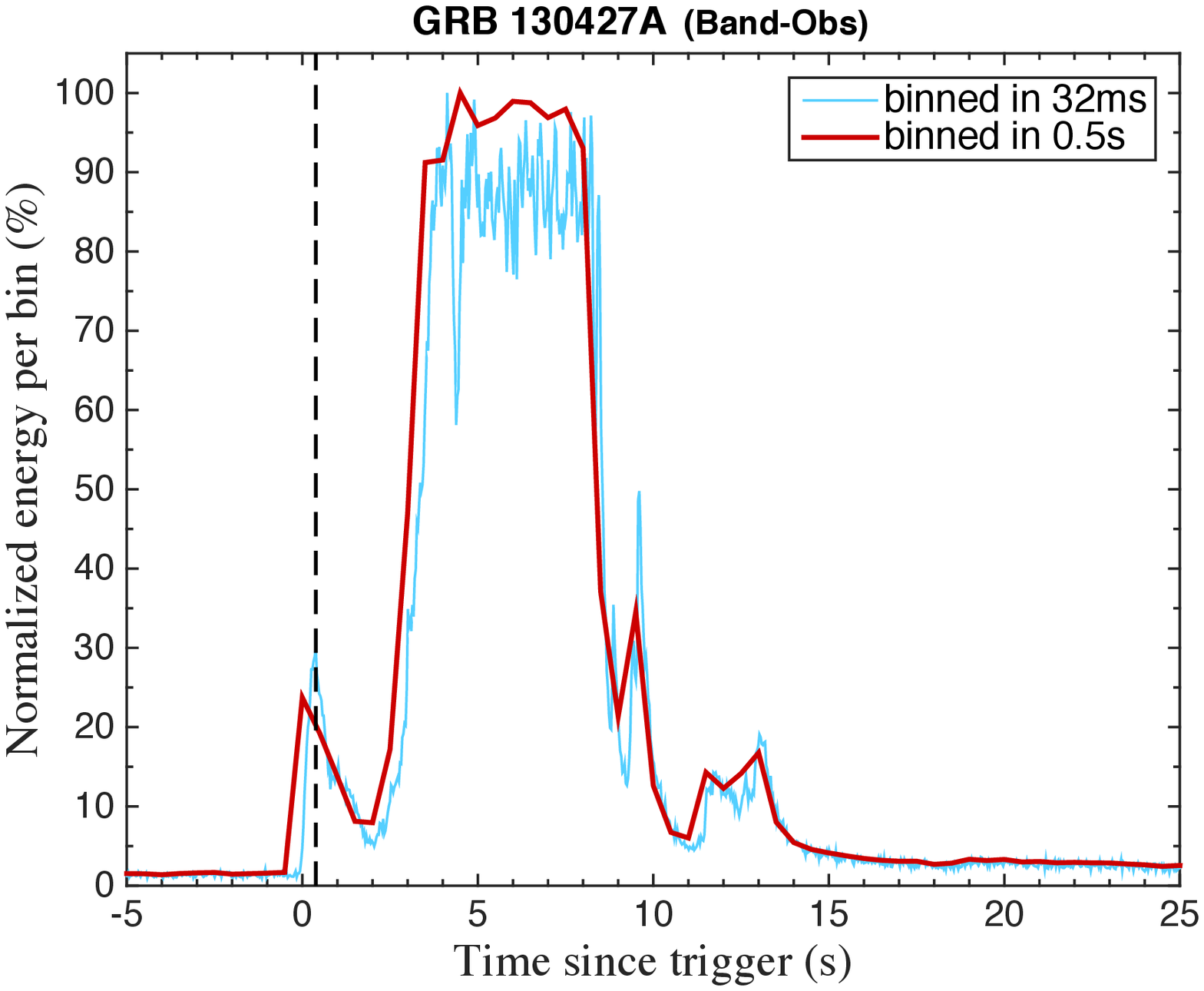}
		\includegraphics[width=0.90\linewidth]{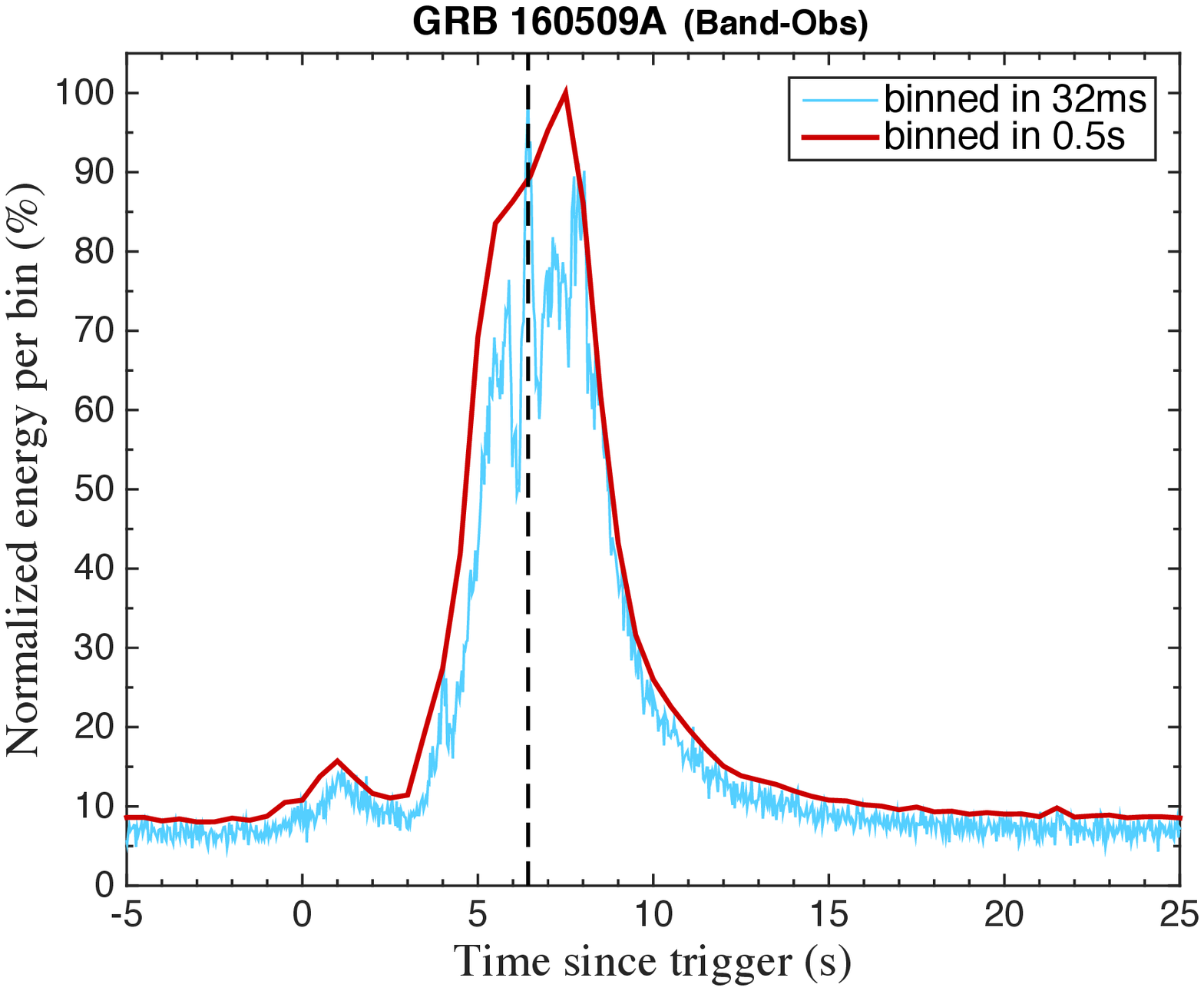}
		
	\end{minipage}
	\caption{The energy curves~(energy received per bin) in Band-Obs of 8 GRBs. The $x$-axis is $\Delta t_{\rm obs}/(1+z)$. The thin curves (blue) are the energy curves binned in 0.032~s and the thick (red) curves are those binned in 0.5~s. The $y$-axis is the normalized energy received, i.e., the energy received in Band-Obs per bin divided by the maximum energy received per bin of the corresponding energy curve. The vertical dashed lines (black) refer to the first main peaks we choose for each GRB.}\label{energyobs}
\end{figure*}

The energy curves (see fig.~\ref{energyobs})show that they are quite similar to the light curves. Thus, the first main peaks of the energy curves have counterparts in the light curves, i.e., these $t_{\rm low,obs}/(1+z)$ determined by this energy criterion (see Table \ref{tab:energycurve}) are similar to those determined by light curves.
It is easy to understand, as more photons are often associated with more energy received.
However, as the energy distribution of photons may change over time, the energy received is not necessarily proportional to the number of photons.

\begin{figure}[htbp]
	\centering
	\includegraphics[width=1\linewidth]{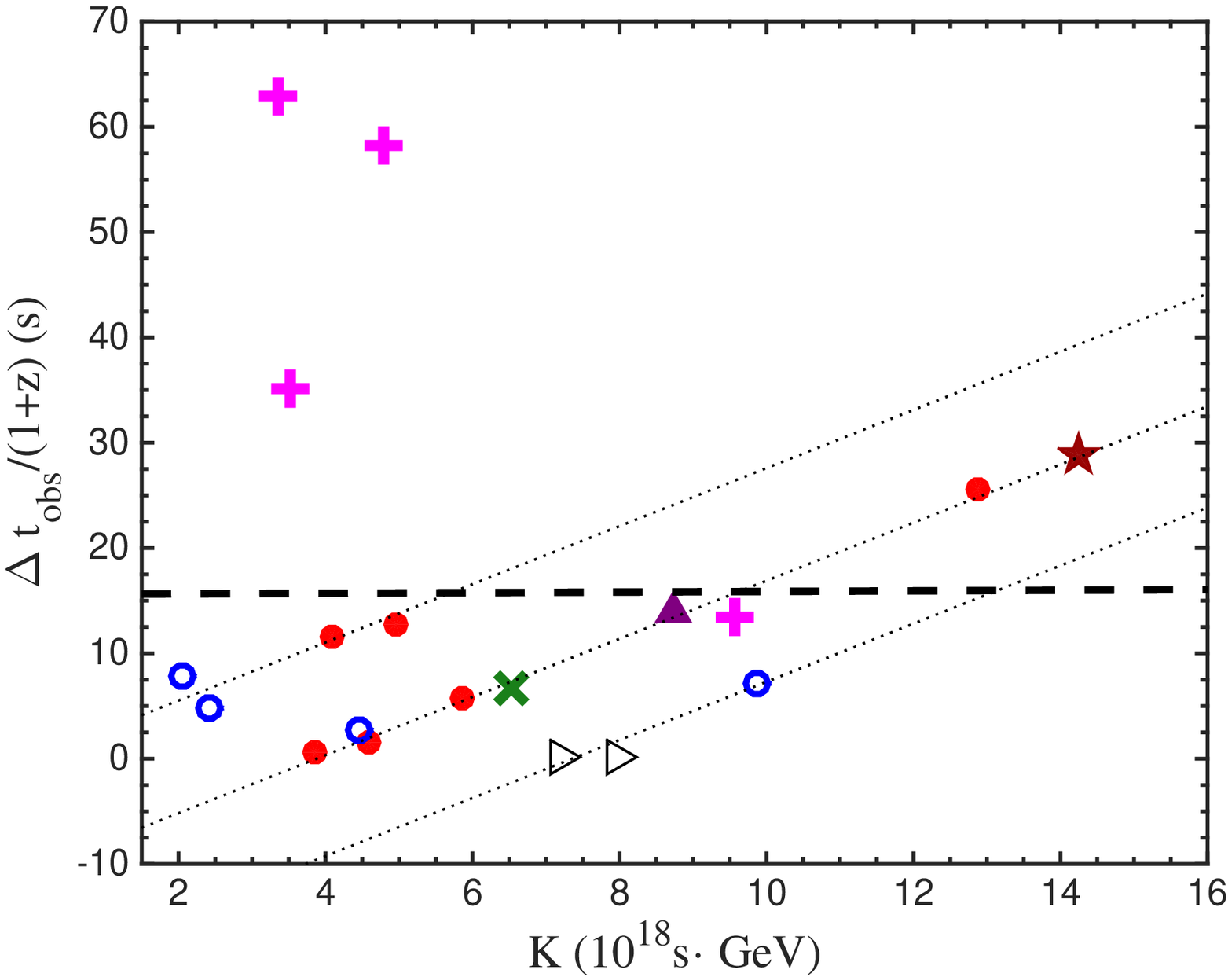}
	\caption{The $\Delta t_{\rm obs}/(1+z)$-$K$ plot for Band-Obs using Criterion 1. The dashed line marks the fit of  all points, while the dotted lines were suggested in refs.~\cite{xhw,xhw160509}.}\label{fig:energy}
\end{figure}

This criterion focuses on the tensity of the GRB.
%The plots of $\Delta t_{\rm obs}/(1+z)$-$K$ (fig.~\ref{fig:energy}) still show a strong linear correlation, and the mainline is still clear. The information of linear fitting and the mainline is listed in table~\ref{tab:energy}, which is very similar to table~\ref{tab:counts}. This indicates that the results are hardly affected when we offer Criterion 1 to replace the former method with light curves.
The fit of points on the plots of $\Delta t_{\rm obs}/(1+z)$-$K$ is shown in fig.~\ref{fig:energy}, whose slope is still $(0\pm 3)\times 10^{-18}\rm~GeV^{-1}$, showing a consistency with fig.~\ref{allphotonsdiag}.

%\begin{table*}[htbp]
%	\centering
%	\caption{Information of the mainline~(fitting 9 events on the mainline under the energy curve criterion).}
%	\begin{tabular}{l|c|c|c|c}
%		\hline
%		\hline
%		Band  & Goodness of fit: $R^2$ & Slope ($10^{18}\rm~GeV^{-1}$) & $y$-intercept (s) & $E_{\rm LV}$ ($10^{17}$~GeV) \\
%		\hline
%		Band-Obs &  0.989174     &   $2.75\pm 0.12$    &   $-10.7\pm 1.1$    & $3.64\pm 0.16$ \\
%		Band-I &    0.990449   &   $2.76\pm 0.11$    &    $-10.7\pm 1.0$   &  $3.62\pm 0.14$\\
%		Band-II &   0.989918    &     $2.75\pm 0.11$  &   $-10.7\pm 1.0$    & $3.64\pm 0.15$ \\
%		\hline
%	\end{tabular}%
%	\label{tab:energy}%
%\end{table*}%

\subsection{Criterion 2: Average energy per photon for low energy photons}\label{c2}
It makes sense to assume that the energy distribution of GRB photons is different from that of the background photons
%**
because of the special mechanism in the GRB source. Therefore, the occurrence of a GRB can cause a significant change to the average energy curve~(average energy per photon during a bin versus time).
Still, we select the photons in a certain energy band from data of two GBM NaI detectors with most detected photons. The average energy of these photons in each time bin is calculated and used for the average energy curve.

\begin{figure*}[htbp]
	\centering
	\begin{minipage}[t]{0.48\linewidth}
		\centering
		\includegraphics[width=0.90\linewidth]{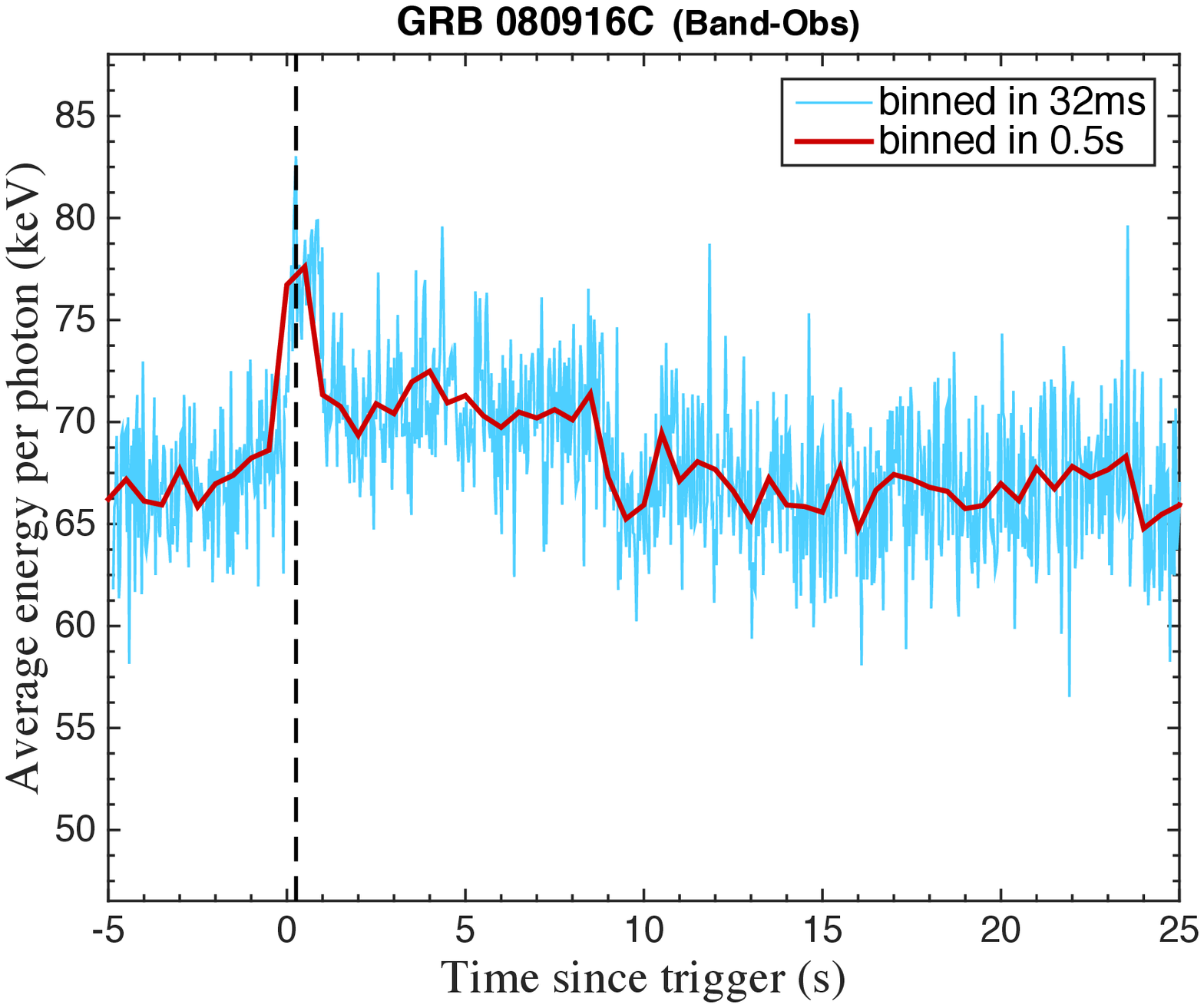}
		\includegraphics[width=0.90\linewidth]{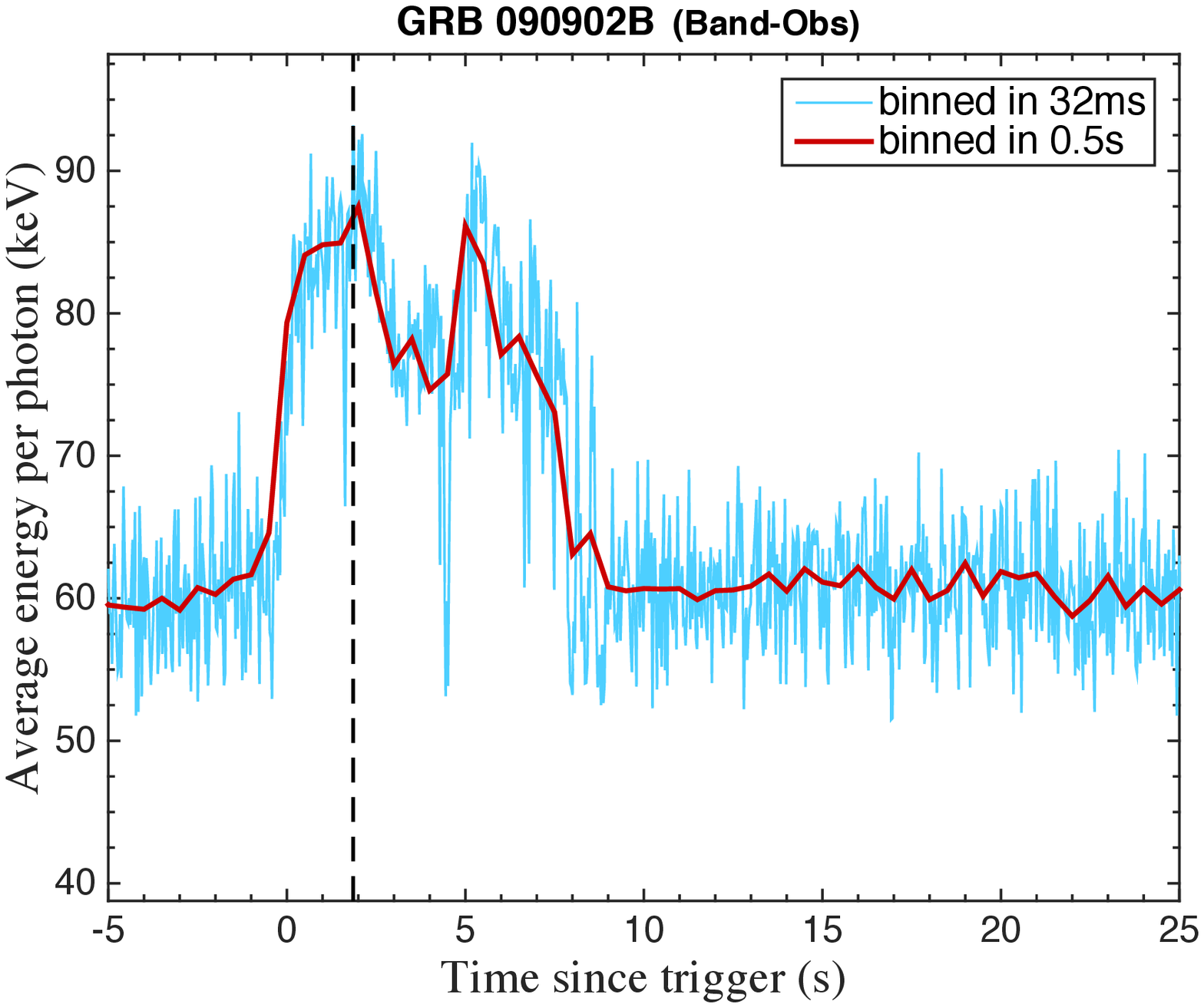}
		\includegraphics[width=0.90\linewidth]{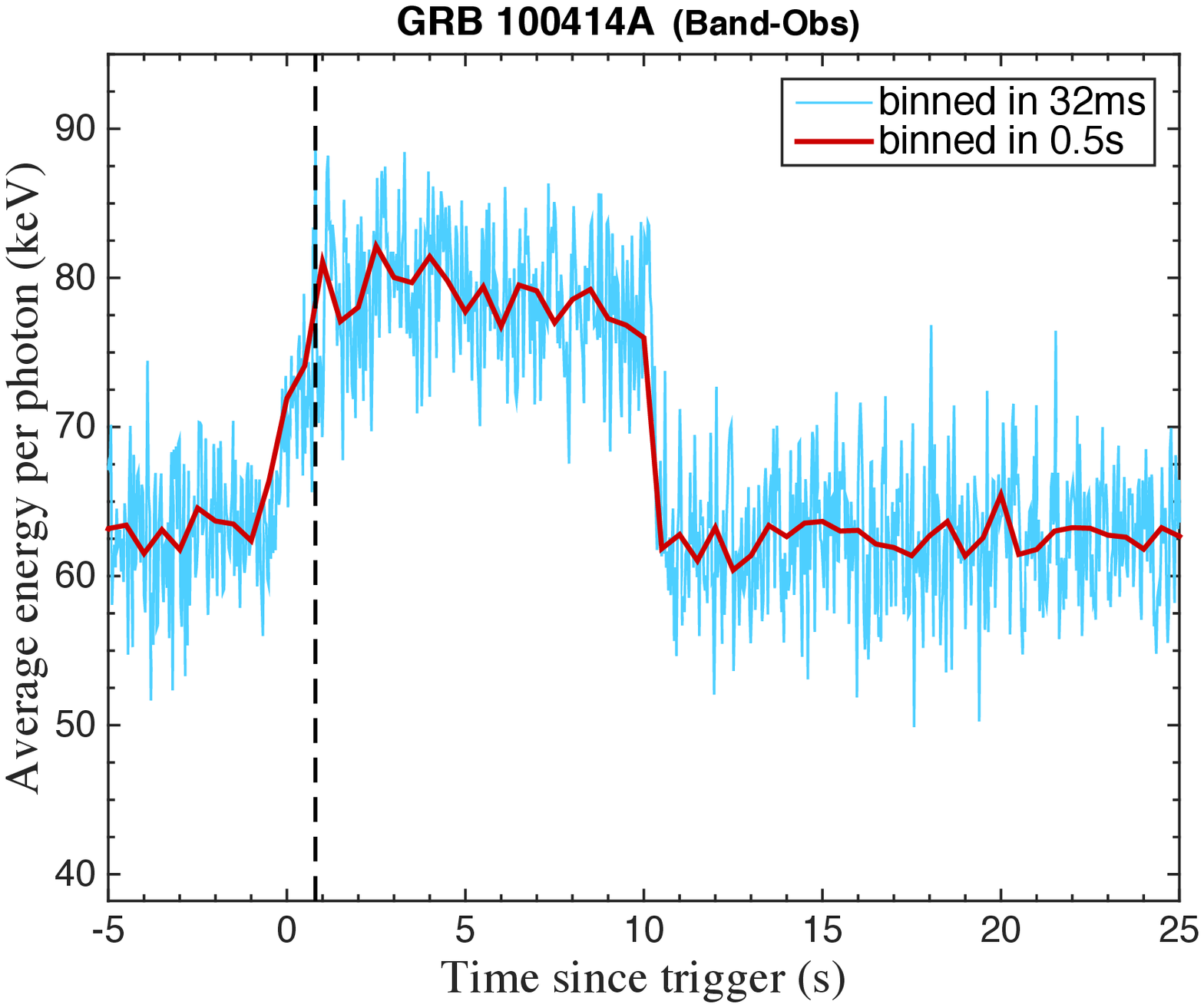}
		\includegraphics[width=0.90\linewidth]{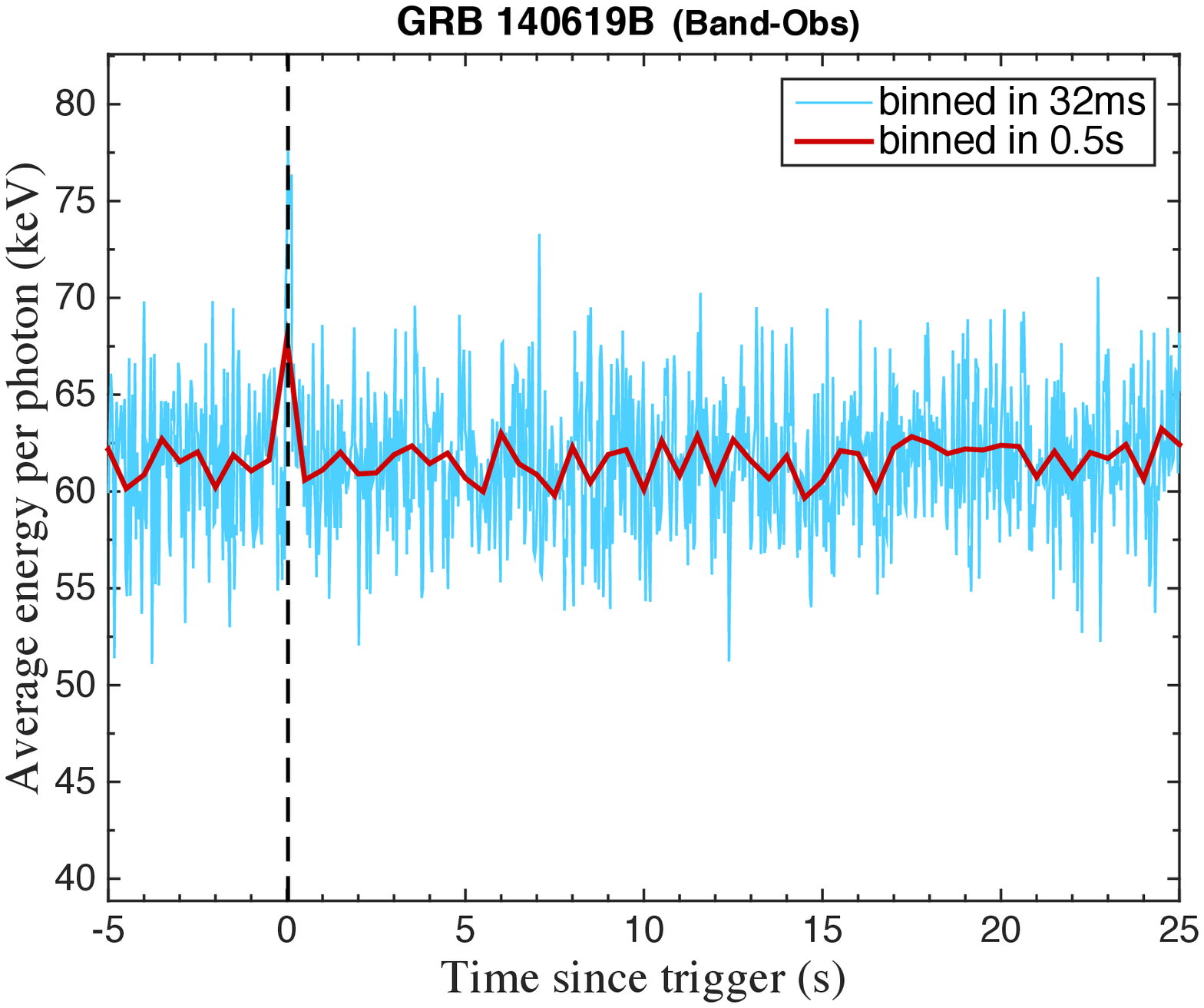}		
	\end{minipage}
	\begin{minipage}[t]{0.48\linewidth}
		\centering
		\includegraphics[width=0.90\linewidth]{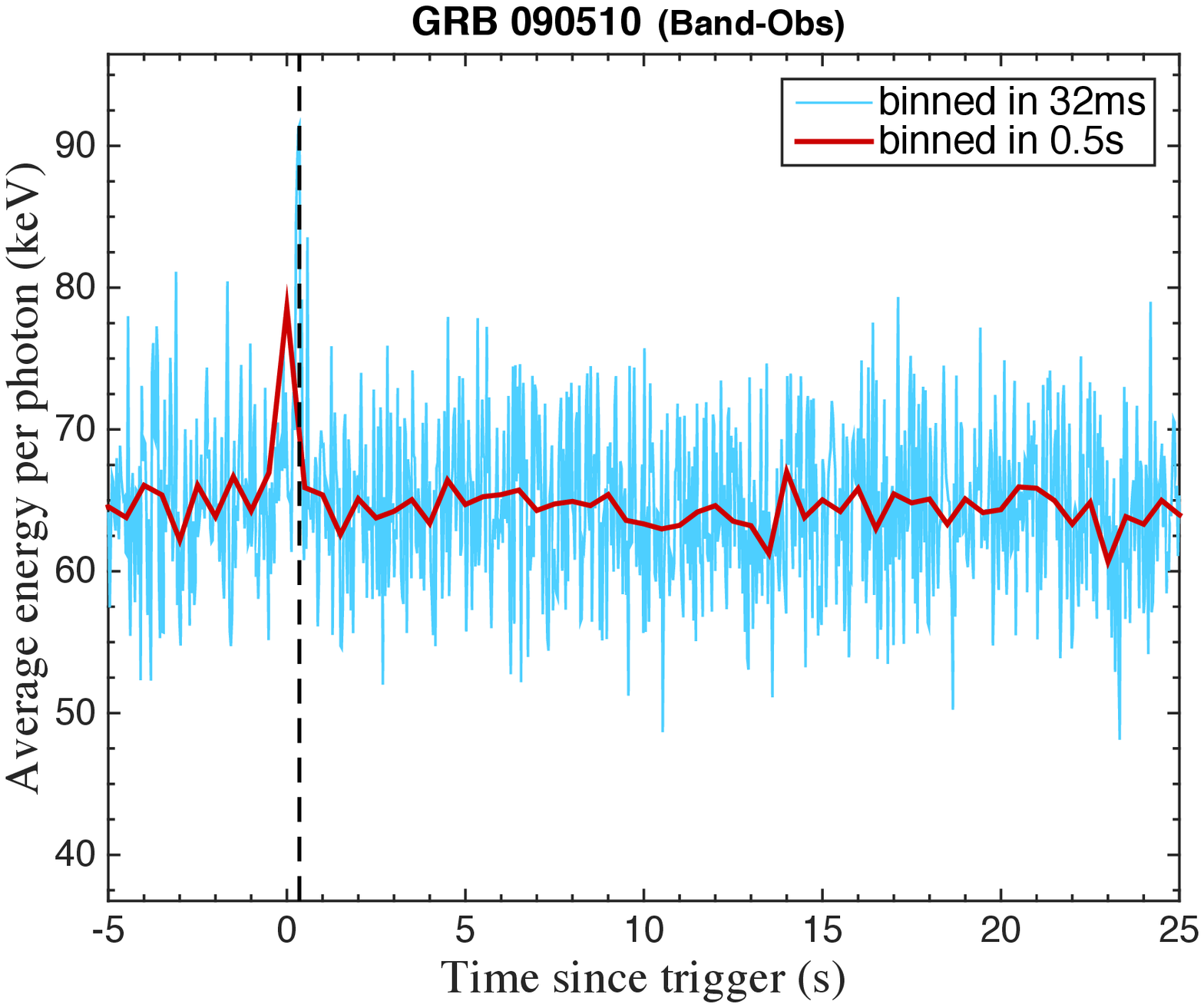}
		\includegraphics[width=0.90\linewidth]{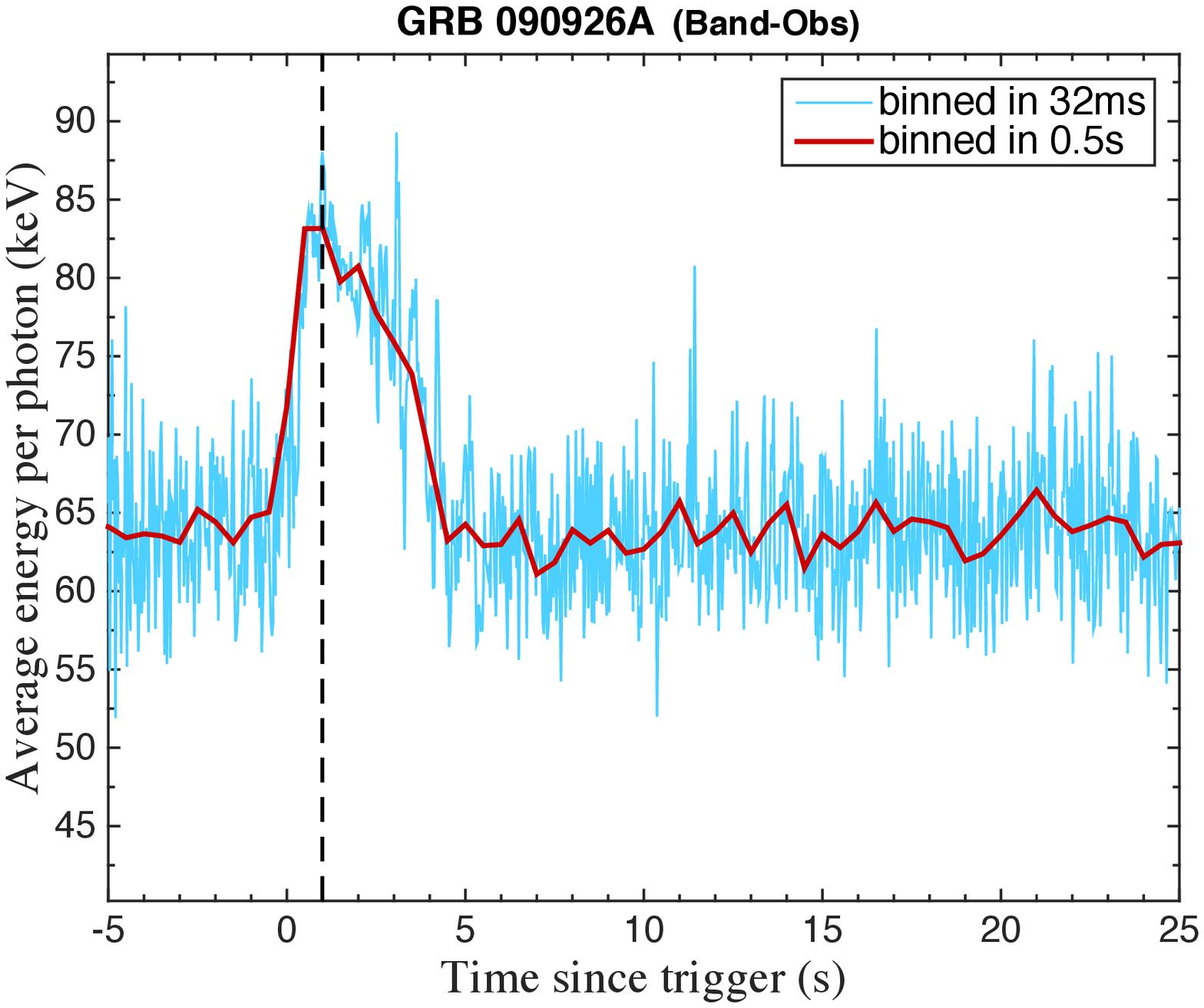}
		\includegraphics[width=0.90\linewidth]{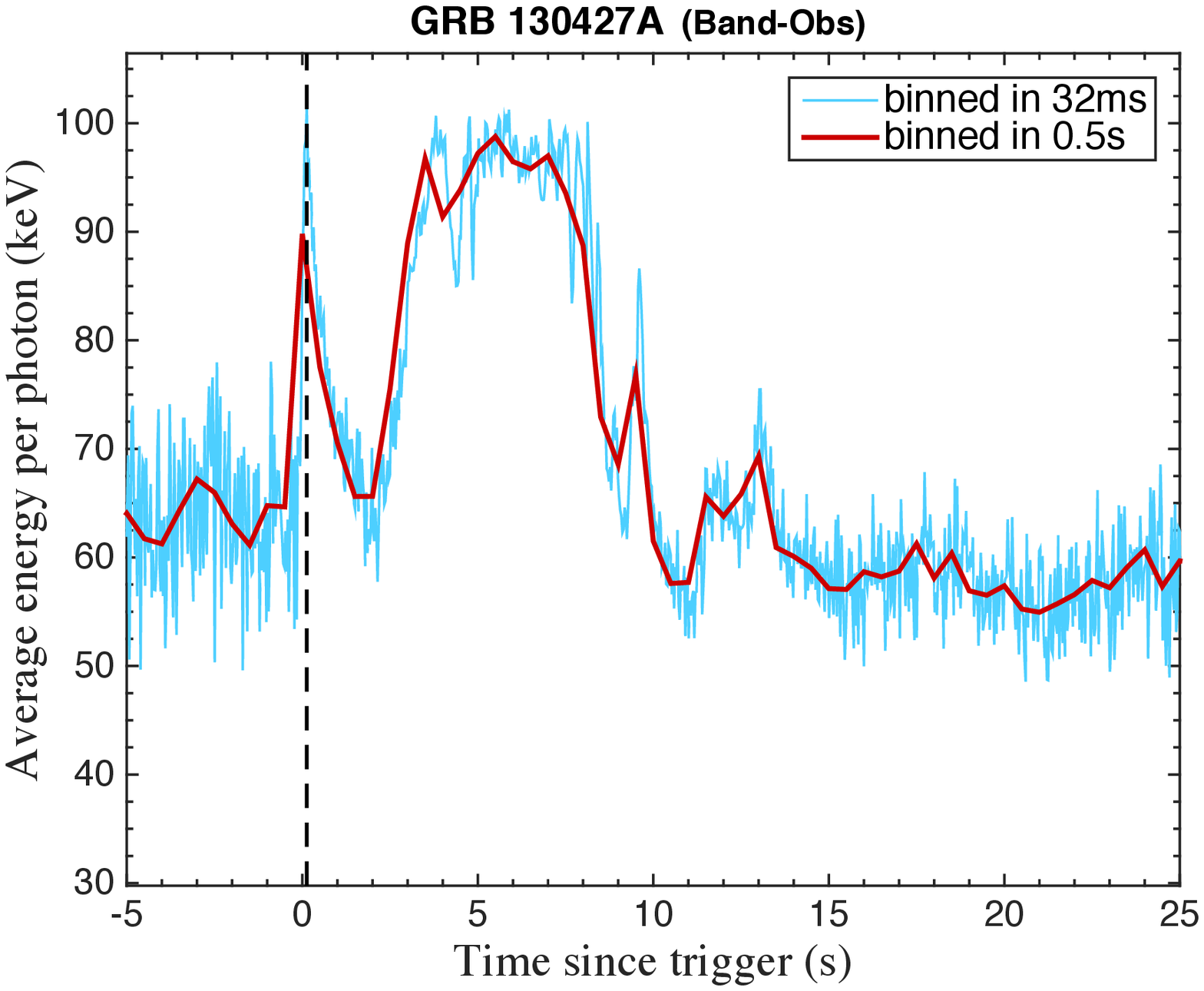}
		\includegraphics[width=0.90\linewidth]{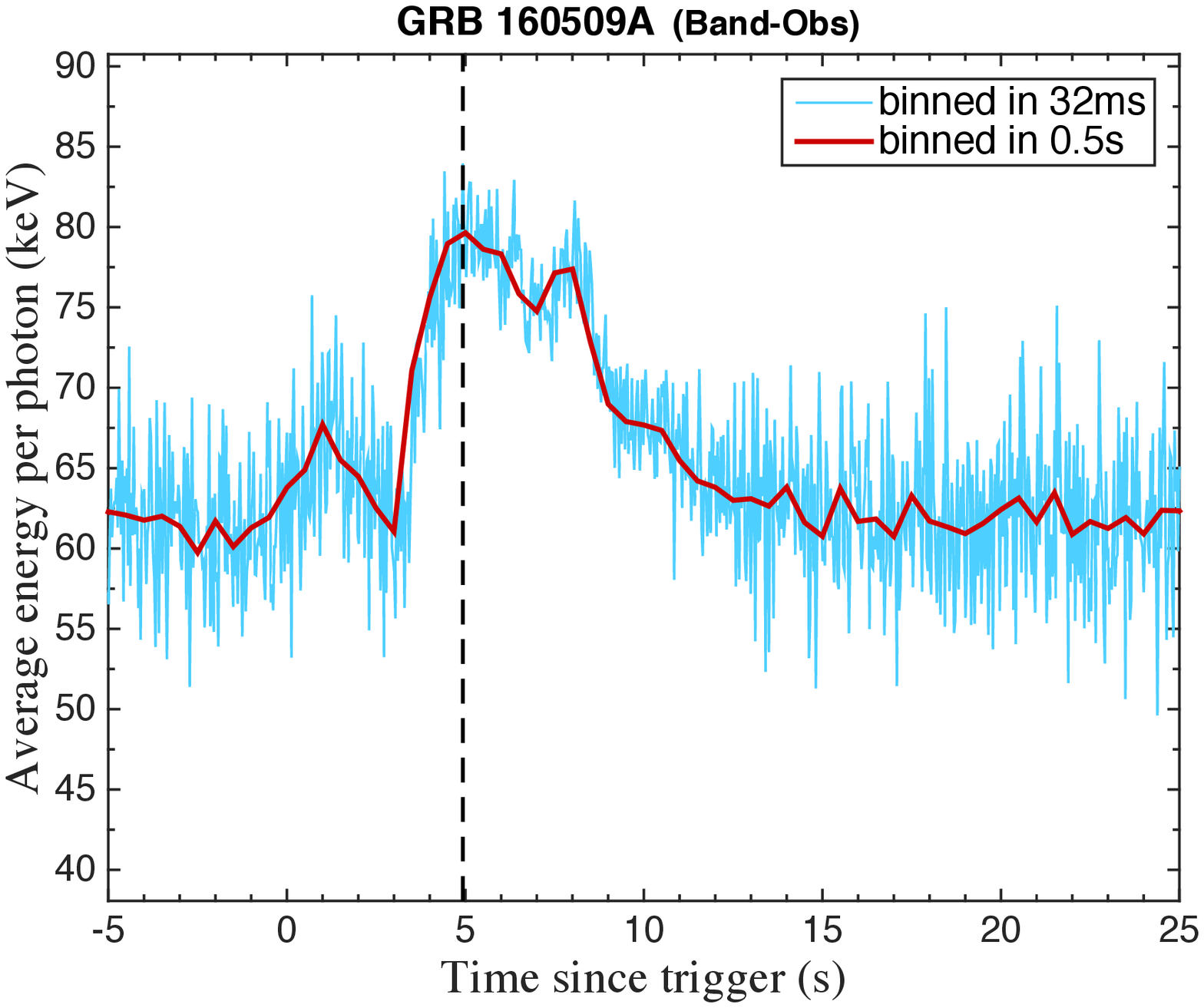}
		
	\end{minipage}
	\caption{The average energy curves in Band-Obs (average energy per photon in Band-Obs) of 8 GRBs. The $x$-axis is $\Delta t_{\rm obs}/(1+z)$. The thin curves (blue) are the energy curves binned in 0.032~s and the thick (red) curves are those binned in 0.5~s. The $y$-axis is the average energy per photon. The vertical dashed lines (black) refer to the significant peaks we choose for each GRB.}\label{fig: AverageEnergy-Obs}%The average energy curves~(average energy per photon during a bin verses time) of 8 GRBs for Band-Obs~($20\sim200$~keV in observer frame). The x-axis is the time since trigger in observer frame. The y-axis refers to the energy in observer frame. The blue curves are the light curves binned in 0.032s~(in source frame, and the same below) and the red curves are those binned in 0.5s. The vertical black dashed lines refer to the characteristic times of low energy photons~($t_{\rm low,src}$) we choose for each GRB.}\label{fig: AverageEnergy-Obs}
\end{figure*}

As the sharp change of the average energy curve ought to be related with some characteristic process, we choose the first significant change of the average energy curve as the characteristic time, $t_{\rm low,obs}/(1+z)$.
As $t_{\rm low,obs}/(1+z)$ ought to be able to represent the intrinsic nature of low energy photons, a low energy band is still needed.
Average energy curves in Band-Obs are plotted in figs.~\ref{fig: AverageEnergy-Obs}.
We bin the curves in 0.5~s and 0.032~s. We first find the first significant peak of the 0.5~s binned curve and then choose the highest point around this peak in the 0.032~s binned curve as the characteristic time.

The photons detected can be divided into two parts: background photons and GRB photons.
The background photons dominate before trigger, so the average energy curve before trigger represents the average energy of background photons.
After the GRB begins, the average energy curve represents the average energy of both background photons and GRB photons.
When the number of GRB photons is much larger than that of background photons (such as 130427A, 090902B, 090926A and 160509A), the average energy curve can approximately represent the average energy of GRB photons.

The average energy curves are quite different from the light curves.
As the number of background photons is small, the fluctuation of these curves is quite large, so the 0.5~s binned curves are useful for finding the peak.
%As the energy distribution of the detected photons is strongly affected by the detecting efficiency on energy, the curves for Band-Obs (chosen in the observer frame, in which the effective area does not change significantly) can represent the average energy per photon in an energy range with less sensitivity to the detector efficiency (shown in the observer frame).
%The counterpart of Band-I or Band-II in the observer frame deviates from Band-Obs as the redshift varies for each GRB, and therefore the average energy curves for the three bands are quite different.
%For example, as the redshift of 130427A ($z = 0.3399$) is far from the average redshift ($\simeq1.86$), the counterpart of Band-I for 130427A differs from Band-Obs, so the curve of 130427A in Band-I is very different from that in Band-Obs.
%Besides, the average energy of background photons of 080916C within Band-II is obviously bigger than that of other GRBs. This can be explained by the fact that the redshift of 080916C is the largest ($z=4.35$), so the counterpart of Band-II in the observer frame is $5.6$-$200$~keV.
%However, the effective area of the detector decreases quickly when the energy is less than 20~keV, so the number of detected photons with lower energy~($<$20~keV in the observer frame) decreases significantly and thus the average energy is larger.

%Despite of these problems,
The relative changes of average energy can be clearly seen.
One example is 130427A, which is quite interesting in many aspects~\cite{130427}.
%The average energy for Band-I or Band-II becomes smaller than the average energy of background photons during $1\sim3$~s (in the source frame, the same below), but then becomes larger during $4\sim7$~s.
This indicates that the GRB process is not a stable process: the energy distribution changes greatly at some time but remains quite stable during other periods of time.
By comparison, the average energy of 100414A is more stable during the GRB, and this implies that the energy distribution may not change much during the whole GRB process of 100414A.
It is interesting that even if the first peak of the light curve of 130427A is lower than the following ``platform", the average energies of the peak and the ``platform"  are similar in the average energy curve.
It shows that though the first peak in the light curve corresponds to a less intensive process, this process may share similar properties with the following intensive process, because they share a similar average energy.

In this subsection, we choose the first significant peak~(or dip) of the average energy curve as the characteristic time, and the results are different from those of sect.~\ref{tpeaksec} in some cases.
These $t_{\rm low,obs}/(1+z)$ corresponding to each GRB for each band are listed in table~\ref{tab:averageenergycurve}.

\begin{table}[htbp]
	\centering
	\caption{$t_{\rm low,obs}/(1+z)$ determined by the average energy criterion~(Criterion 2).% The analysis is done in the source frame. The 3 columns of $t_{\rm low,src}=t_{\rm low,obs}/(1+z)$ refer to the characteristic times since trigger, $t_{\rm low,src}$, for Band-Obs~(20-200~keV in the observer frame), Band-I~(60-570~keV in the source frame) and Band-II~(30-1070~keV in the source frame).
		}
	\label{tab:averageenergycurve}%
	\begin{tabular*}{0.48\textwidth}{@{\extracolsep{\fill}}ccc}
		\hline
		\hline
		GRB&$z$& $t_{\rm low,obs}/(1+z)$ (s) \\
		\hline
		080916C & 4.35  & 0.256 \\
		090510 & 0.903 & 0.352  \\
		090902B & 1.822 & 1.856 \\
		090926A & 2.1071 & 0.992  \\
		100414A & 1.368 & 0.800 \\
		130427A & 0.3399 & 0.128  \\
		140619B & 2.67  & 0.032 \\
		160509A & 1.17  & 4.928 \\
		\hline
	\end{tabular*}%
\end{table}%

\begin{figure}[htbp]
	\centering
	\includegraphics[width=1\linewidth]{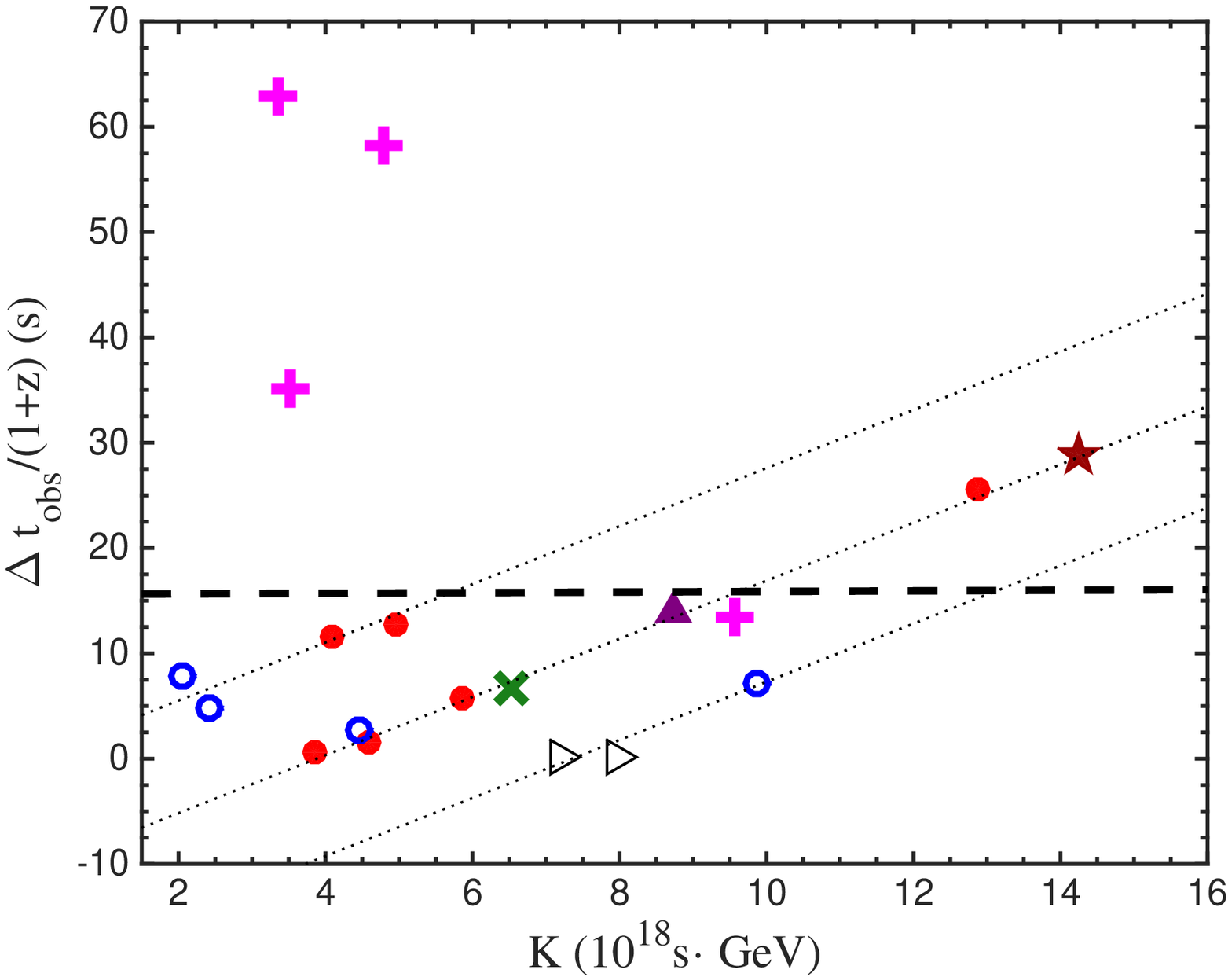}
	\caption{The $\Delta t_{\rm obs}/(1+z)$-$K$ plot for Band-Obs using Criterion 2. The dashed line marks the fit of all points, while the dotted lines were suggested in refs.~\cite{xhw,xhw160509}.%The plots for Band-I and Band-II are nearly the same as this figure. Two triangles refer to two events from short GRBs and the other points refer to events from long bursts. The middle dashed line refers to the mainline.
		}\label{fig:aveenergy}
\end{figure}

The $\Delta t_{\rm obs}/(1+z)$-$K$ plot for this criterion (fig.~\ref{fig:aveenergy}) is different from those in former discussion, as shown in figs.~\ref{allphotonsdiag} and \ref{fig:energy}.
The points near the ``mainline" become a little more scattered mainly because four events of GRB 090902B moves up along the $y$ direction compared to figs.~\ref{allphotonsdiag} and \ref{fig:energy}.
%We fit all the 9 events~(080916C(2), 090902B(1), 090902B(2), 090902B(4), 090902B(6), 090926A, 100414A, 130427A(1) and 160509A), and the results are listed in table \ref{tab:aveenergy1}.
The fit of all points
%still shows no concrete evidence for LV effect. The
gives a slope $(0\pm3)\times 10^{-18}\rm~GeV^{-1}$, which is still consistent with previous results.

%%\begin{table*}[htbp]
%%	\centering
%%	\caption{Information of the mainline (fitting 9 events on the mainline under the average energy curve criterion).}
%%	\begin{tabular}{l|c|c|c|c}
%%		\hline
%%		\hline
%%		Band  & Goodness of fit: $R^2$ & Slope ($10^{18}\rm~GeV^{-1}$) & $y$-intercept (s) & $E_{\rm LV}$ ($10^{17}$ GeV) \\
%%		\hline
%%		Band-Obs &  0.981603     &   $2.81\pm 0.16$    &   $-10.3\pm 1.4$    & $3.56\pm 0.20$ \\
%%		Band-I &    0.982743   &   $2.76\pm 0.14$    &    $-10.9\pm 1.4$   &  $3.62\pm 0.18$\\
%%		Band-II &   0.982283    &     $2.75\pm 0.14$  &   $-9.8\pm 1.2$    & $3.64\pm 0.19$ \\
%%		\hline
%%	\end{tabular}%
%%	\label{tab:aveenergy1}%
%%\end{table*}%

%The goodness of fit, $R^2$, decreases slightly compared to that in sect.~\ref{tpeaksec} and sect.~\ref{c1}, but the linear relation still exists. This illustrates that the correlation between $K$ and $\Delta t_{\rm obs}/(1+z)$ still exists.
%The difference between $E_{\rm LV}$ determined by this criterion and former criteria is less than the uncertainty of $E_{\rm LV}$. This fact supports the conclusion and the value of $E_{\rm LV}$ in refs.~\cite{xhw,xhw160509}.

In this subsection, we offer an alternative viewpoint on the signal for low energy photons.
We focus on the change of energy distribution but not the tensity of GRB.
However the results are similar.
%The linear correlation in the $\Delta t_{\rm obs}/(1+z)$-$K$ plot still exists and the value of $E_{\rm LV}$ is in accordance with former results within uncertainty.

\subsection{Criterion 3: Combine photon numbers and their average energy}\label{c3}

\begin{figure*}[htbp]
	\centering
	\begin{minipage}[t]{0.48\linewidth}
		\centering
		\includegraphics[width=0.98\linewidth]{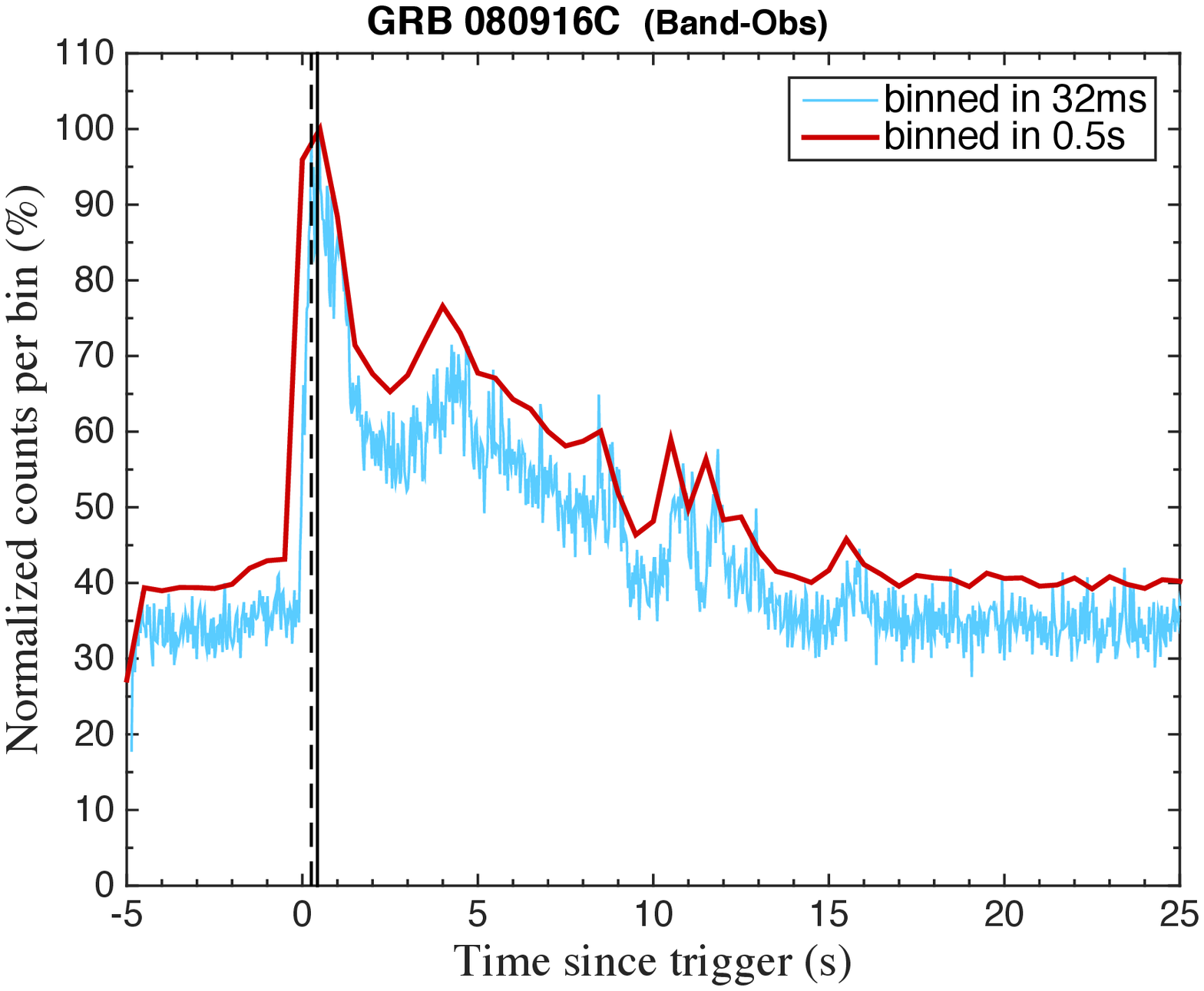}
		
		The light curve of GRB 080916C.
	\end{minipage}
	\begin{minipage}[t]{0.48\linewidth}
		\centering
		\includegraphics[width=0.98\linewidth]{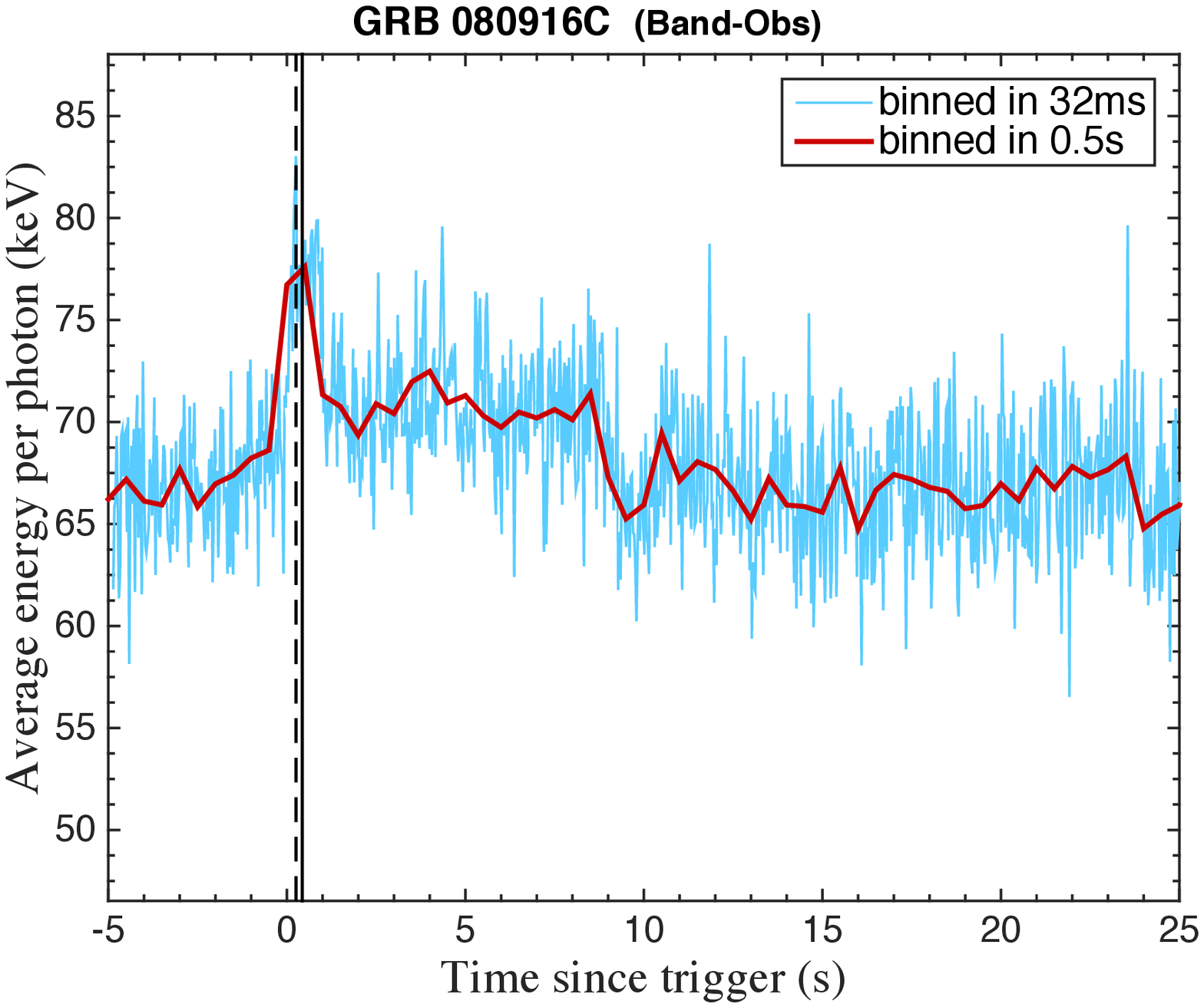}
		
		The average energy curve of GRB 080916C.
	\end{minipage}
	\caption{The light curve and average energy curve of GRB 080916C in Band-Obs. The $x$-axis refers to $\Delta t_{\rm obs}/(1+z)$. The vertical solid lines refer to $t_{\rm obs}/(1+z)$ chosen by Criterion 0~(the first main peak of the light curve) and the vertical dashed lines refer to $t_{\rm obs}/(1+z)$ chosen by Criterion 2~(the significant peak of the average energy curve).}\label{compare-080916}
\end{figure*}

\begin{figure*}[htbp]
	\centering
	\begin{minipage}[t]{0.48\linewidth}
		\centering
		\includegraphics[width=0.98\linewidth]{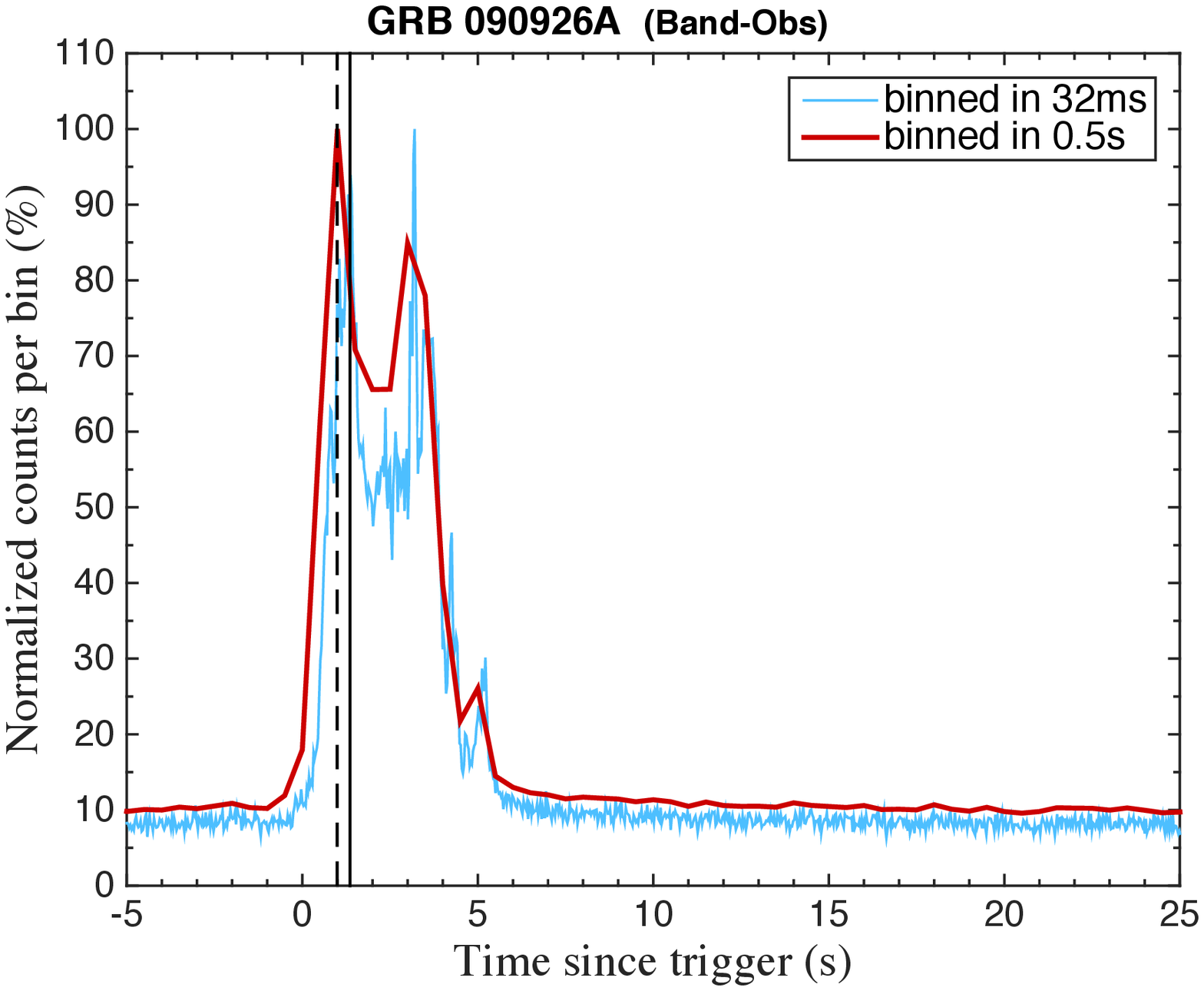}
		
		The light curve of GRB 090926A.
	\end{minipage}
	\begin{minipage}[t]{0.48\linewidth}
		\centering
		\includegraphics[width=0.98\linewidth]{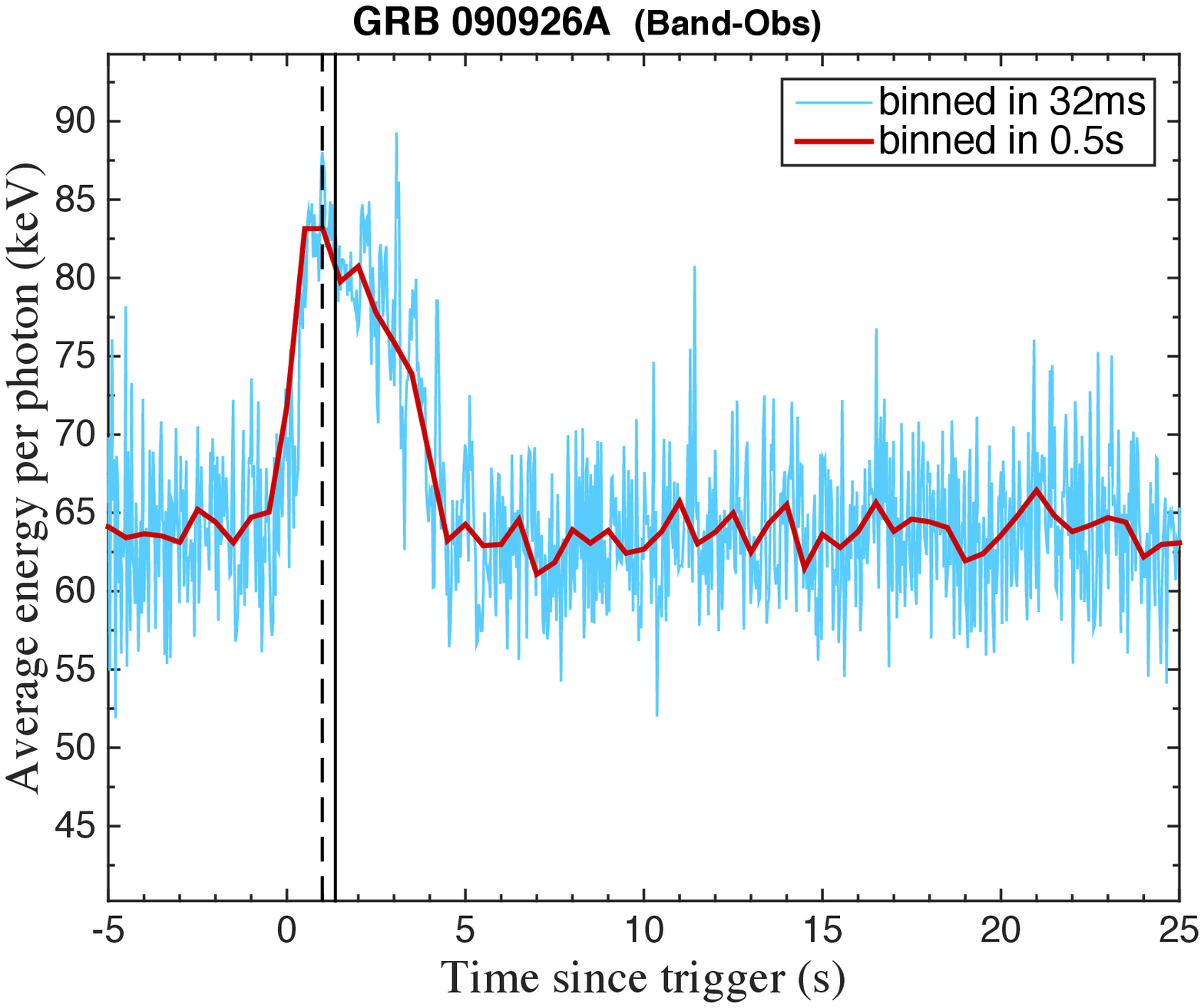}
		
		The average energy curve of GRB 090926A.
	\end{minipage}
	\caption{The light curve and average energy curve of GRB 090926A in Band-Obs. The vertical solid lines refer to $t_{\rm obs}/(1+z)$ of Criterion 0~(the first main peak of the light curve) and the vertical dashed lines refer to $t_{\rm obs}/(1+z)$ of Criterion 2~(the significant peak of the average energy curve).}\label{compare-090926}
\end{figure*}

\begin{figure*}[htbp]
	\centering
	\begin{minipage}[t]{0.48\linewidth}
		\centering
		\includegraphics[width=0.98\linewidth]{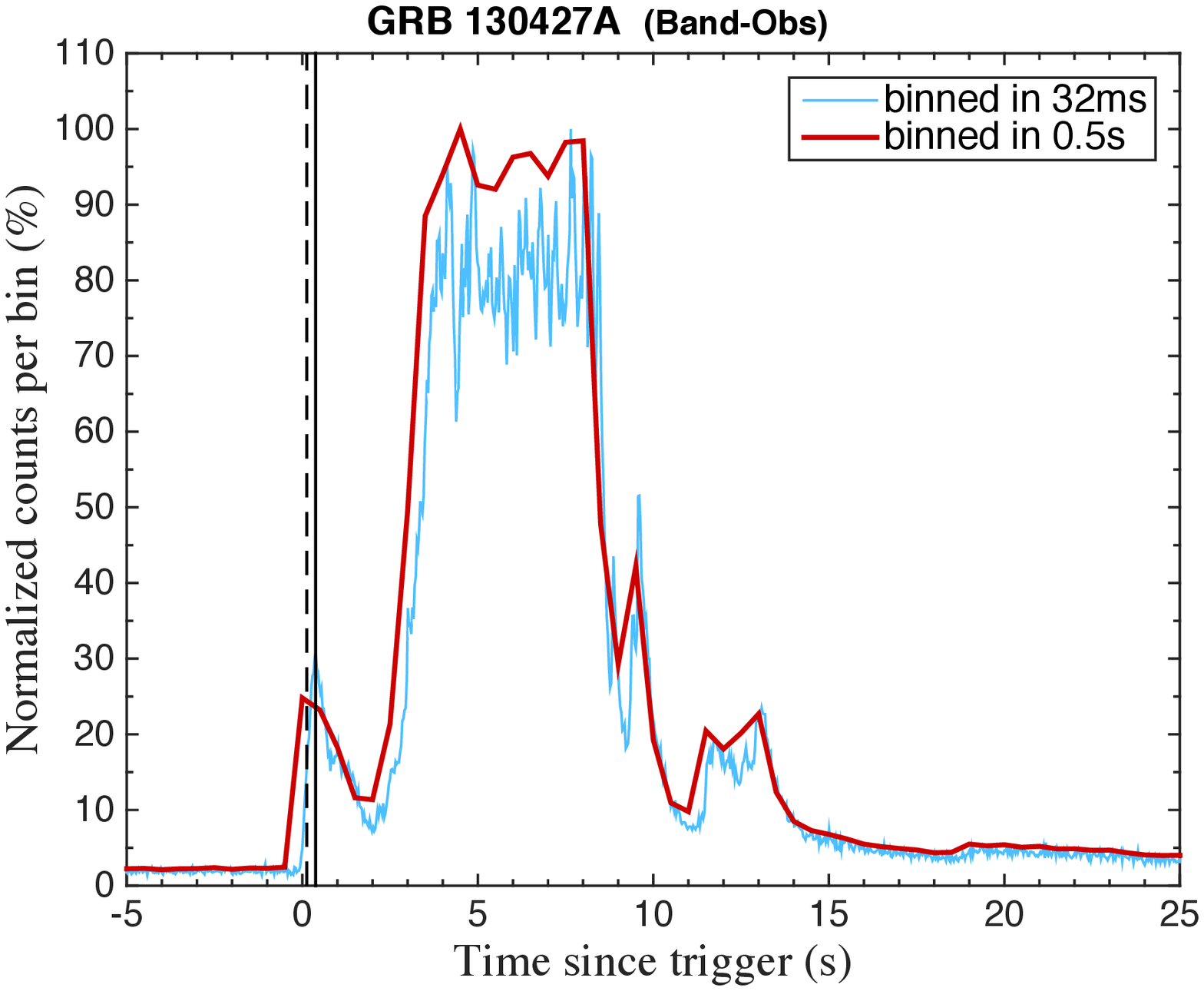}
		
		The light curve of GRB 130427A.
	\end{minipage}
	\begin{minipage}[t]{0.48\linewidth}
		\centering
		\includegraphics[width=0.98\linewidth]{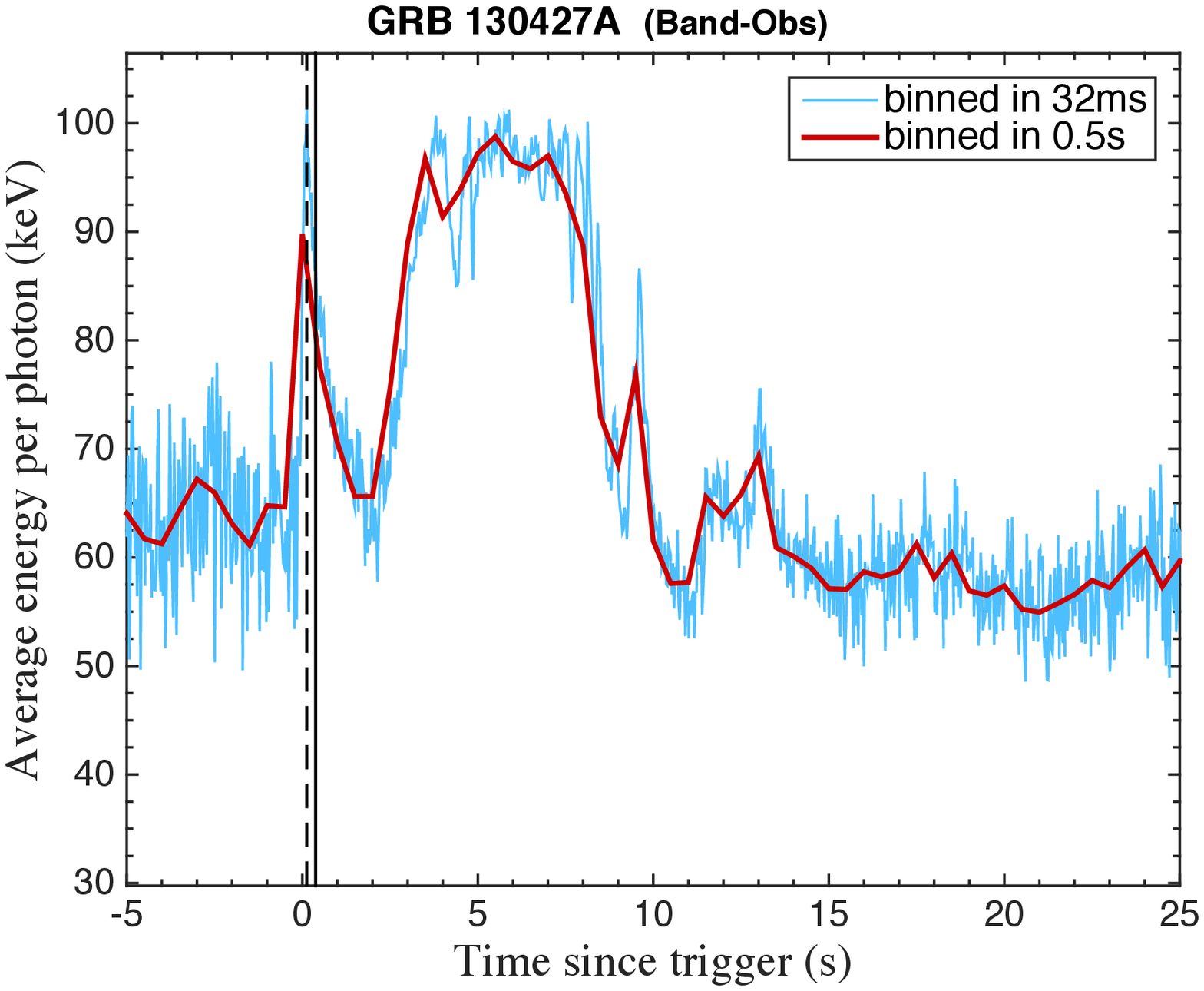}
		
		The average energy curve of GRB 130427A.
	\end{minipage}
	\caption{The light curve and average energy curve of GRB 130427A in Band-Obs. The vertical solid lines refer to $t_{\rm obs}/(1+z)$ of Criterion 0~(the first main peak of the light curve) and the vertical dashed lines refer to $t_{\rm obs}/(1+z)$ of Criterion 2~(the significant peak of the average energy curve).}\label{compare-130427}
\end{figure*}

\begin{figure*}[htbp]
	\centering
	\begin{minipage}[t]{0.48\linewidth}
		\centering
		\includegraphics[width=0.98\linewidth]{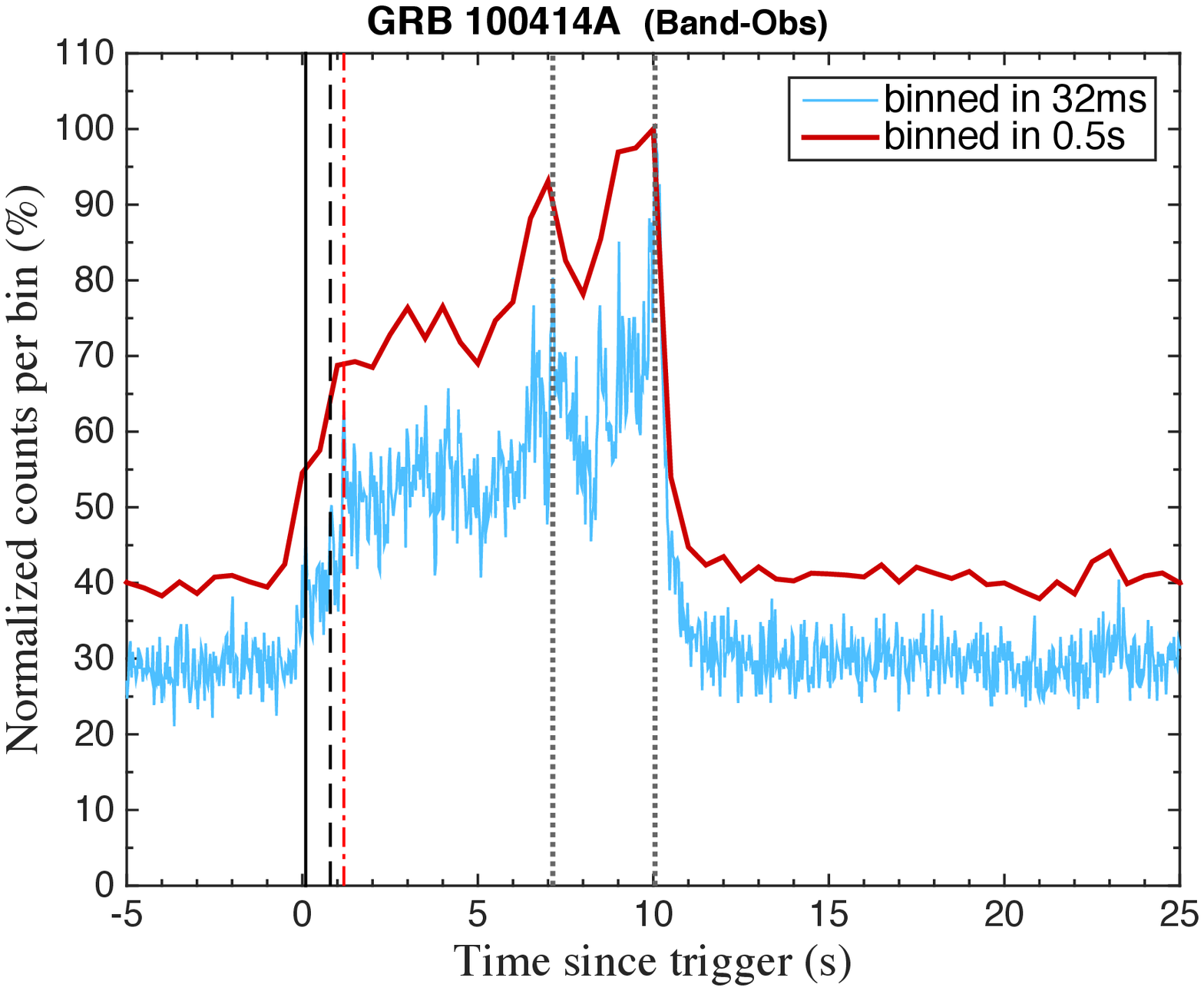}
		
		The light curve of GRB 100414A.
	\end{minipage}
	\begin{minipage}[t]{0.48\linewidth}
		\centering
		\includegraphics[width=0.98\linewidth]{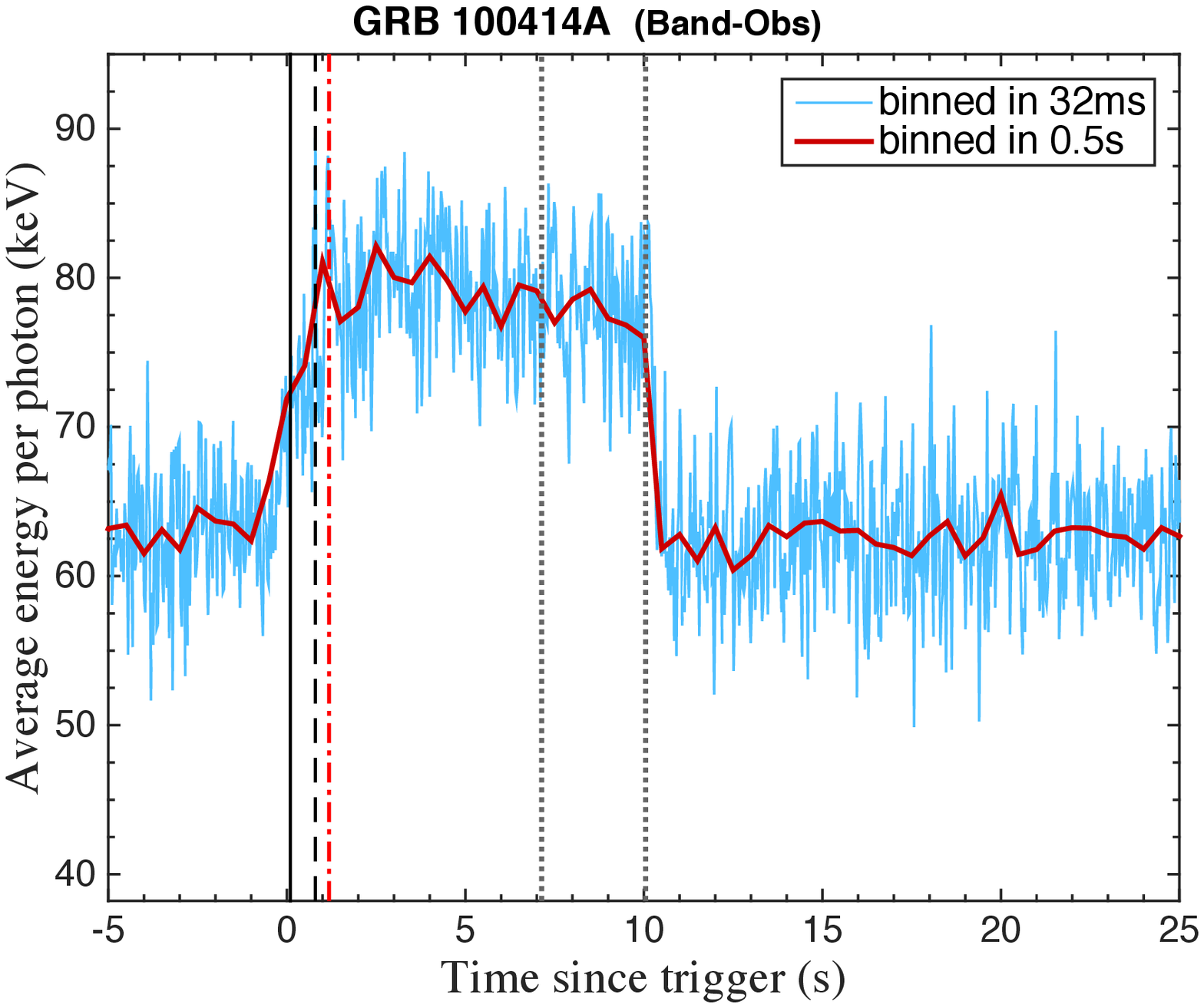}
		
		The average energy curve of GRB 100414A.
	\end{minipage}
	\caption{The light curve and average energy curve of GRB 100414A in Band-Obs. The vertical solid lines refer to $t_{\rm obs}/(1+z)$ of Criterion 0~(the first main peak of the light curve) and the vertical dashed lines refer to $t_{\rm obs}/(1+z)$ of Criterion 2~(the significant peak of the average energy curve). The vertical dash-dotted line~(red) refers to $t_{\rm obs}/(1+z)$ of Criterion 3 and the dotted lines refer to other peaks in the light curve.}\label{compare-100414}
\end{figure*}

\begin{figure*}[htbp]
	\centering
	\begin{minipage}[t]{0.48\linewidth}
		\centering
		\includegraphics[width=0.98\linewidth]{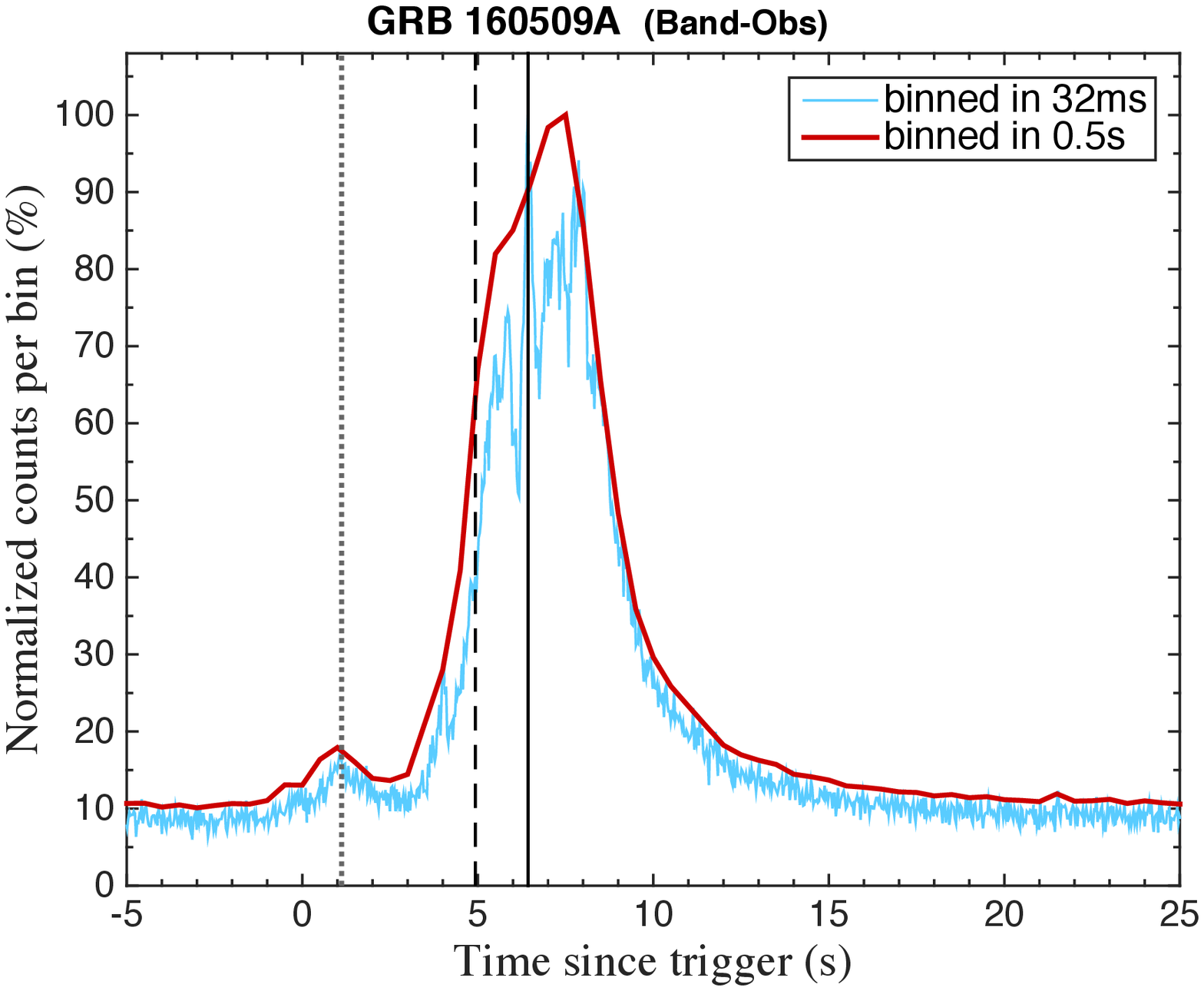}
		
		The light curve of GRB 160509A.
	\end{minipage}
	\begin{minipage}[t]{0.48\linewidth}
		\centering
		\includegraphics[width=0.98\linewidth]{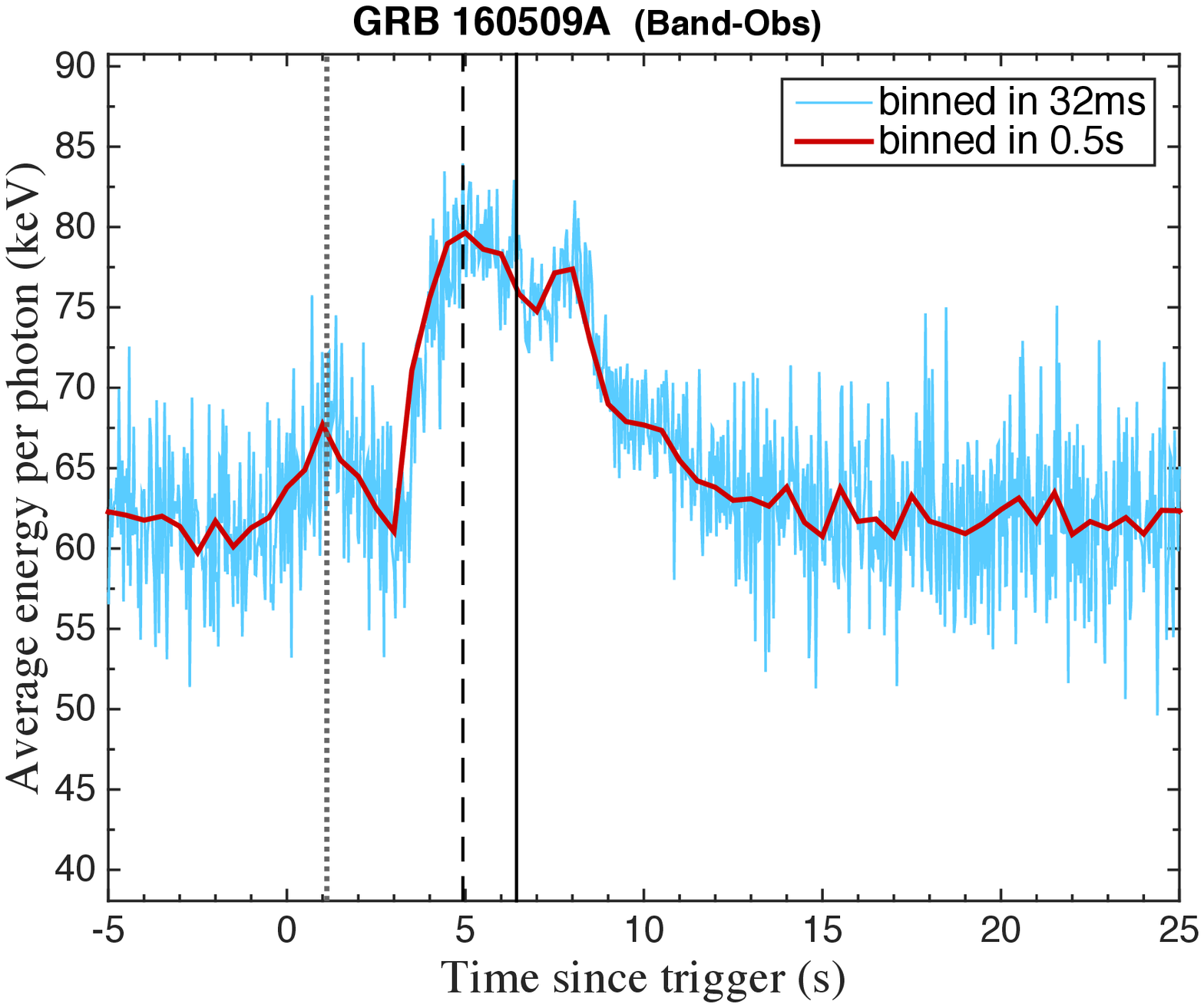}
		
		The average energy curve of GRB 160509A.
	\end{minipage}
	\caption{The light curve and average energy curve of GRB 160509A in Band-Obs. The vertical solid lines refer to $t_{\rm obs}/(1+z)$ of Criterion 0~(the first main peak of the light curve) and the vertical dashed lines refer to $t_{\rm obs}/(1+z)$ of Criterion 2~(the significant peak of the average energy curve). The dotted lines refer to a smaller peak in the light curve.}\label{compare-160509}
\end{figure*}

\begin{figure*}
	\centering
	\begin{minipage}[t]{0.48\linewidth}
		\centering
		\includegraphics[width=0.98\linewidth]{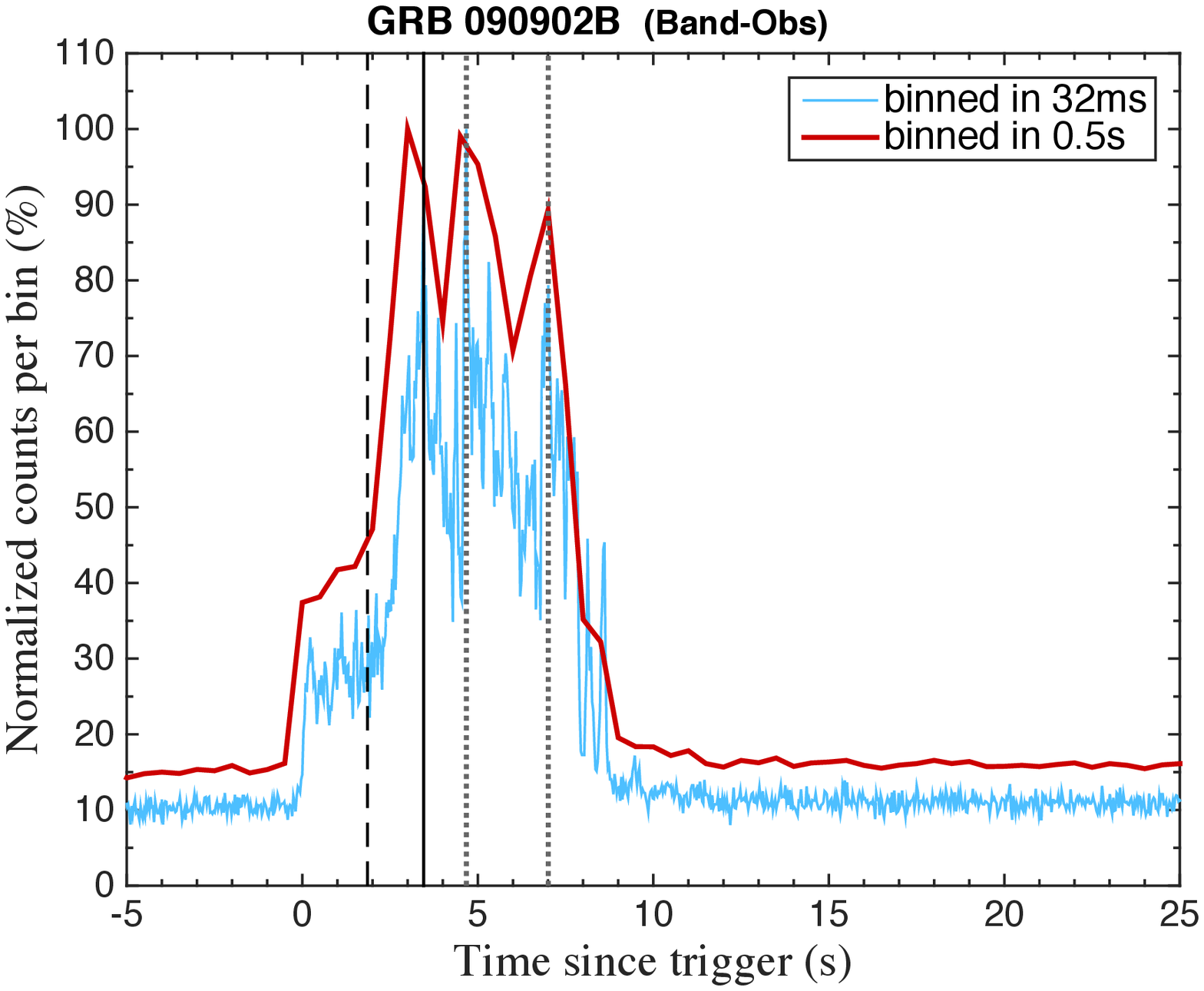}
		
		The light curve of GRB 090902B.
	\end{minipage}
	\begin{minipage}[t]{0.48\linewidth}
		\centering
		\includegraphics[width=0.98\linewidth]{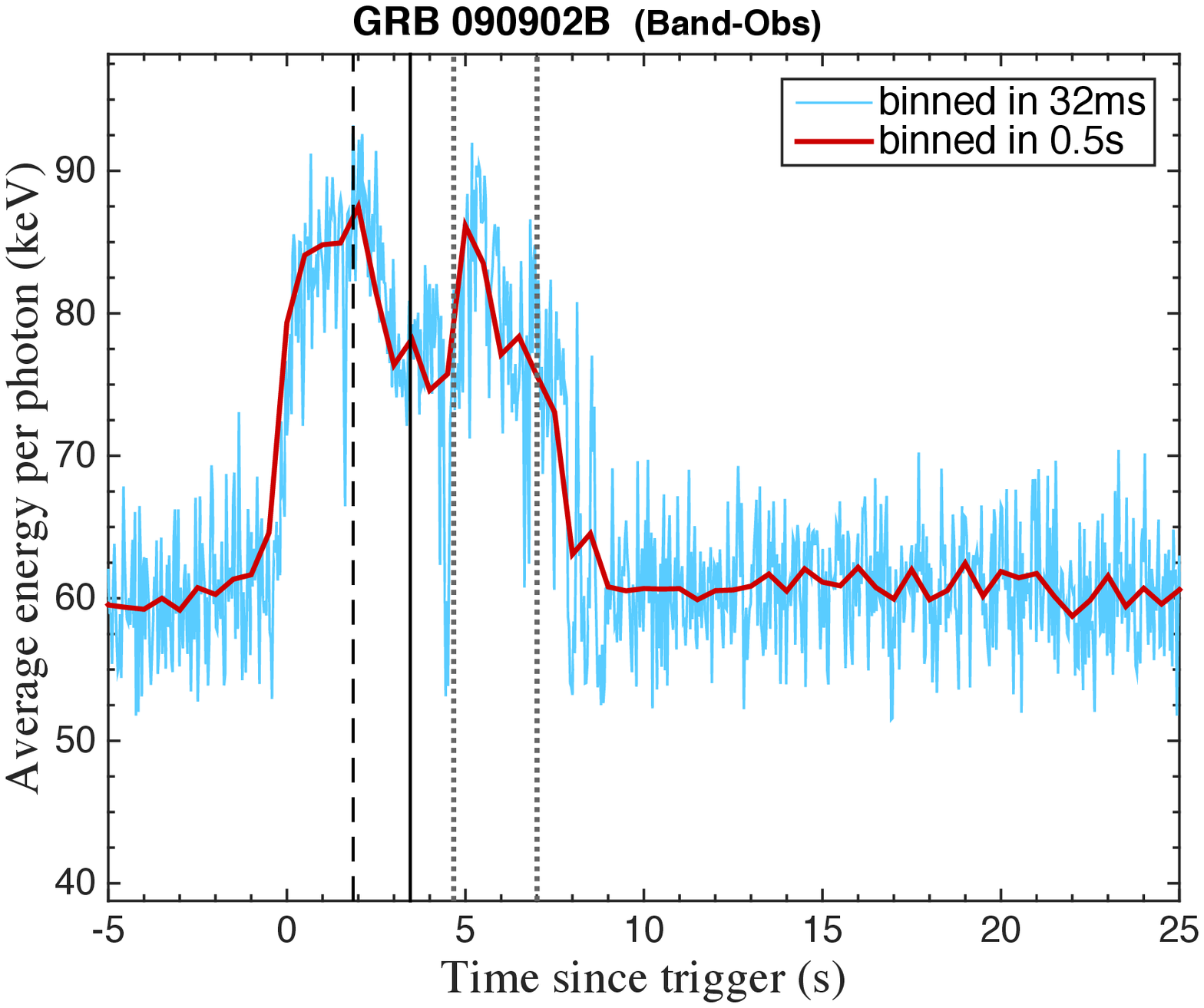}
		
		The average energy curve of GRB 090902B.
	\end{minipage}
	\caption{The light curve and average energy curve of GRB 090902B in Band-Obs. The vertical solid lines refer to $t_{\rm obs}/(1+z)$ of Criterion 0~(the first main peak of the light curve) and the vertical dashed lines refer to $t_{\rm obs}/(1+z)$ of Criterion 2~(the significant peak of the average energy curve). The dotted lines refer to the other peaks in the light curve.}\label{compare-090902}
\end{figure*}

In the following discussion, we compare the results from Criterion 2~(criterion of the average energy curve) with that from sect.~\ref{tpeaksec}.
Short bursts only have one significant peak in their light curves, energy curves and average energy curves; therefore
the characteristic times, $t_{\rm low,obs}/(1+z)$, for the 3 criteria are nearly the same.
Therefore we only discuss long bursts below.
For long bursts, the results of sect.~\ref{c1} and sect.~\ref{tpeaksec} show that the obtained $t_{\rm low,obs}/(1+z)$ are almost the same for the light curve criterion~(we call it Criterion 0, and the same below) and Criterion 1~(criterion of the energy curve).
The peaks chosen by Criterion 0 and Criterion 1 are of nearly the same position.
Therefore, we mainly compare the peaks chosen by Criterion 0 and Criterion 2.
%Also, as the results are not sensitive to the energy band, we use Band-Obs for discussion below and omit the cases of Band-I and Band-II.

We notice that for GRB 080916C, GRB 090926A and GRB 130427A, the peaks chosen by Criterion 0 and Criterion 2 are very close to each other, as shown in figs.~\ref{compare-080916}, \ref{compare-090926} and \ref{compare-130427}.
Their $t_{\rm low,obs}/(1+z)$ refer to significant peaks of both the 0.5~s binned light curve and 0.5~s binned average energy curve.
This shows that the characteristic times for these 3 GRBs are not only the times when the number of photons per bin sharply changes but also the times when the average energy per bin significantly changes.
This strongly implies that there exists some special physical mechanisms around these $t_{\rm low,obs}/(1+z)$.
Before now, we only know that these $t_{\rm low,obs}/(1+z)$ match the time with maximum number of photons.
Now with the help of the average energy curve, we can distinguish the peak that also matches the time of the change of energy distribution from other peaks that do not match this mechanism.
This helps to reduce artificial factors especially when choosing the first significant peak for the GRBs with multiple peaks.
As the inner process of a GRB is very complex, one may want to choose the characteristic low energy event from more perspectives. The main peak in a light curve refers to a process with many photons emitted. One can be more confident that this is a characteristic process, if the energy distribution also changes significantly.
Therefore, we offer a balanced criterion that considers both light curve and average energy curve:
\begin{quote}
	%Criterion 3: to choose the first significant peak of the light curve in Band-Obs and the peak also matches a significant change in the average energy curve,
	Criterion 3: choose the first significant peak of the light curve in Band-Obs that also matches a significant change in the average energy curve
\end{quote}
where we have considered the fact that the results for different bands do not vary much.

Now we offer Criterion 3 to the other long bursts~(GRB 090902B, GRB 100414A and GRB 160509A).

For GRB 100414A~(see fig.~\ref{compare-100414}), the peak chosen with the criterion in sect.~\ref{tpeaksec}~(and in refs.~\cite{xhw,xhw160509}) does not match a peak in the average energy curve.
Using Criterion 3, we suggest a new characteristic time marked by the dash-dotted line~(red) in fig.~\ref{compare-100414} ($t_{\rm low,obs}/(1+z)=1.184$~s).
It marks a significant peak of the light curve and it also falls around the peak of the average energy curve.
The 0.5~s binned curves are used to show the main processes of GRB, and from fig.~\ref{compare-100414}, we notice that the new $t_{\rm low,obs}/(1+z)$ coincides with the significant peaks in both light curve and average energy curve binned in 0.5~s.
On the contrast, the former $t_{\rm low,obs}/(1+z)$~(marked by the vertical solid line) does not mark a significant peak in the average energy curve.
Actually the former $t_{\rm low,obs}/(1+z)$ does not mark a significant peak in the 0.5~s binned light curve.
This suggests that the new $t_{\rm low,obs}/(1+z)$ chosen by Criterion 3 may reflect a physical mechanism and thus a more convincing characteristic time for low energy photons.
Therefore, we may suggest that $t_{\rm low,obs}/(1+z)=1.184$~s be used in later discussion.
Besides, we can see that other peaks in the light curves~(marked by vertical dotted lines) do not match the significant peaks in the average energy curve, and this means that these peaks may not be related with a special mechanism that changes the average energy.
Therefore, these peaks are not chosen as the characteristic time.
We can see that by considering the average energy curve, Criterion 3 helps to restrict the possible peaks chosen from the light curve, and this helps to reduce the arbitrariness in choosing $t_{\rm low,obs}/(1+z)$. We therefore recommend to use Criterion 3 for the choice of $t_{\rm low,obs}/(1+z)$ as the characteristic time for low energy photons.

Now we come to GRB 160509A.
160509A~\cite{xhw160509} is considered to be a strong evidence to support the prediction of ref.~\cite{xhw}, so it is important to check whether $t_{\rm low,obs}/(1+z)$ for 160509A still holds for the new Criterion 3.
In the 0.5~s binned light curve of 160509A (see fig.~\ref{compare-160509}), there is a small peak at around 1.120~s~(marked by dotted line). We do not choose this peak because it is too small compared to the next peak in sect.~\ref{tpeaksec}.
Now we can give more evidence by using Criterion 3.
From the average energy curve of 160509A, we find that the bigger peak in the light curve (marked by solid line, which we choose as $t_{\rm low,obs}/(1+z)$) can match the main peak in the 0.5~s binned average energy curve.
However the smaller peak in the light curve (at around 1.120~s) does not match the main peak in the average energy curve. There is a smaller peak but this peak is also too small compared to the next peak.
Therefore, we do not treat the smaller peak as the first significant peak and suppose that the next peak represents more reasonable significant time for low energy photons.

It becomes more complicated to apply the criterion to GRB 090902B (see fig.~\ref{compare-090902}).
%As the peaks in the light curve and average energy curve match with each other in other GRBs, the case becomes worse for 090902B.
The 0.5~s binned light curve has 3 main peaks while the 0.5~s binned average energy curve only has two. It seems that the 3 peaks of the light curve do not match the significant peak in the average energy curve.
We notice that the first peak of the light curve is significant and is the nearest to the first main peak of the average energy curve. Therefore, we assume that the first peak of the light curve is related with the physical mechanism shown by the average energy curve.

\begin{figure}[htbp]
	\centering
	\includegraphics[width=1\linewidth]{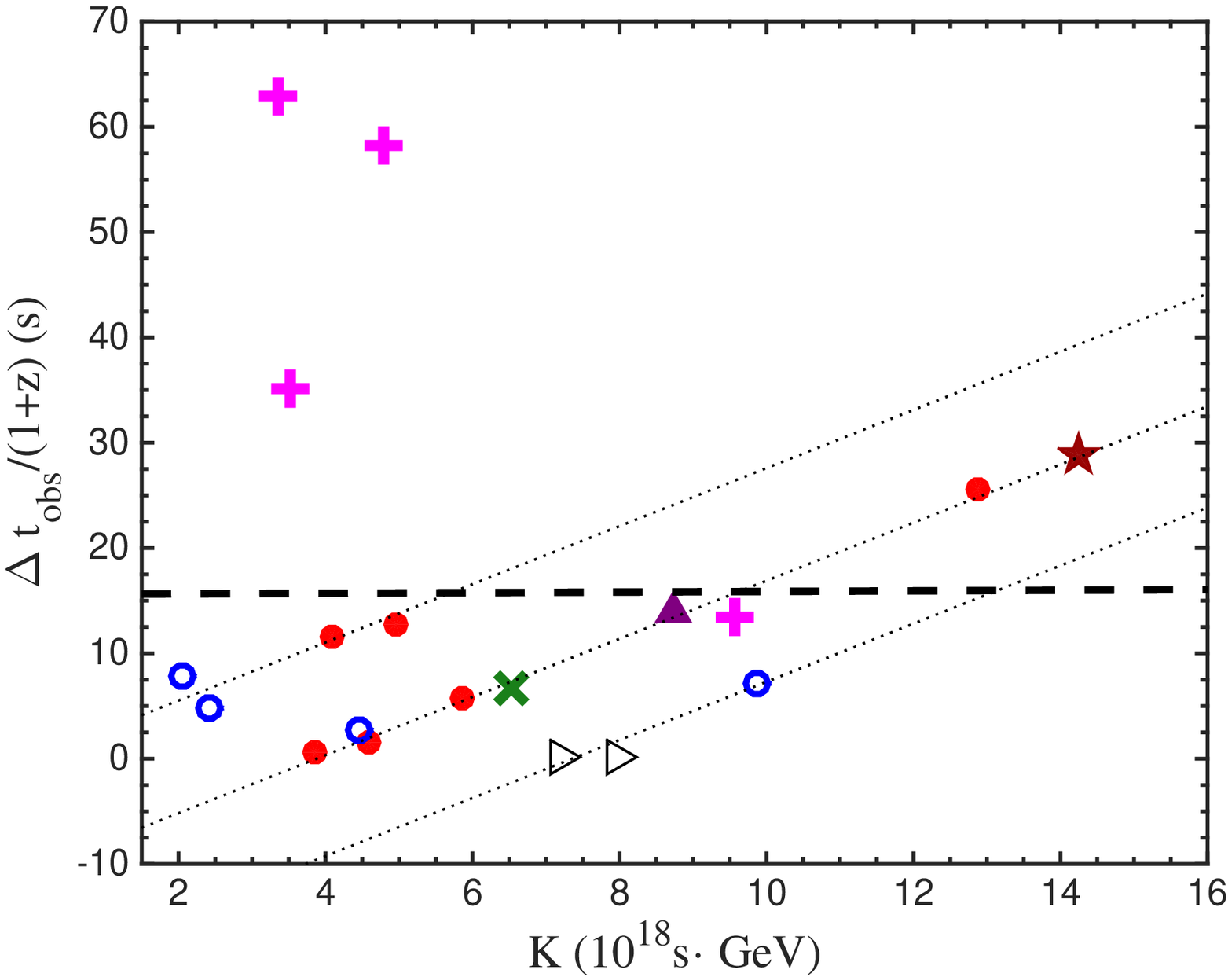}
	\caption{The $\Delta t_{\rm obs}/(1+z)$-$K$ plot for Band-Obs using Criterion 3. Two triangles refer to two events from short GRBs and the other points are events from long bursts. The dashed line marks the fit of all points, while the dotted lines were suggested in refs.~\cite{xhw,xhw160509}.}\label{fig:c3}
\end{figure}

\begin{table}[htbp]
	\centering
	\caption{The low energy characteristic times we recommend (using Criterion 3) for 8 GRBs.}
	\label{tab:tlowfinal}
	\begin{tabular}{cccc}
		\hline
		\hline
		GRB   &$z$& $t_{\rm low,obs}/(1+z)$~(s) & $t_{\rm low,obs}$~(s) \\
		\hline
		080916C &4.35& 0.480  &  2.568\\
		090510&0.903&0.288&0.548\\
		090902B &1.822& 3.456  &9.753\\
		090926A &2.1071& 1.376 &4.275\\
		100414A &1.368& 1.184  &2.804 \\
		130427A &0.3399& 0.384 &0.515 \\
		140619B &2.67& -0.032  &-0.117 \\
		160509A &1.17& 6.432 &13.957 \\
		\hline
		\hline
	\end{tabular}%
\end{table}%

%\begin{table*}
%	\centering
%	\caption{Information of the mainline (fitting 9 events on the mainline under Criterion 3).}
%	\begin{tabular}{l|c|c|c|c}
%		\hline
%		\hline
%		Band  & Goodness of fit: $R^2$ & Slope ($10^{18}\rm~GeV^{-1}$) & $y$-intercept (s) & $E_{\rm LV}$ ($10^{17}$ GeV) \\
%		\hline
%		Band-Obs &  0.990387     &   $2.76\pm 0.11$    &   $-10.9\pm 1.0$    & $3.62\pm 0.14$ \\
%		\hline
%	\end{tabular}%
%	\label{tab:c3}%
%\end{table*}%

In summary, Criterion 3 combines the effect of maximum number of photons and the effect of changing the energy distribution.
We recommend this criterion because more aspects of the GRB are considered and thus arbitrariness is reduced in this way.
Considering the fact that different bands for low energy photons do not affect the results much,
%and the fact that it seems more reasonable to use the time in the source reference frame
we recommend the method of determining $t_{\rm low,obs}/(1+z)$ below:
\begin{itemize}
	\item[1. ] First bin the light curve and the average energy curve for Band-Obs in 0.5~s.
	\item[2. ] Then find the first significant peak of the light curve that also matches a significant change of the average energy curve.
	\item[3. ] Use the time of the highest point around this peak in the 32~ms binned light curve as the characteristic time for low energy photons $t_{\rm low,obs}/(1+z)$.
\end{itemize}
To conclude this section, we list $t_{\rm low,obs}/(1+z)$ for 8 GRBs we recommend~(using the method above) in table~\ref{tab:tlowfinal} and draw the $\Delta t_{\rm obs}/(1+z)$-$K$ plot in fig.~\ref{fig:c3}.
%By fitting the mainline, we obtain the results in table~\ref{tab:c3}. The results agree with those
%in refs.~\cite{xhw,xhw160509}.
By fitting all the points, we obtain the slope $(0\pm 3)\times 10^{-18}~\rm GeV^{-1}$, still consistent with previous results.

\section{Application to FERMI detectors of low energy photons }\label{instruments}

\subsection{Application to NaI, BGO, and LLE data}

We have illustrated in former sections that, for GBM NaI data, the changes in binnings and energy bands do not bring significant changes to $t_{\rm low,obs}/(1+z)$~(the first main peak). In order to use the FERMI data comprehensively, we now apply the method and Criterion 3 to other low energy photon detectors of the FERMI telescope.
The FERMI telescope has two parts: GBM and LAT. GBM consists of 12 NaI detectors and 2 BGO detectors whose effective energy bands are approximately 8-1000~keV and 150~keV - 30~MeV respectively, while LAT approximately covers the photon energy range 20~MeV - 300~GeV.
To make full use of the FERMI data, we use the following methods to plot three sets of light curves and average energy curves for the 8 GRBs discussed above:
\begin{enumerate}
	\item Set-NaI: For each GRB, draw the light curve and average energy curve by using all the photon events of the two GBM NaI detectors with most number of detected photons.
	\item Set-BGO: For each GRB, draw the light curve and average energy curve by using all the photon events of the two GBM BGO detectors.
	\item Set-LLE: For each GRB, draw the light curve by using the LAT Low-Energy~(LLE) events~\cite{fermi-3} and draw the average energy curve by using the LLE events with energies less than 1 GeV. The number of photons with energies larger than 1 GeV is quite small and thus causes little influence to the light curve. However, as the count per bin is also small, these high-energy events lead to notable changes of the average energy curves. Therefore, we cut off the high energy events in the average energy curves. %Ackermann et al. 2013
	Since the 1-$\sigma$ angular resolution of LAT is about $3.5^\circ$~\cite{latoverview}, a $12^{\circ}$ region of interest~(ROI) is chosen as in ref.~\cite{xhw160509} in order to include more photons from the GRB source.
	%%ROI or without ROI
\end{enumerate}
As the energies for most LLE events are less than $\sim 100$ MeV, which is negligible compared to the energy of high energy photons~($>$ 30 GeV), the LV effect of LLE~(and also BGO, NaI) photons, if really exists, must produce tiny arrival time shift compared to that of high energy events. Thus the LLE events can still be treated as low energy photons and we can ignore the LV effect on them. Because it is important to test whether the $t_{\rm low}$ we choose really represent significant low energy process, we do this analysis to look for evidence from data of other detectors.

All the curves are binned in 0.5~s and 0.032~s in the source frame~($t_{\rm obs}/(1+z)$) as in the previous section. For each set of curves, we find the first main peak in the 0.5s-binned curves for each GRB and choose the highest point around this peak as the characteristic time. The light curves and average energy curves of NaI, BGO and LLE data
\footnote{We download the data from \url{https://fermi.gsfc.nasa.gov/ssc/observations/types/grbs/lat_grbs/table.php}. We find that the LLE (LAT) data for GRB 100414A are not available. Therefore, we do not plot the LLE curves for GRB 100414A.}
are plotted respectively in figs.~\ref{nai-1}-\ref{nai-2} (Set-NaI), figs.~\ref{bgo-1}-\ref{bgo-2} (Set-BGO), and figs.~\ref{lle-1}-\ref{lle-2} (Set-LLE).
The characteristic times we choose for 8 GRBs for each curve set are listed in table~\ref{tlow}.
For all these figures, the left column refers to the light curves and the right column refers to the average energy curves of the corresponding GRB. The $x$-axes refer to the time since trigger in the source frame, i.e. $t_{\rm obs}/(1+z)$. The thin curves (blue) are binned in 32~ms and the thick curves (red) are binned in 0.5~s. The $y$-axes of the light curves refer to the normalized counts, i.e., the counts per bin divided by the maximum counts per bin of the corresponding light curve. The $y$-axes of the average energy curves refer to the average energy per photon (measured in the observer reference frame). The vertical lines (black) refer to the characteristic times we choose for each GRB.
%tables.

% table generated by Excel2LaTeX from sheet '???1'
\begin{table*}[htbp]
	\centering
	\caption{The characteristic times (since trigger) for low energy photons from NaI, BGO and LLE data. As we have mentioned, we do not have $t_{\rm low}$ for GRB 100414A from LLE data because of the lack of data.}
	\begin{tabular*}{1\textwidth}{@{\extracolsep{\fill}}cccccccc}
		\hline
		\hline
		\multirow{2}{*}{GRB}    & \multirow{2}{*}{$z$} &  \multicolumn{3}{c}{$t_{\rm low,obs}/(1+z)$ (s)}   &       \multicolumn{3}{c}{$t_{\rm low,obs} (s)$}   \\
		&       & NaI & BGO & LLE & NaI & BGO & LLE\\
		\hline
		080916C & 4.35  & 0.480  & 0.864  & 0.864  & 2.568  & 4.622  & 4.622  \\
		090510 & 0.903 & 0.288  & 0.288  & 0.384  & 0.548  & 0.548  & 0.731  \\
		090902B & 1.822 & 3.456  & 3.456  & 2.624  & 9.753  & 9.753  & 7.405  \\
		090926A & 2.1071 & 1.376  & 1.344  & 1.920  & 4.275  & 4.176  & 5.966  \\
		100414A & 1.368 & 1.184  & 1.184  & - & 2.804  & 2.804  & - \\
		130427A & 0.3399 & 0.352  & 0.224  & -0.032  & 0.472  & 0.300  & -0.043  \\
		140619B & 2.67  & 0.096  & 0.000  & 0.096  & 0.352  & 0.000  & 0.352  \\
		160509A & 1.17  & 6.432  & 5.536  & 5.504  & 13.957  & 12.013  & 11.944  \\
		\hline
		\hline
	\end{tabular*}%
	\label{tlow}%
\end{table*}%

In former sections, for each GRB, the characteristic peak in the light curve is required to match a peak in the corresponding average energy curve, but this time, the peak in the light curve matches a dip in the average energy curve for NaI and BGO data. This suggests that, within a broad band, the average energy of GRB photons is lower than that of background photons.

\begin{figure*}[htbp]
	\centering
	\includegraphics[width=0.45\linewidth]{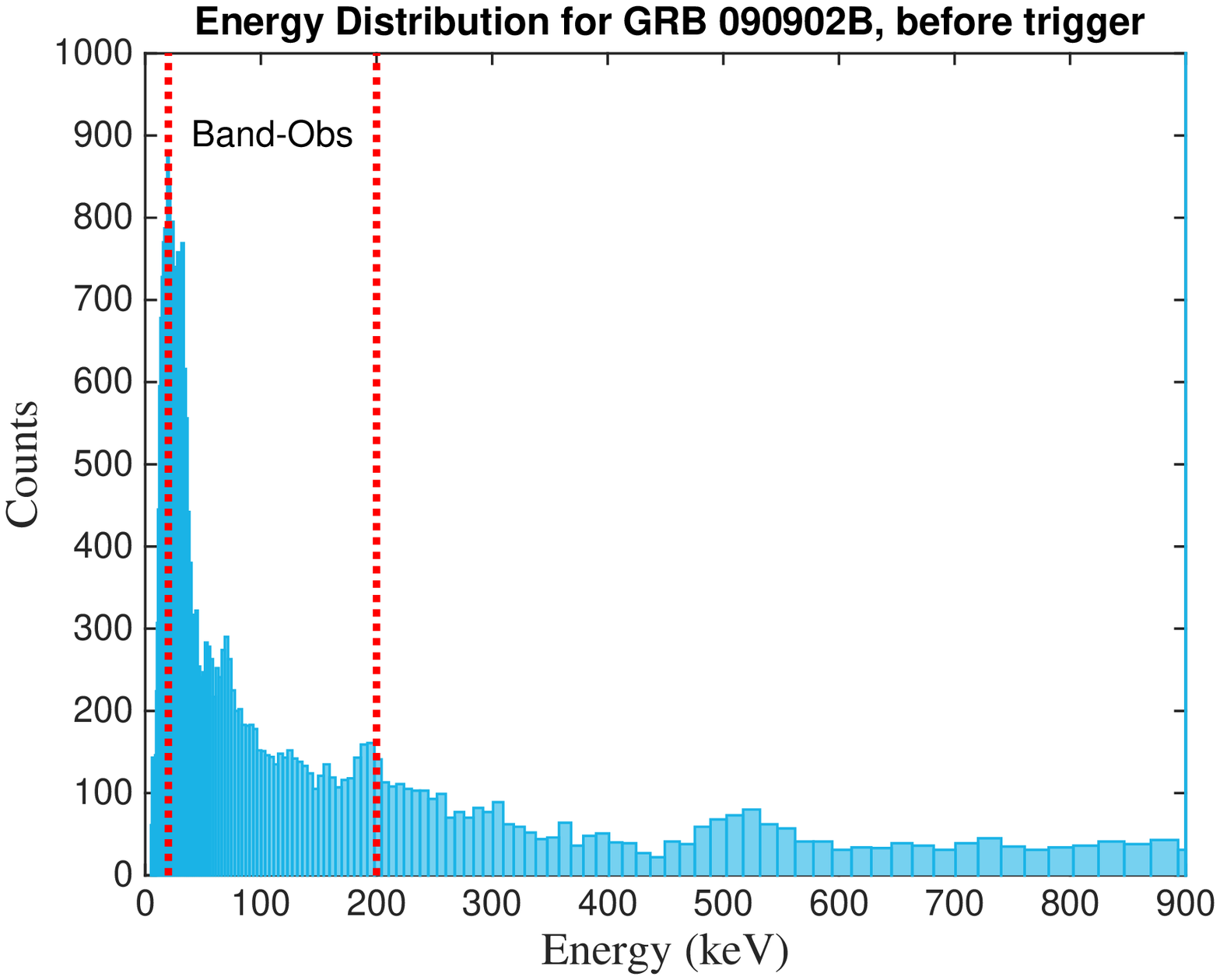}
	\includegraphics[width=0.45\linewidth]{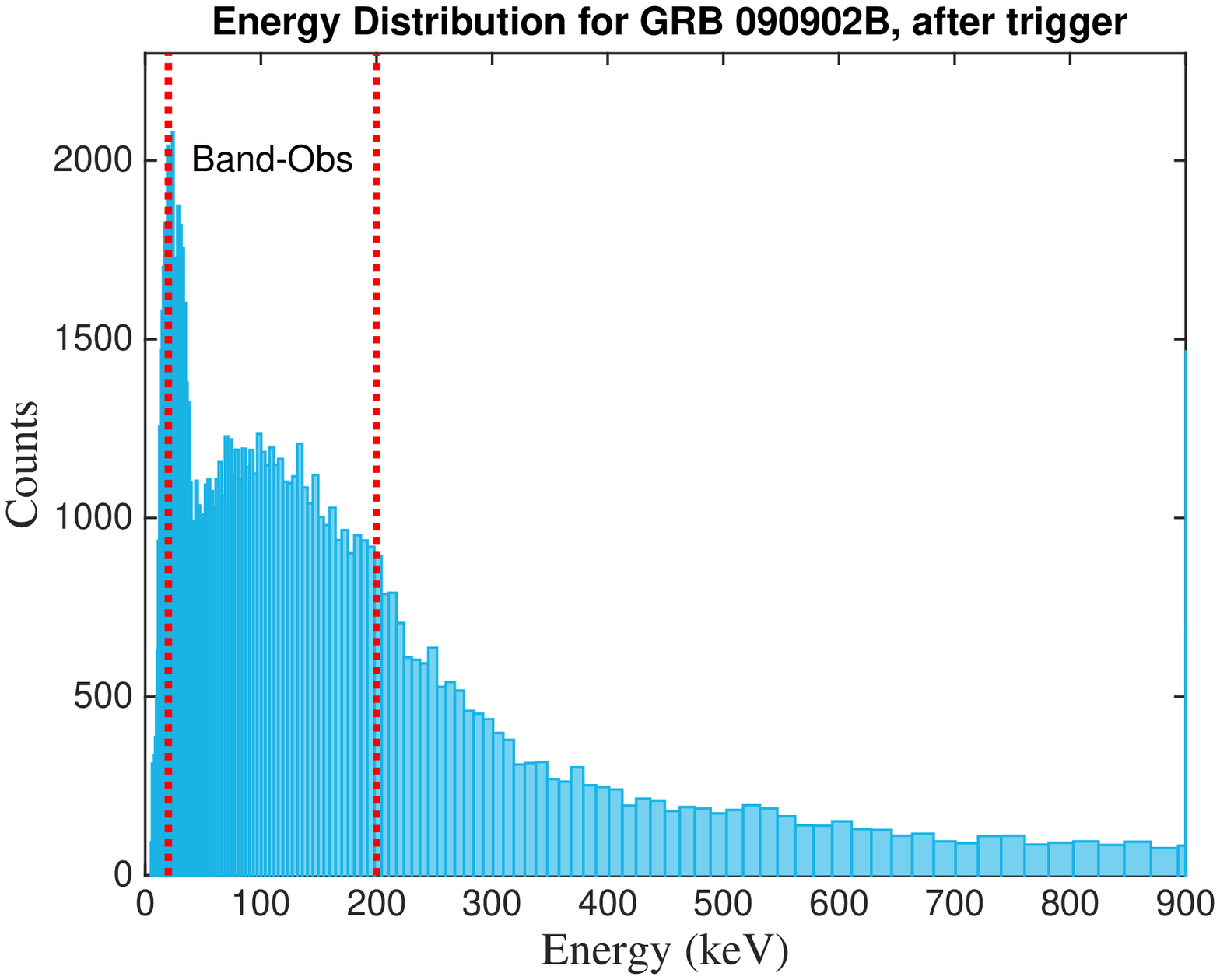}
	\caption{The energy distributions of Set-NaI photons of GRB 090902B within [-20,0]s and [0,20]s since trigger. The red dashed lines mark Band-Obs.}\label{fig:ed090902}
\end{figure*}

To clarify this, energy distributions of Set-NaI of GRB 090902B before and after trigger are plotted in fig.~\ref{fig:ed090902}. In previous sections, Band-Obs~(marked by red dashed lines) is applied. However, in this section, all the photons~(Set-NaI) are used. As shown in fig.~\ref{fig:ed090902}, if we only consider photons within Band-Obs, obviously the average energy increase after trigger~(i.e. approximately after GRB begins). However, if we focus on photons from the whole band of NaI detectors, relatively more lower energy photons makes the average energy of whole band after trigger smaller. This explains why there is a peak when we choose Band-Obs and why there is a dip if we use Set-NaI.

As mentioned before, the energy distribution within Band-Obs can be treated as the real energy distribution of GRB and background photons, while the energy distribution outside Band-Obs is distorted by the performance of detectors. However, as we focus on the changes of energy distribution, Set-NaI is used in this section.

For Set-NaI and Set-BGO, the peaks we choose in the light curves still match significant changes (the sharp decreasing of average energy per photon) in the corresponding average energy curves, especially for long bursts (e.g. 080916C, 090926A, 130427A, 160509A, etc.). This indicates that, around the characteristic time, there may be some certain mechanism that causes significant changes in aspects of both observed counts and energy distribution of photons. Moreover, this mechanism affects the photons within the NaI and BGO bands (about 8 keV - 30 MeV), which are much broader than Band-Obs.
%The consistence of the $t_{\rm low,src}$s determined from NaI, BGO and LLE data indicates that the $t_{\rm low,src}$ may reflect a certain mechanism that affect the photons within about 8 keV - 30 MeV.

\begin{figure}[ht]
	\centering
	
	Case 1
	\includegraphics[width=0.90\linewidth]{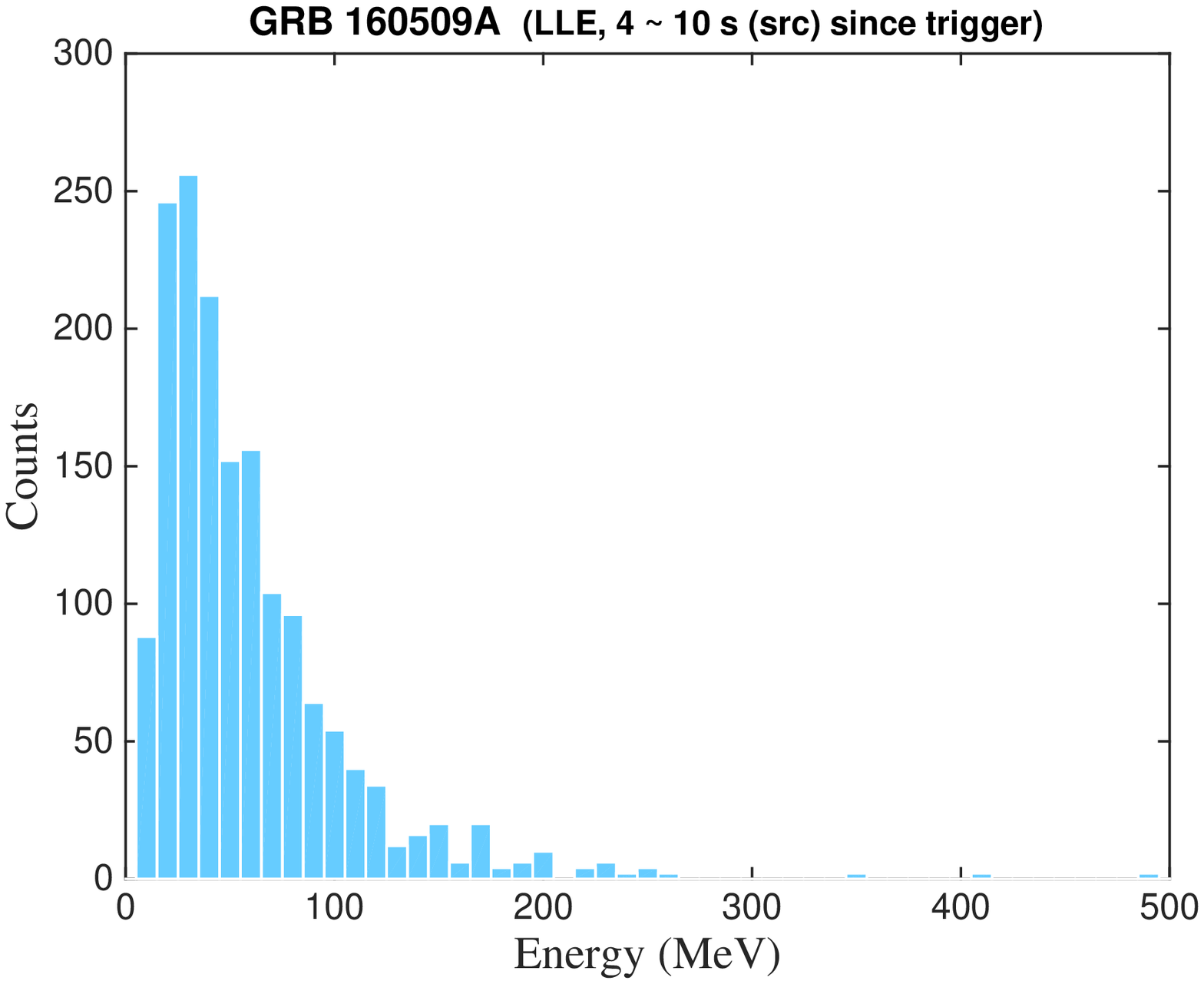}
	
	Case 2
	\includegraphics[width=0.90\linewidth]{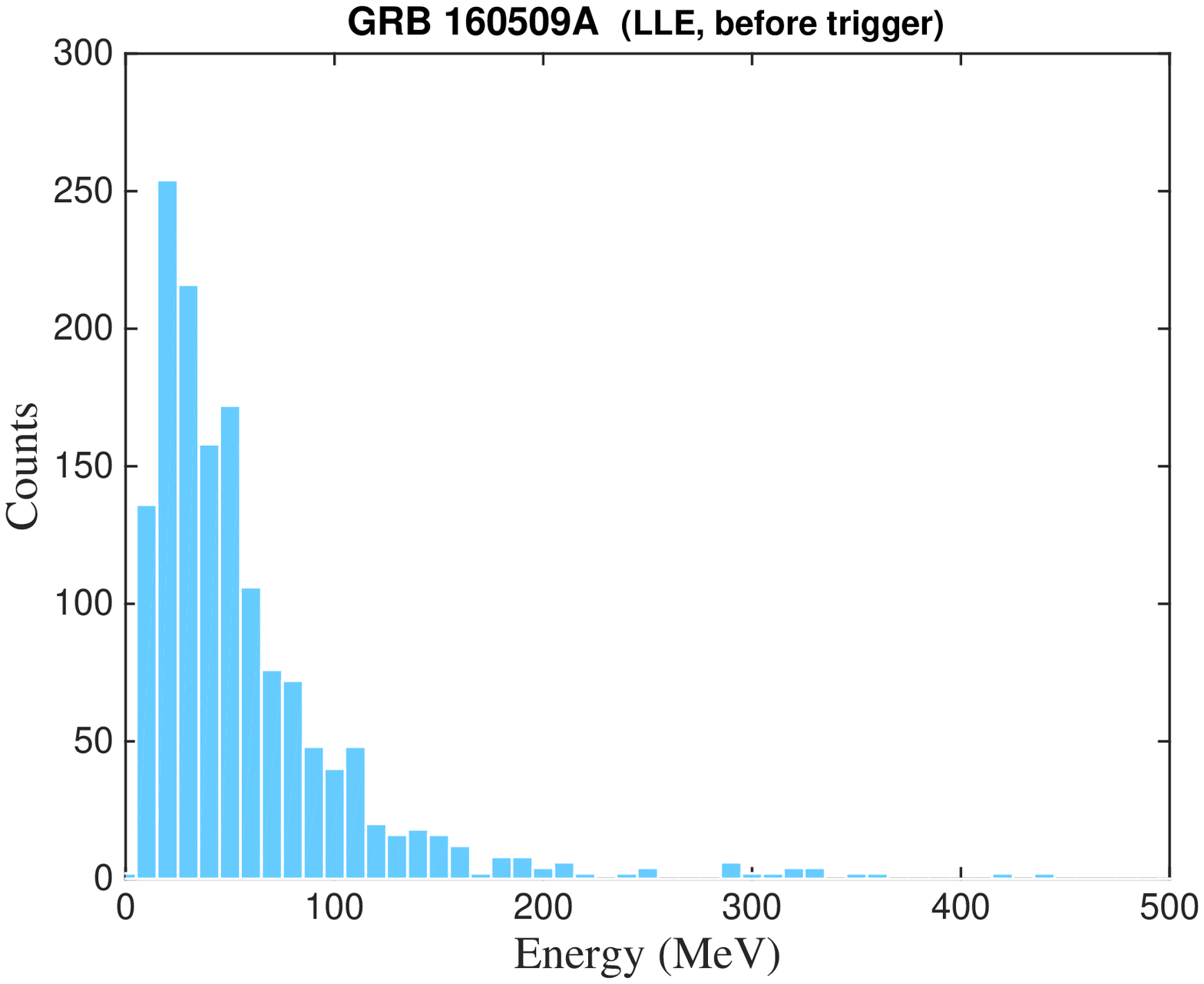}
	\caption{The energy distribution histograms for GRB 160509A from LLE data. The upper one uses the photons within 4-15~s (with $t_{\rm obs}/(1+z)$) since trigger. The lower one uses the photons before trigger. The $y$-axis refers to the counts of certain rectangle. We can see that most of LLE photons are $\lesssim$ 100 MeV.}\label{16ED-LLE}
\end{figure}

\begin{figure}[h!]
	\centering
	Case 1
	\includegraphics[width=0.9\linewidth]{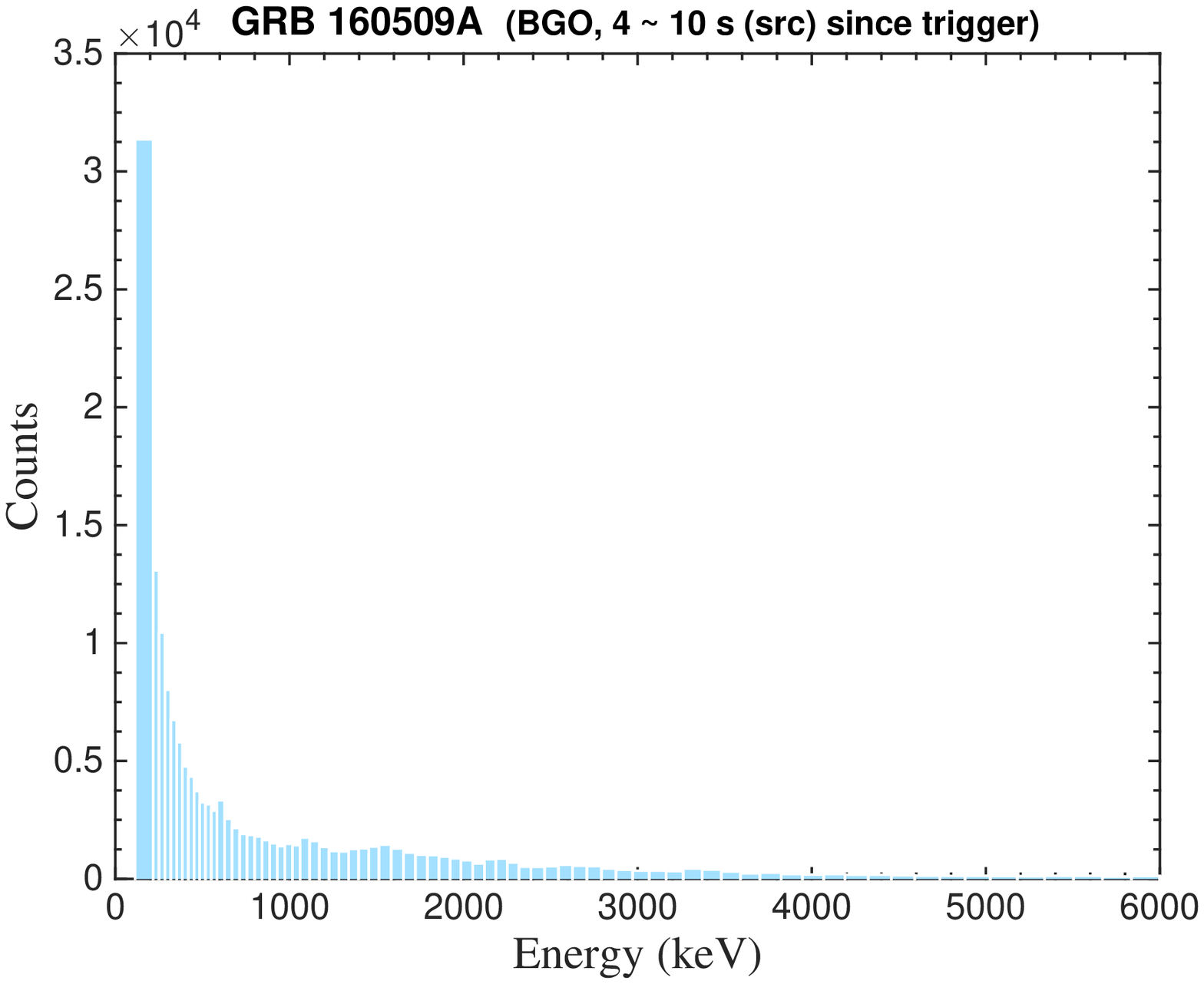}
	
	Case 2
	\includegraphics[width=0.9\linewidth]{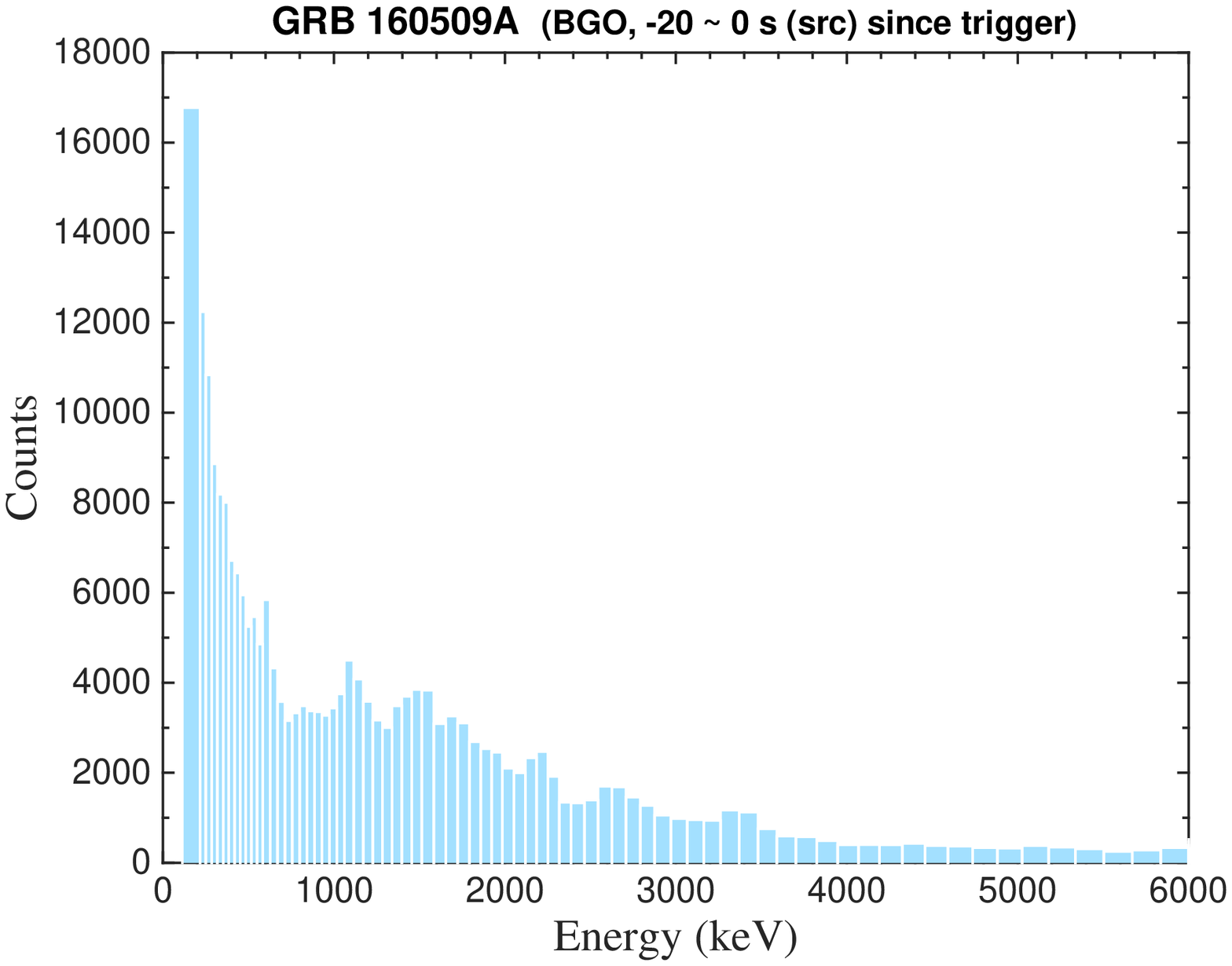}
	
	\caption{The energy distribution histograms for GRB 160509A from BGO data. The upper one uses the photons within 4-15~s (with $t_{\rm obs}/(1+z)$) since trigger. The lower one uses the photons within $-20$-0~s (with $t_{\rm obs}/(1+z)$) since trigger. The $y$-axis refers to the counts of certain rectangle. The bounds of the rectangles are chosen according to the energy channels of the instrument.}\label{16ED-BGO}
\end{figure}

\begin{figure*}[h]
	\centering
	\begin{minipage}[t]{0.44\linewidth}
		\centering
		\includegraphics[width=0.90\linewidth]{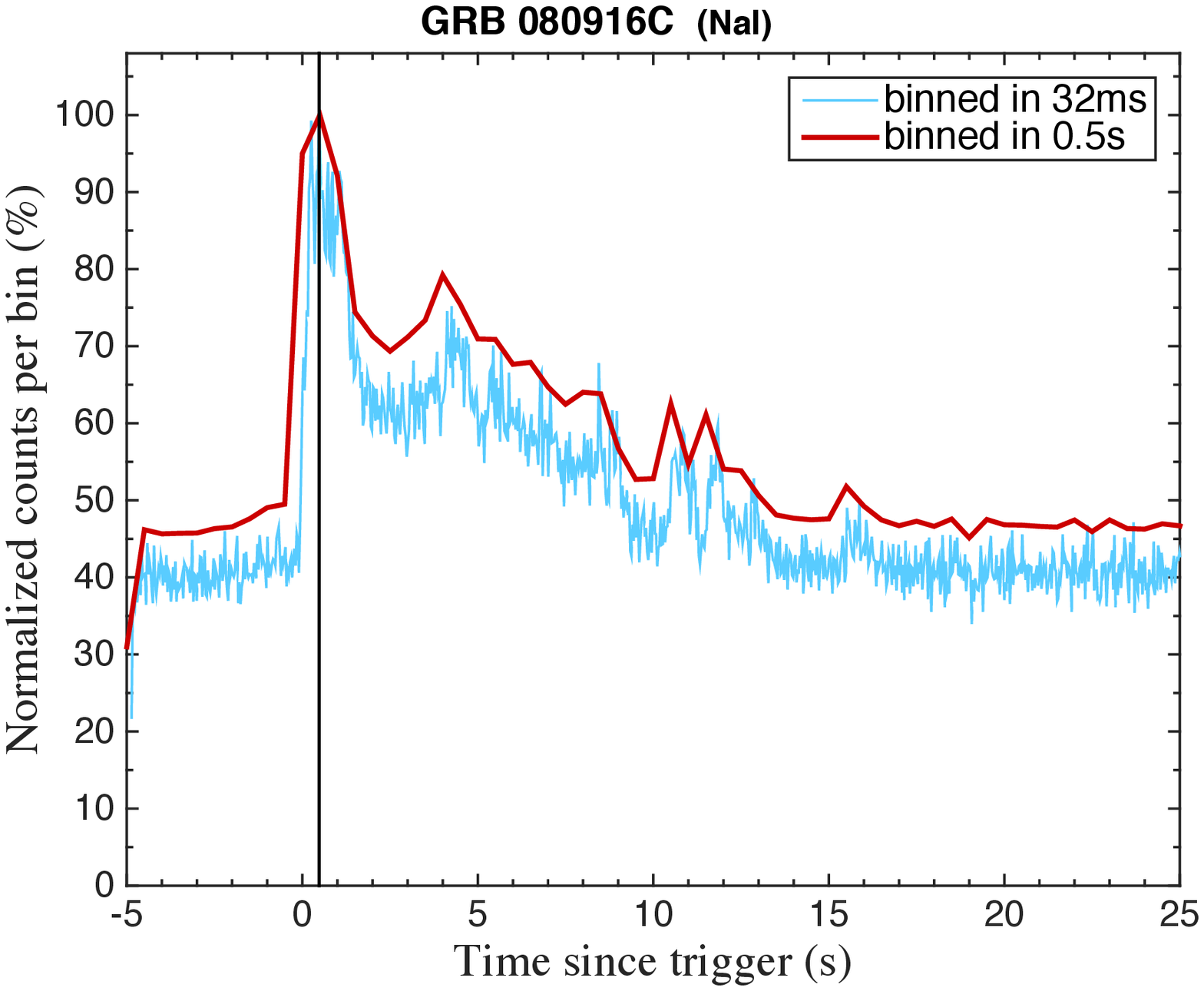}
		\includegraphics[width=0.90\linewidth]{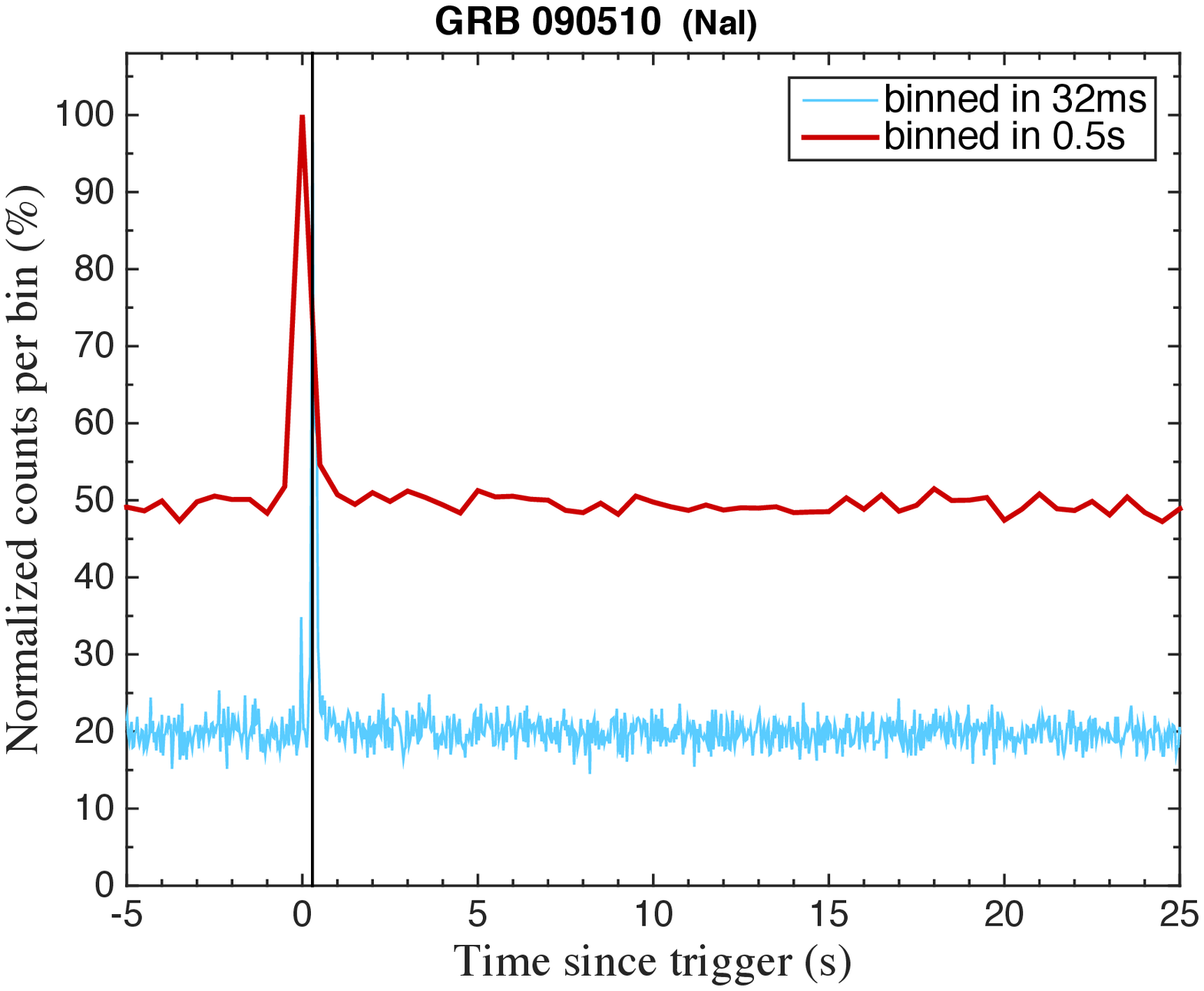}
		\includegraphics[width=0.90\linewidth]{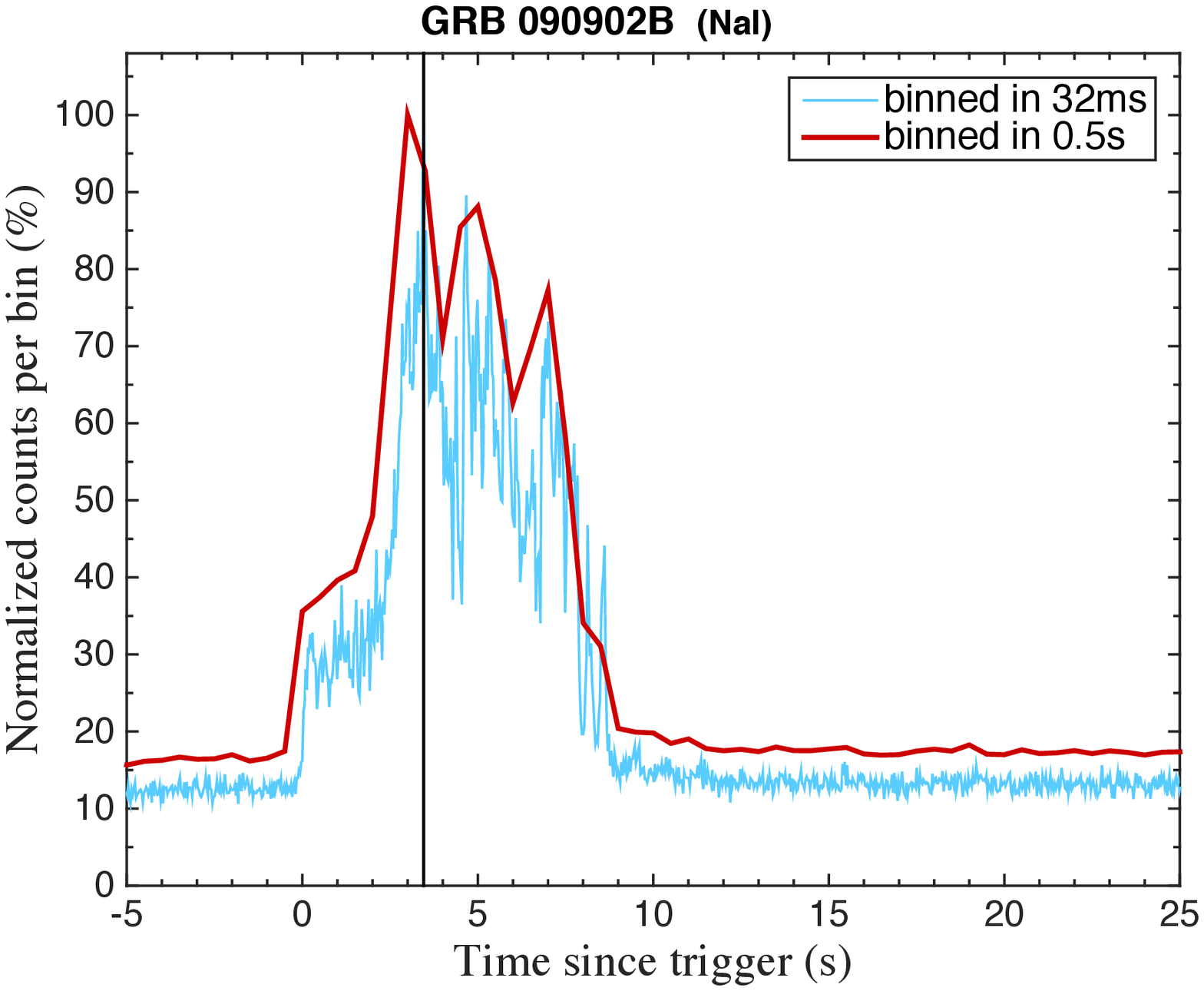}
		\includegraphics[width=0.90\linewidth]{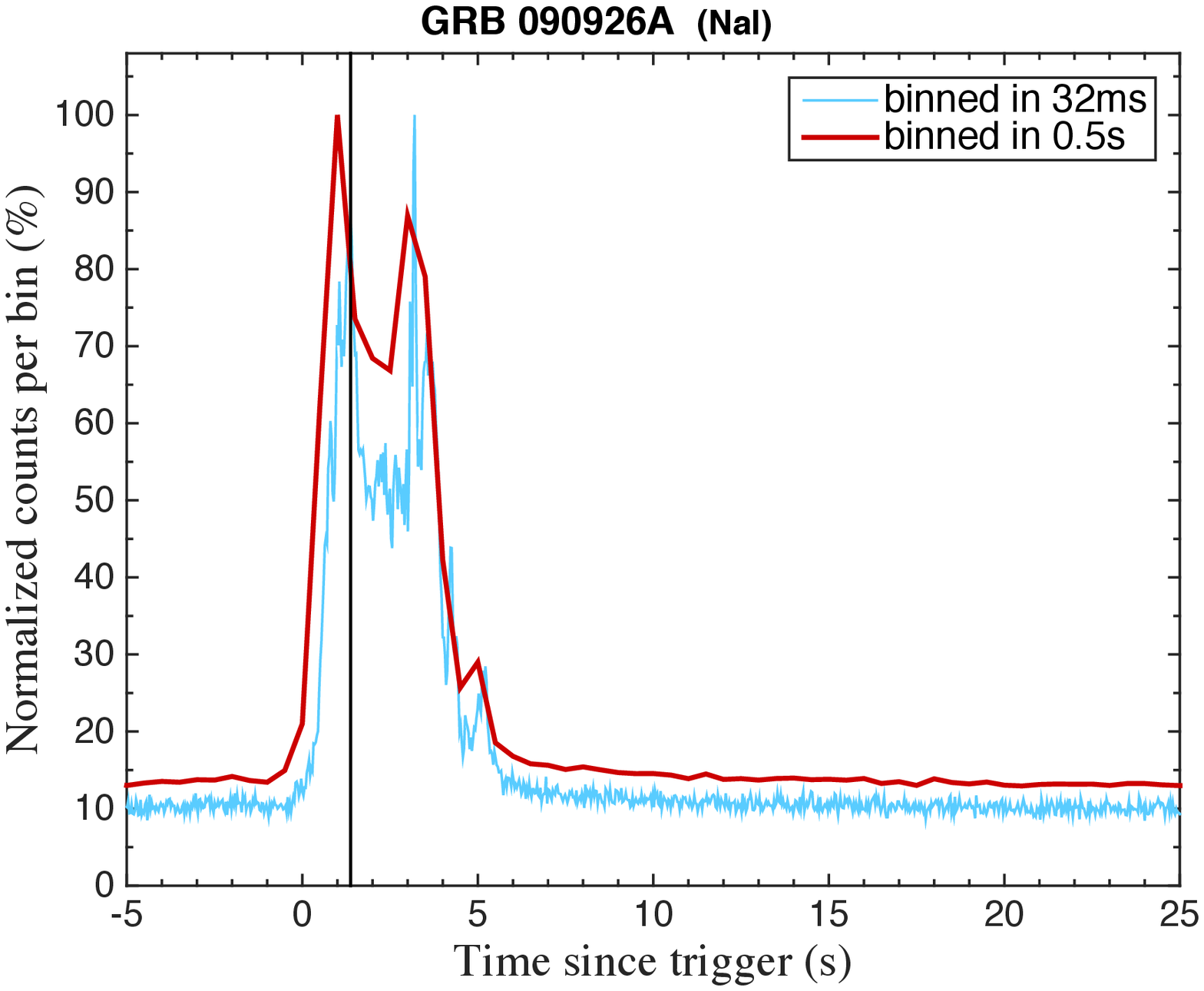}
	\end{minipage}
	\begin{minipage}[t]{0.44\linewidth}
		\centering
		\includegraphics[width=0.90\linewidth]{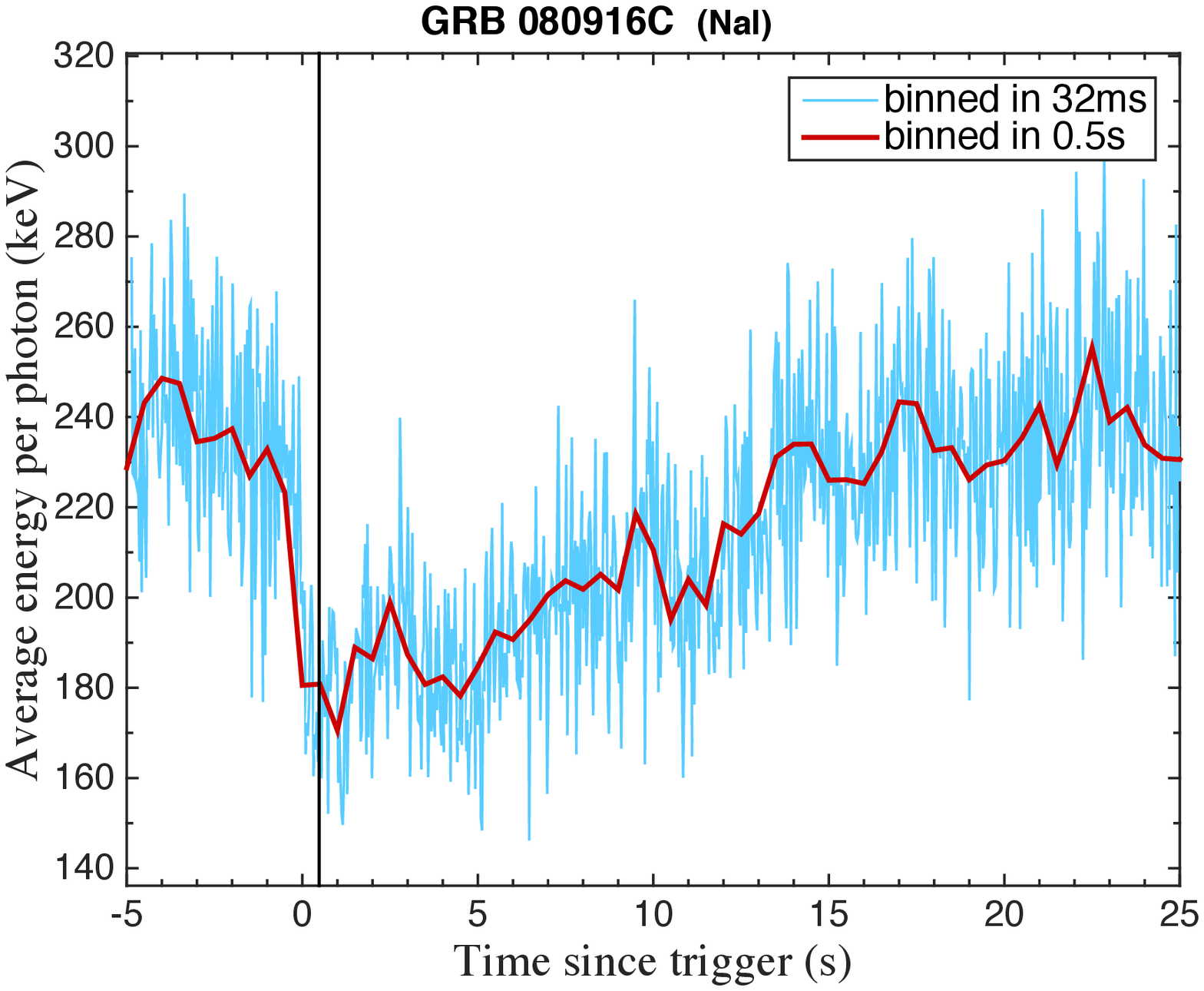}
		\includegraphics[width=0.90\linewidth]{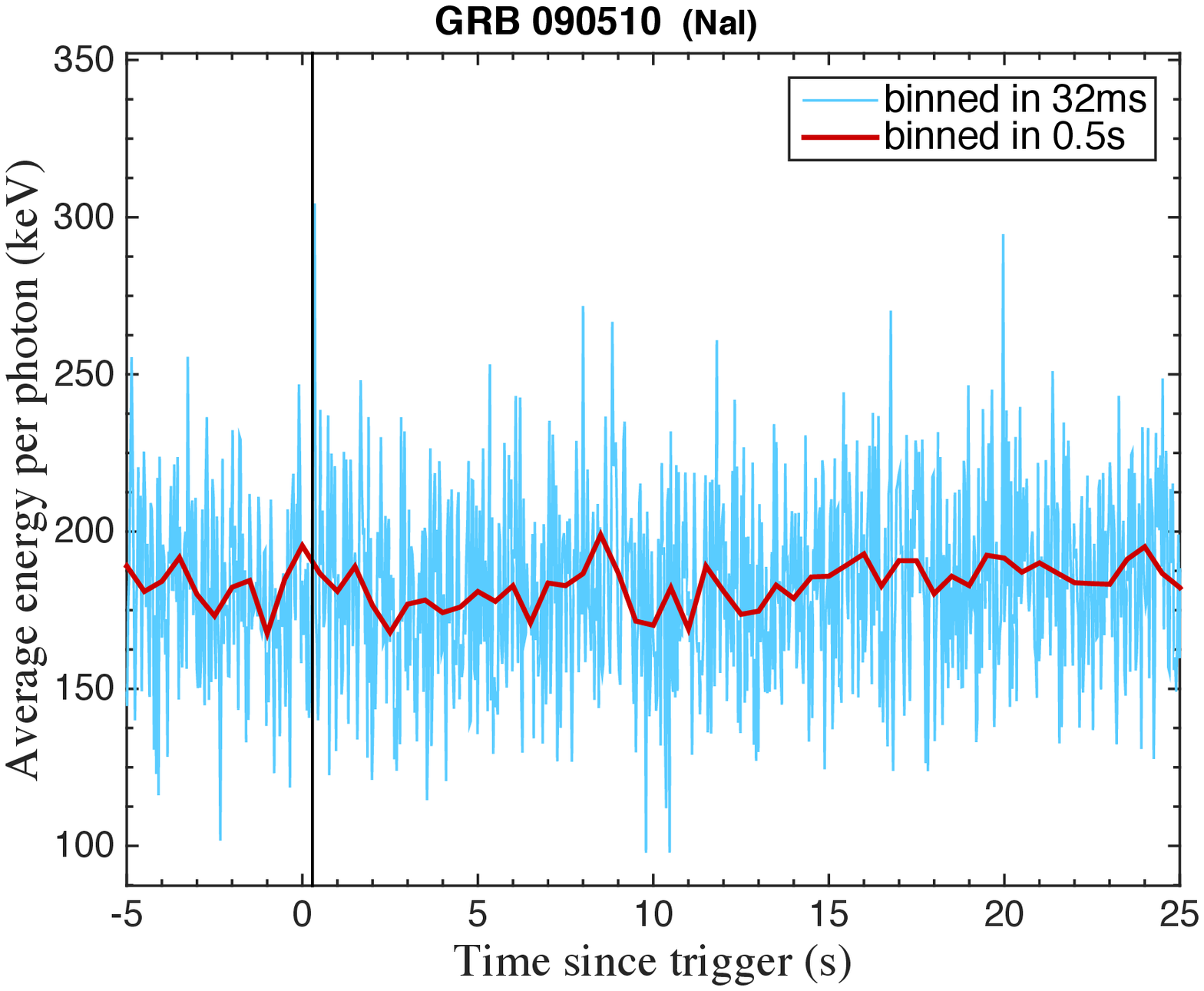}
		\includegraphics[width=0.90\linewidth]{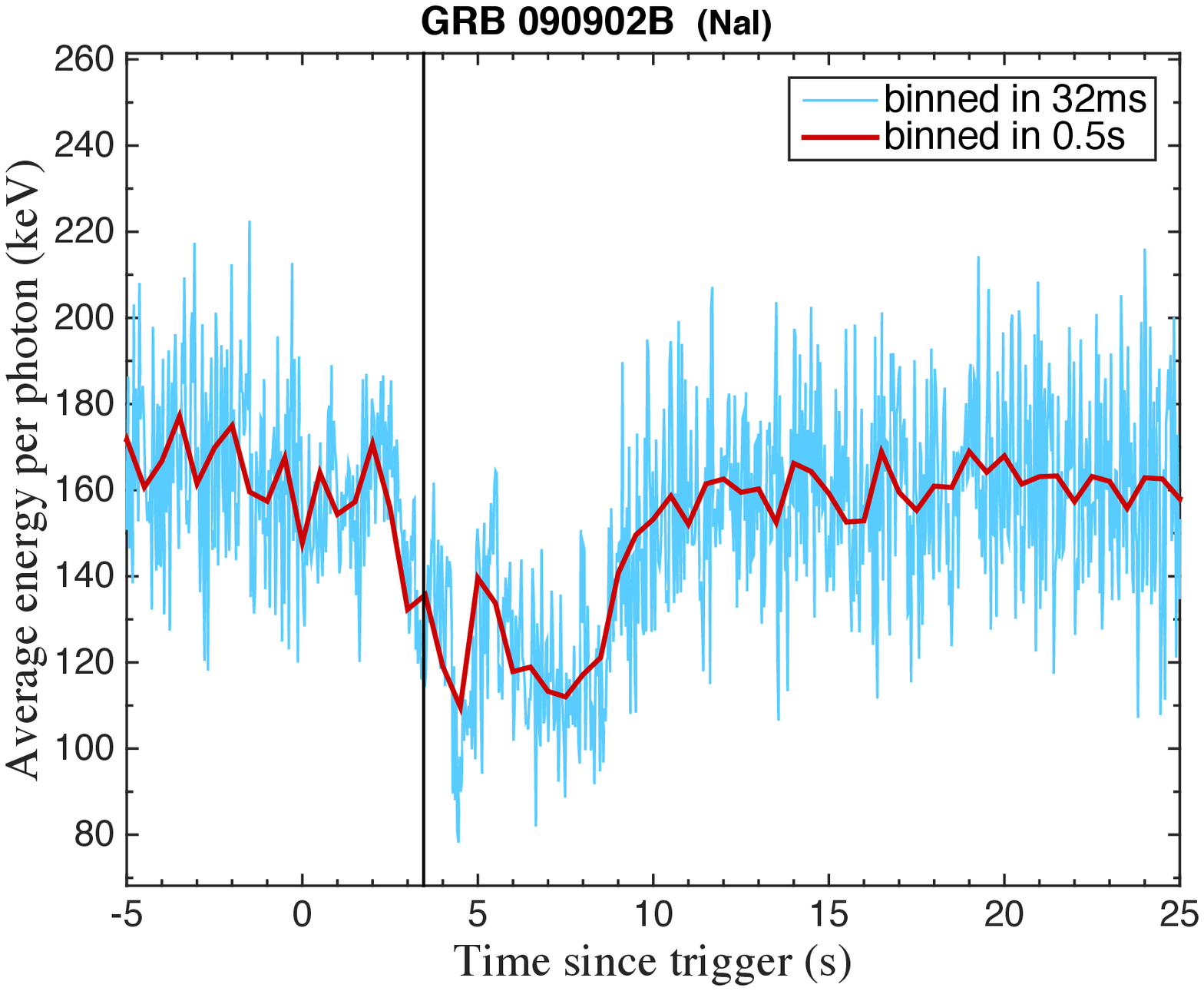}
		\includegraphics[width=0.90\linewidth]{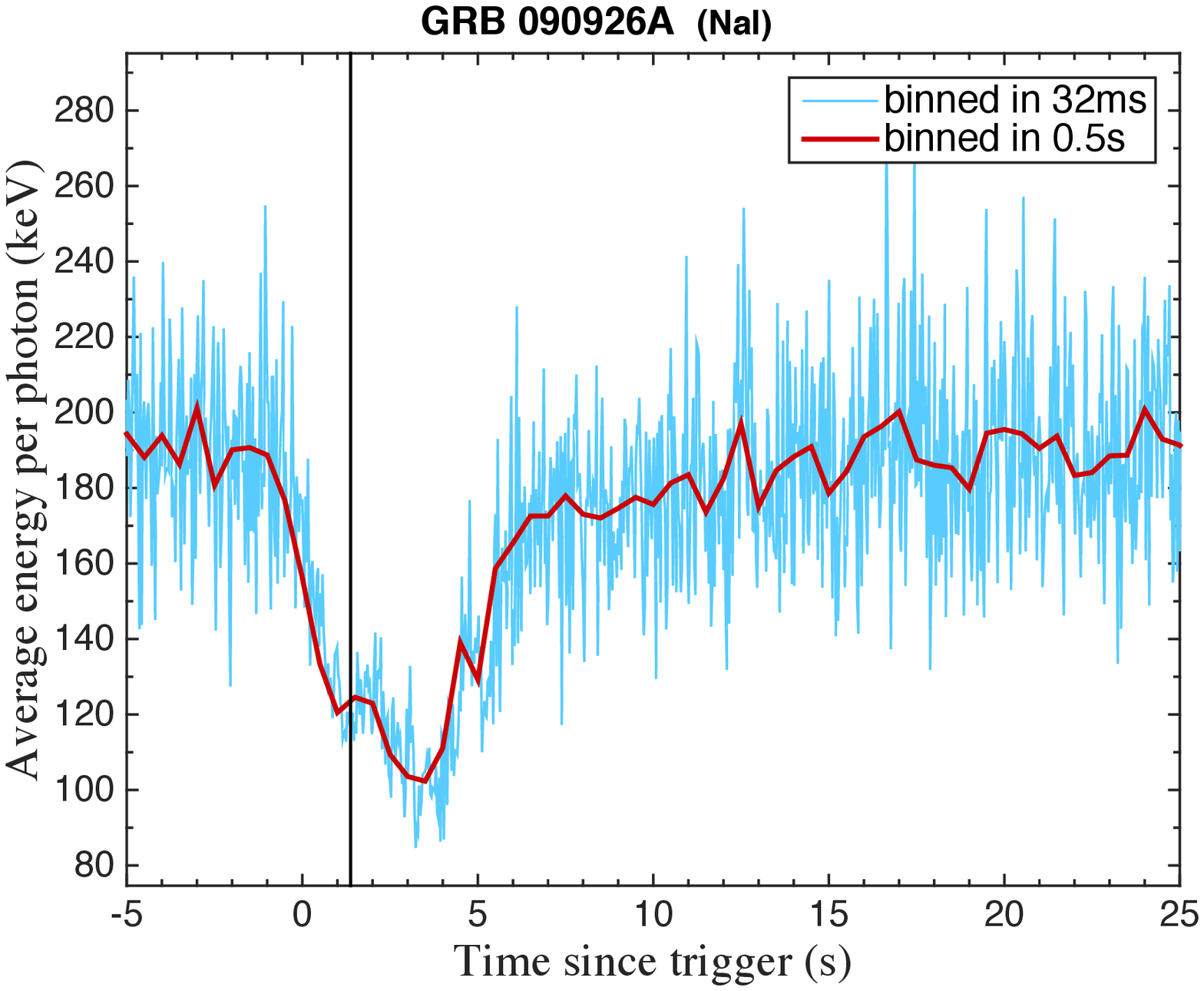}
	\end{minipage}
	\caption[Curve Set I-a]{Curves in Set-NaI (part 1): light curves and average energy curves of GBM NaI data.
		%The light column refers to the light curves for GRB 080916C, 090510, 090902B
	}\label{nai-1}
\end{figure*}

\begin{figure*}[htbp]
	\centering
	\begin{minipage}[t]{0.44\linewidth}
		\centering
		\includegraphics[width=0.90\linewidth]{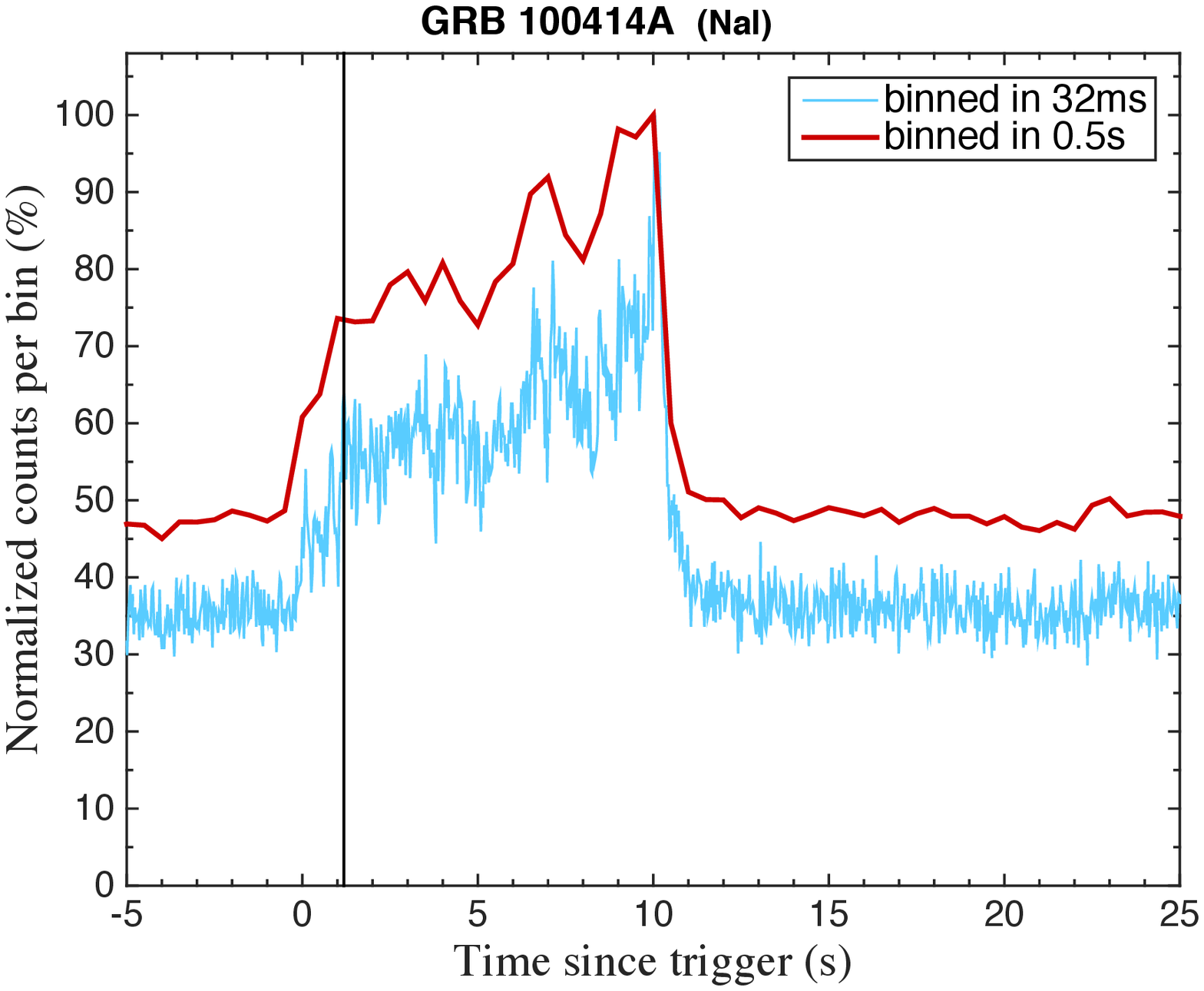}
		\includegraphics[width=0.90\linewidth]{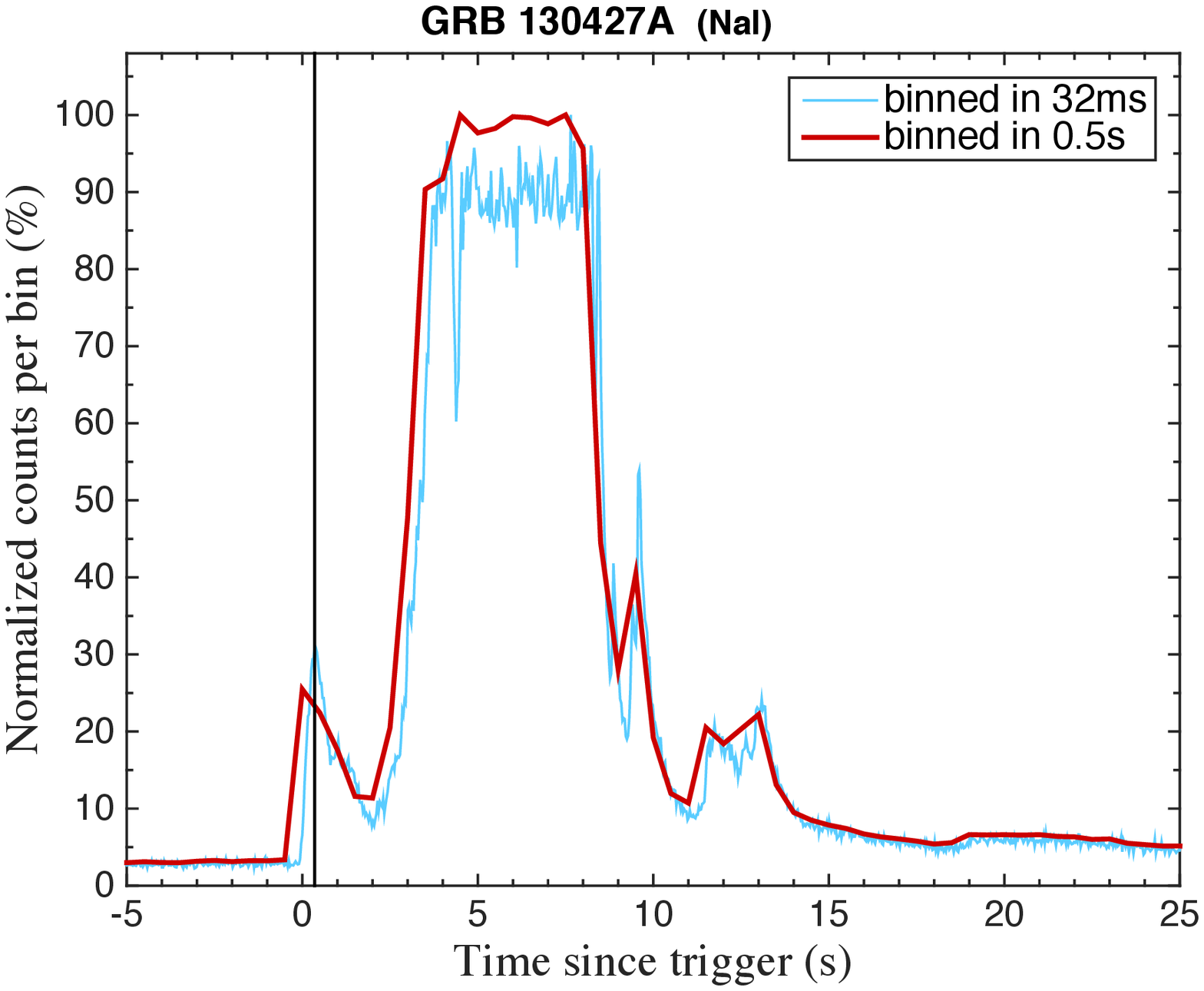}
		\includegraphics[width=0.90\linewidth]{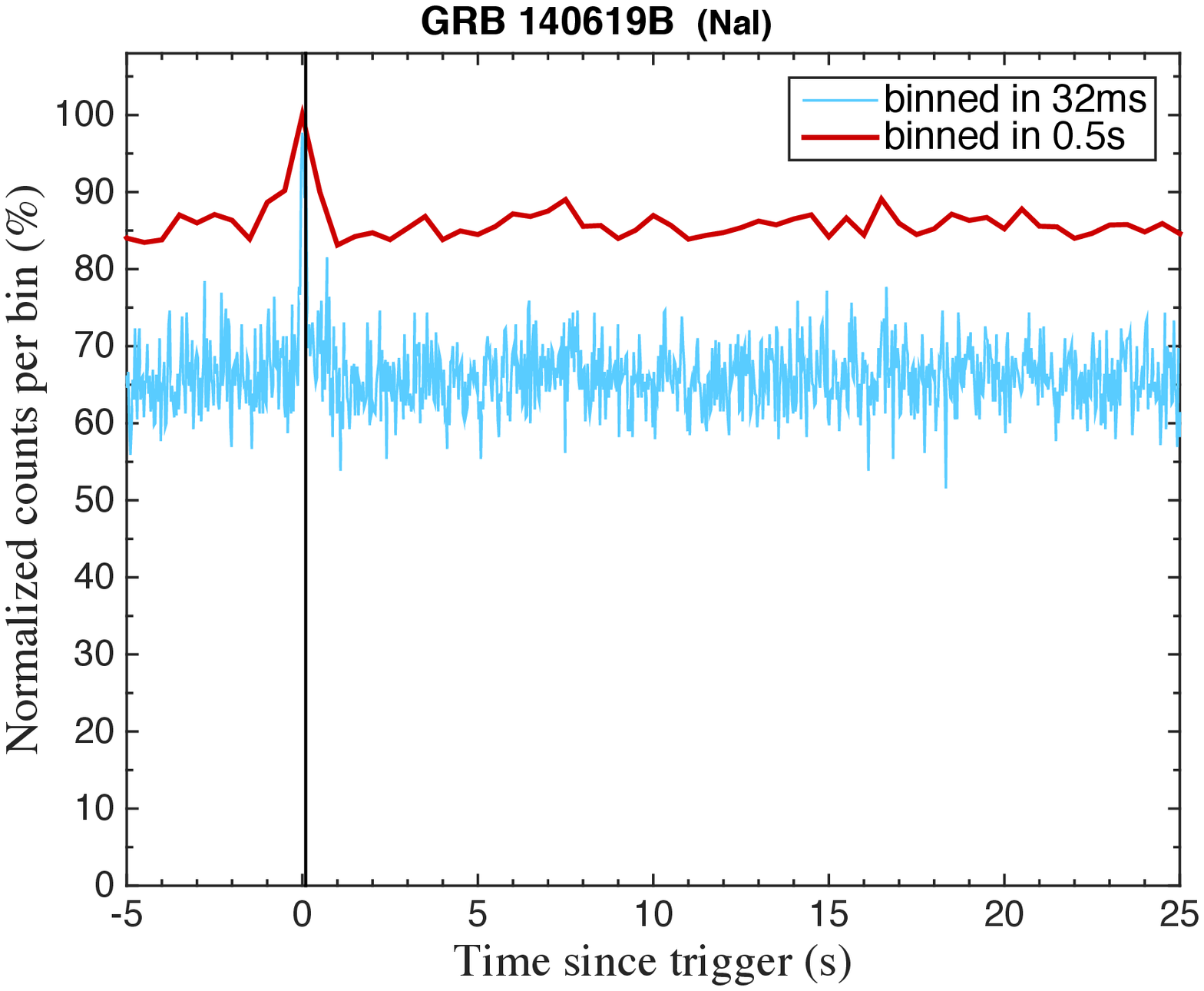}
		\includegraphics[width=0.90\linewidth]{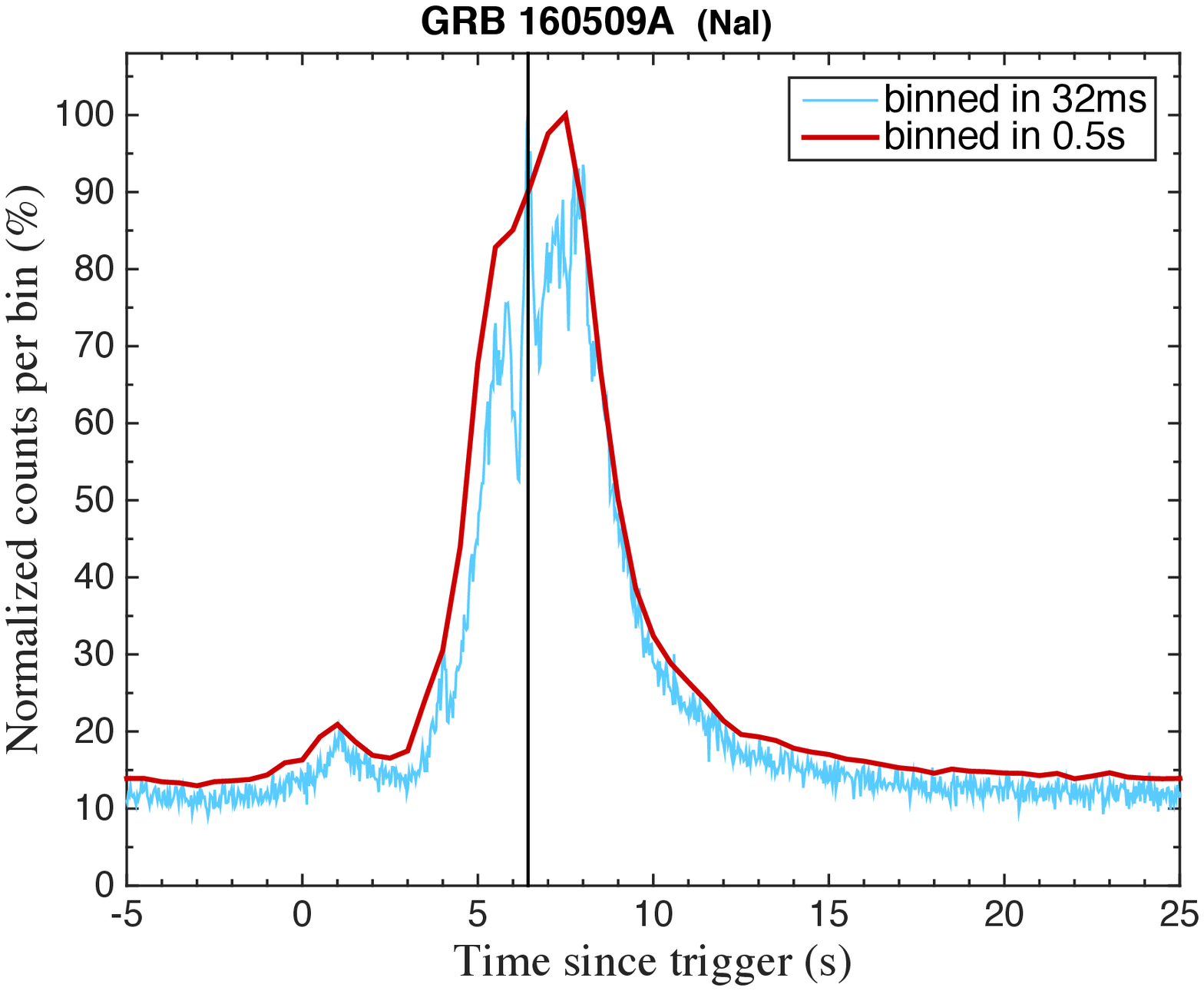}
	\end{minipage}
	\begin{minipage}[t]{0.44\linewidth}
		\centering
		\includegraphics[width=0.90\linewidth]{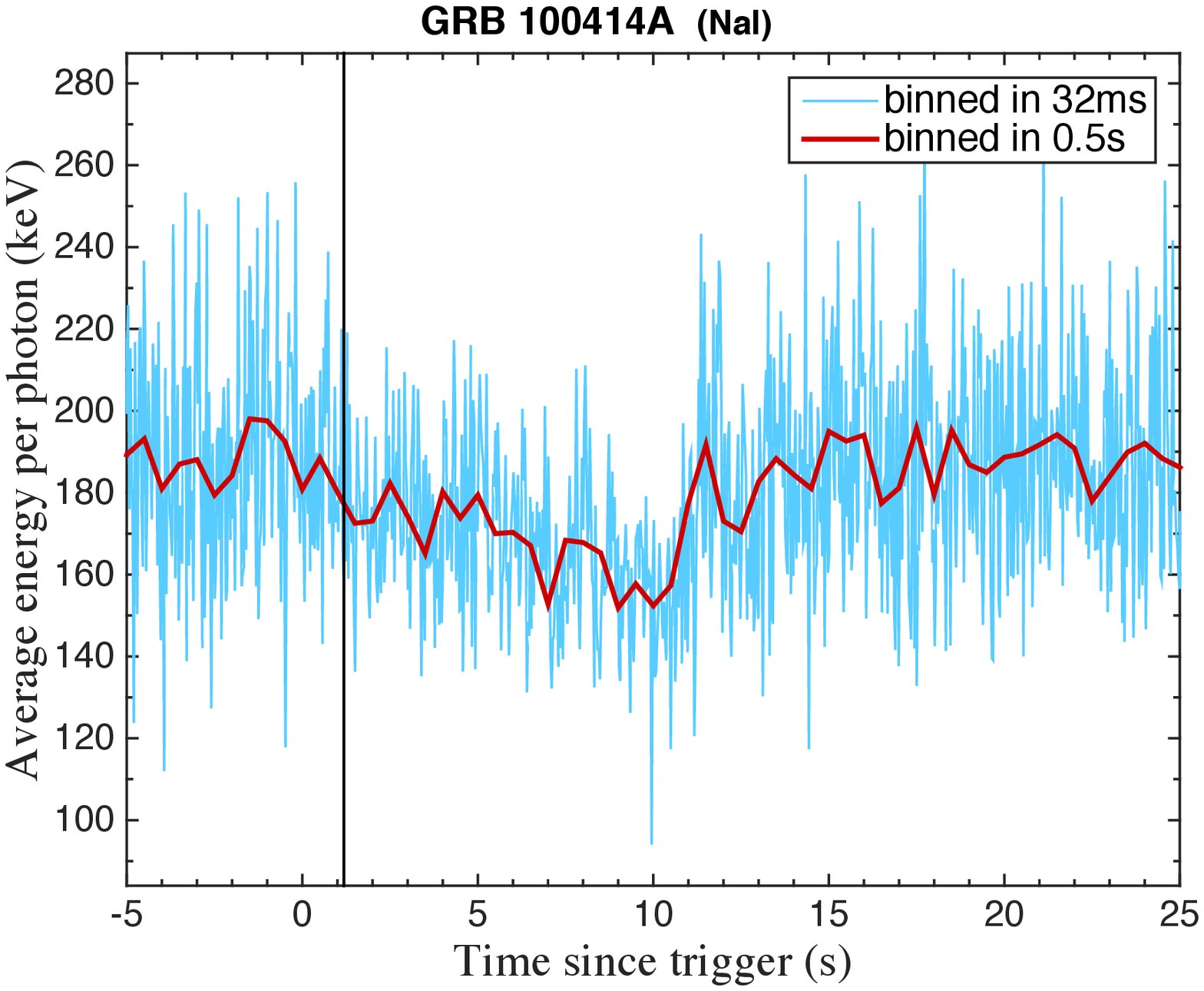}
		\includegraphics[width=0.90\linewidth]{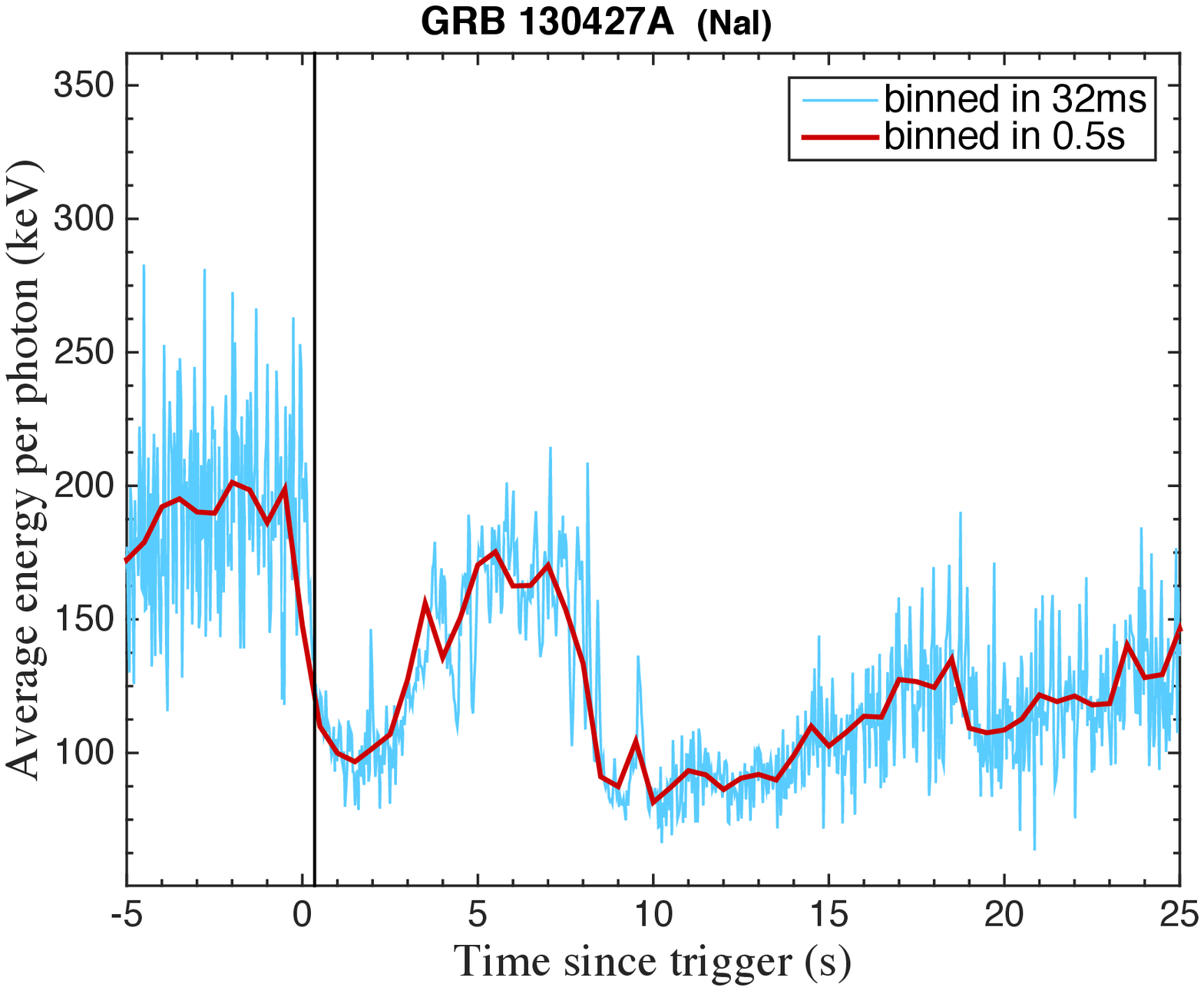}
		\includegraphics[width=0.90\linewidth]{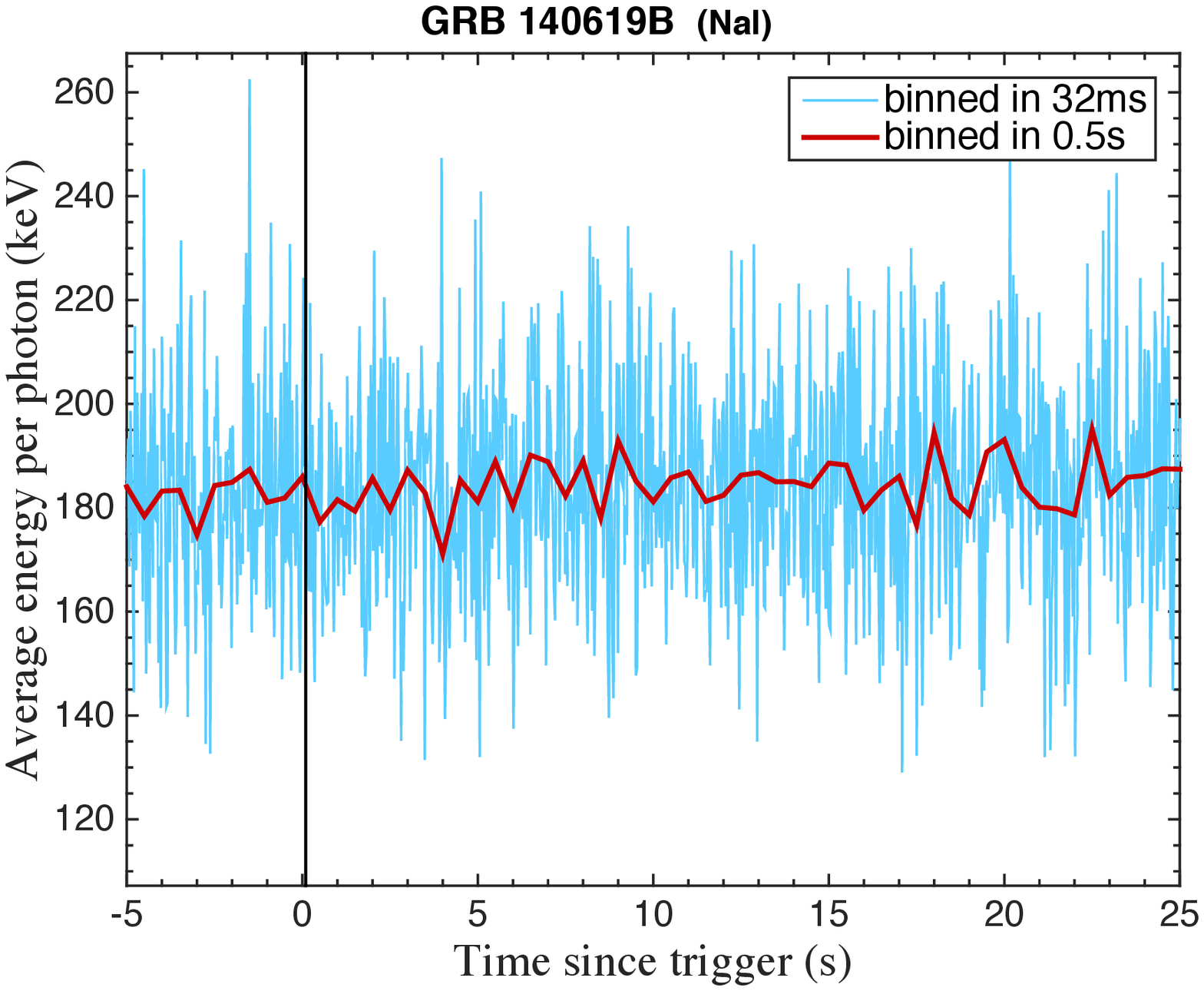}
		\includegraphics[width=0.90\linewidth]{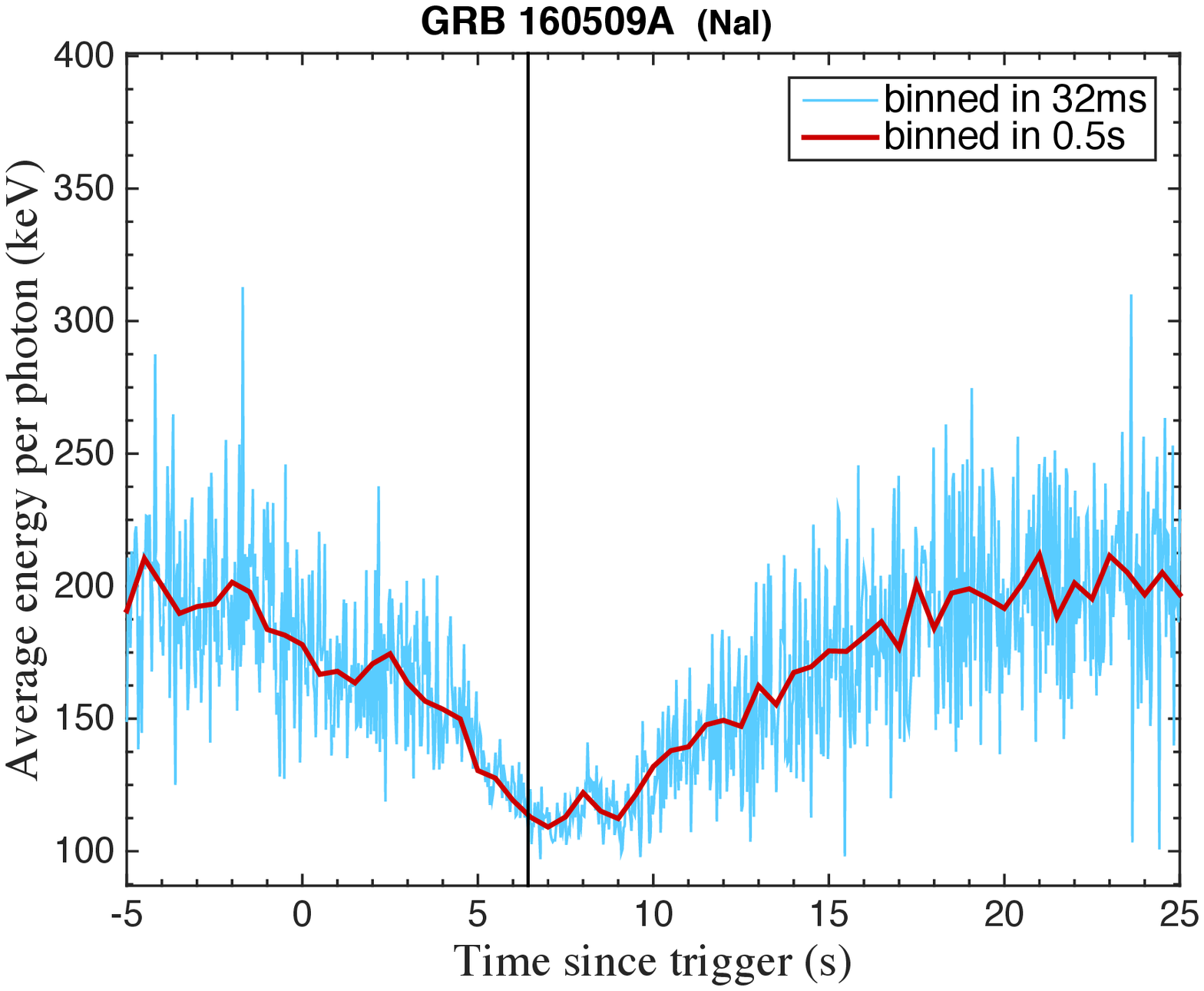}
	\end{minipage}
	\caption[Curve Set I-b]{Curves in Set-NaI (part 2): light curves and average energy curves of GBM NaI data.}\label{nai-2}
\end{figure*}

\begin{figure*}[htbp]
	\centering
	\begin{minipage}[t]{0.44\linewidth}
		\centering
		\includegraphics[width=0.90\linewidth]{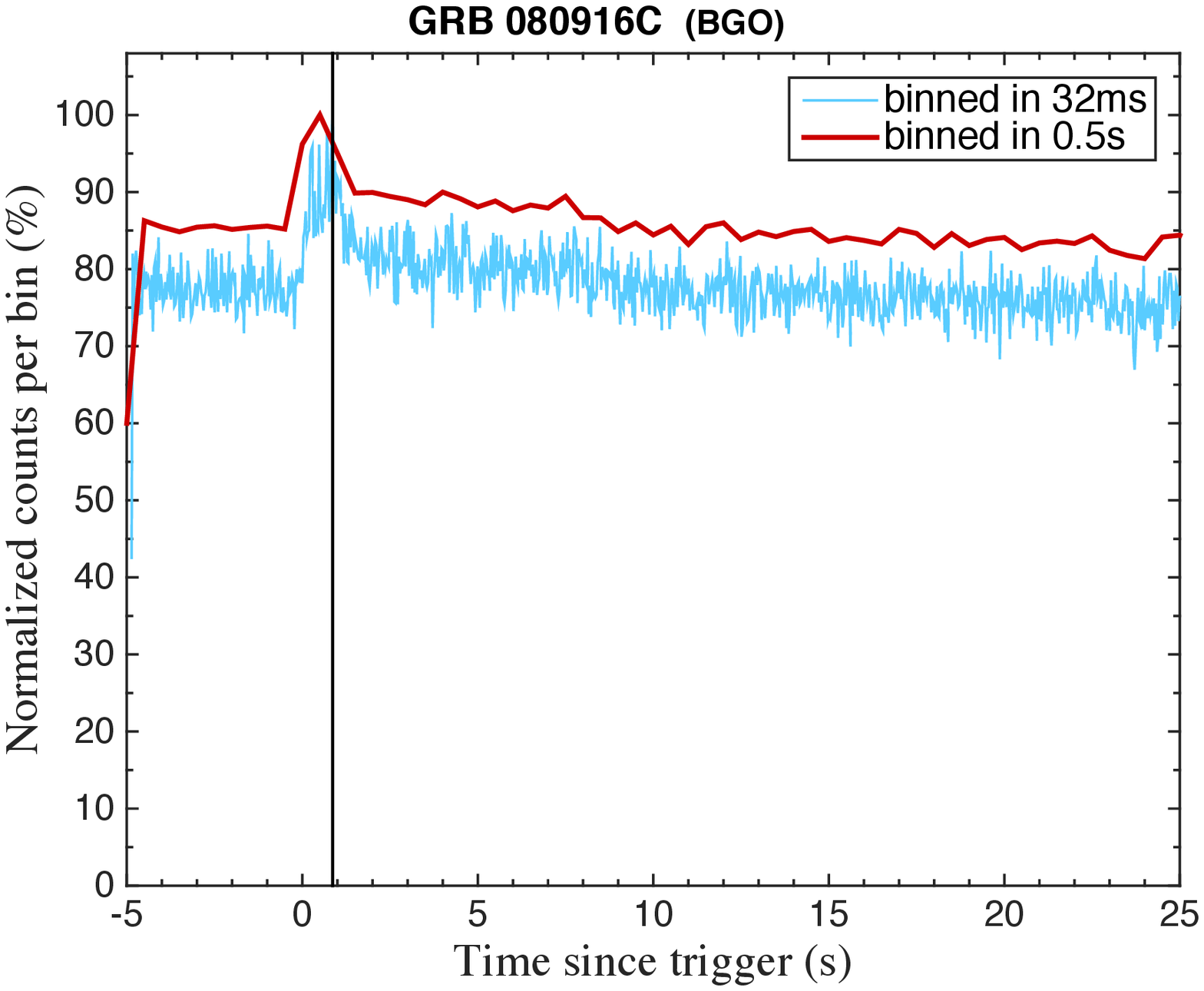}
		\includegraphics[width=0.90\linewidth]{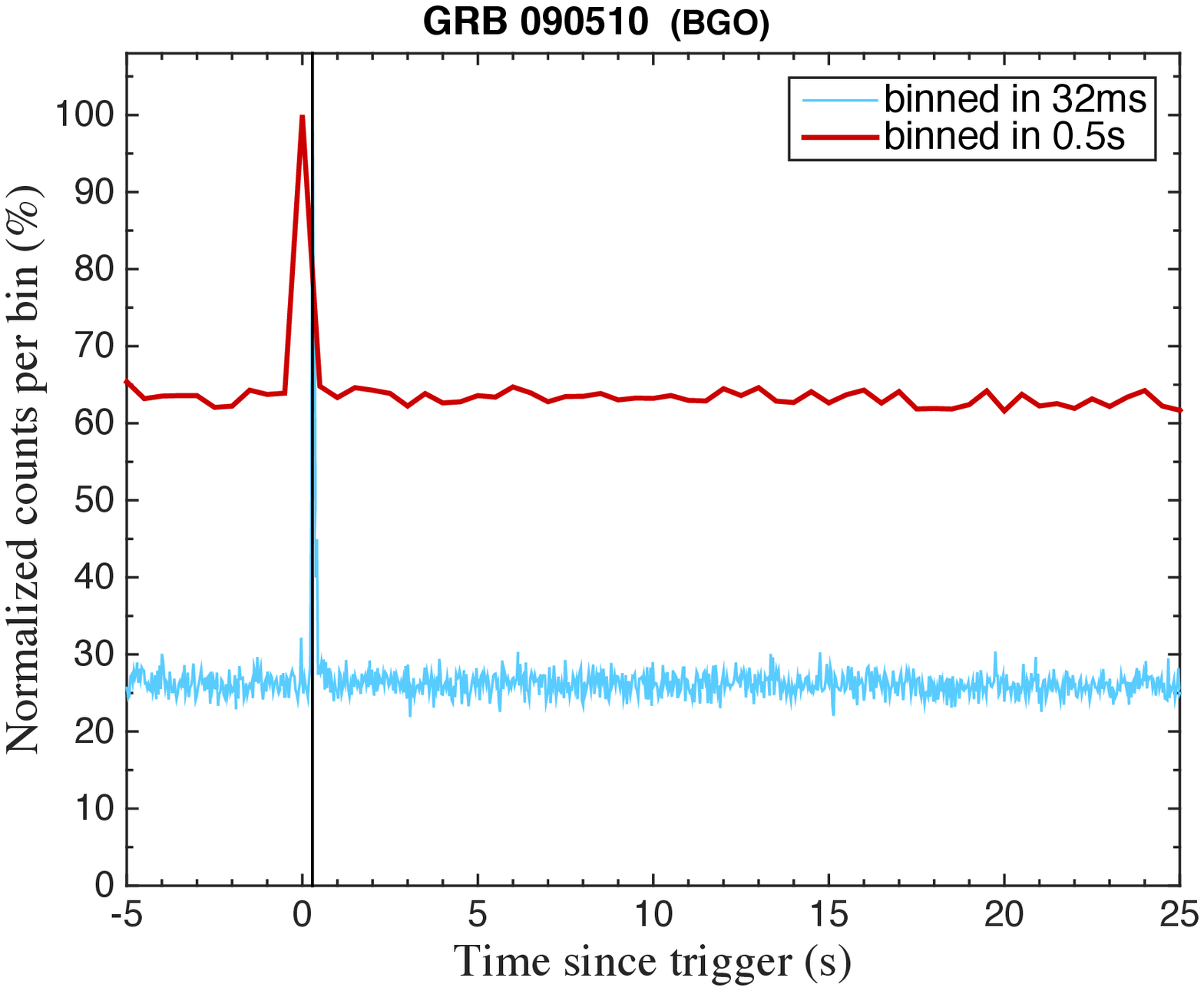}
		\includegraphics[width=0.90\linewidth]{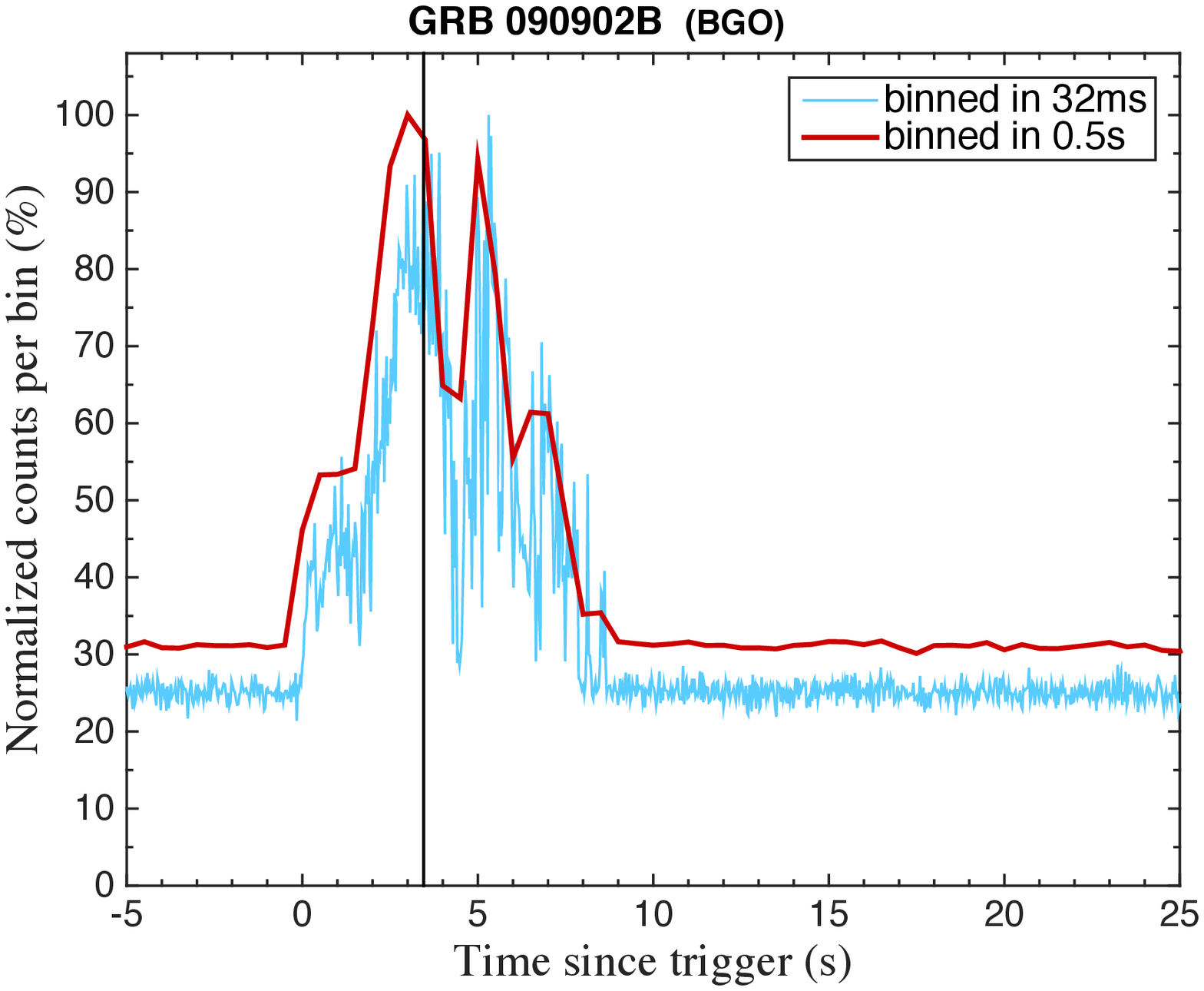}
		\includegraphics[width=0.90\linewidth]{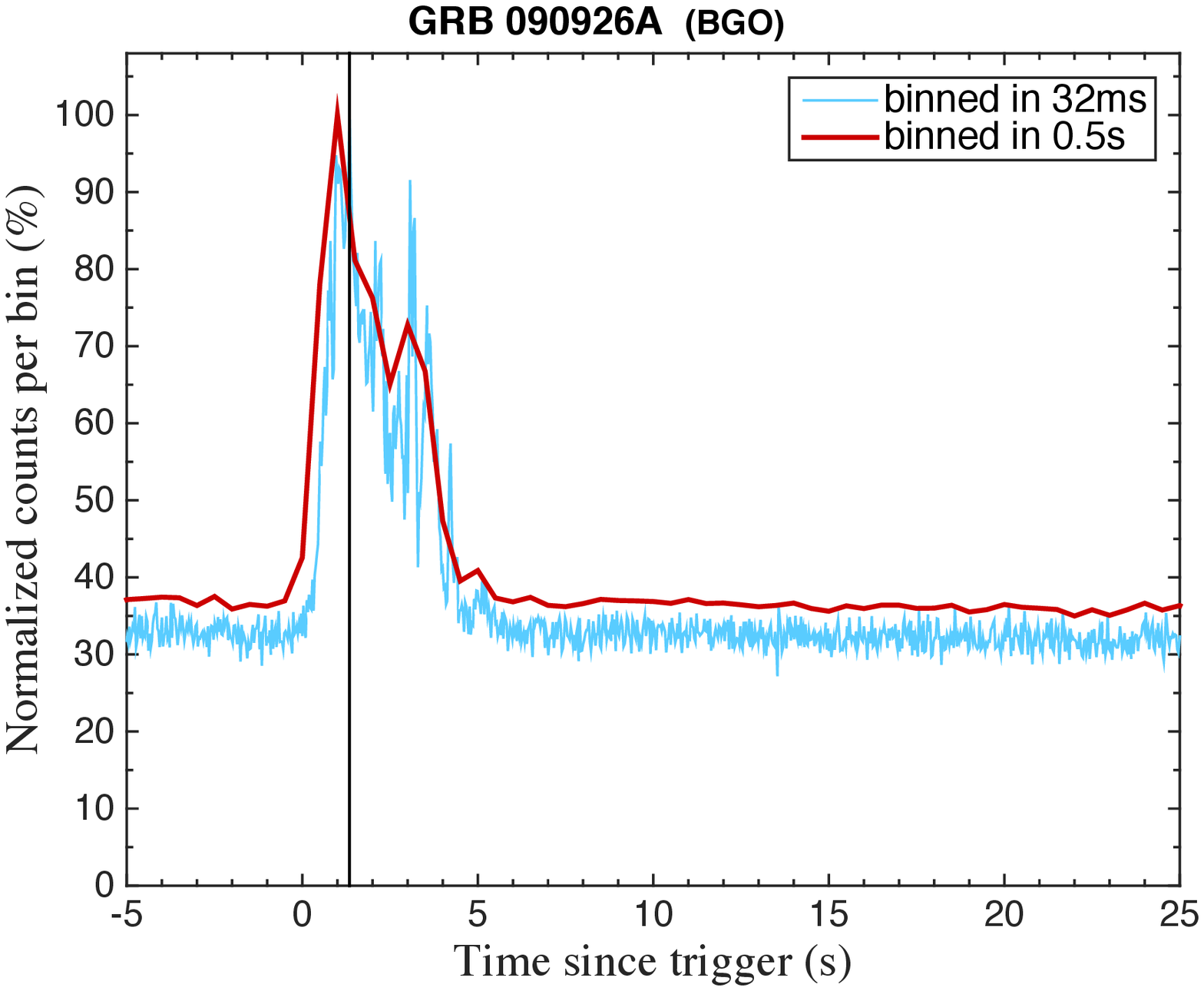}
	\end{minipage}
	\begin{minipage}[t]{0.44\linewidth}
		\centering
		\includegraphics[width=0.90\linewidth]{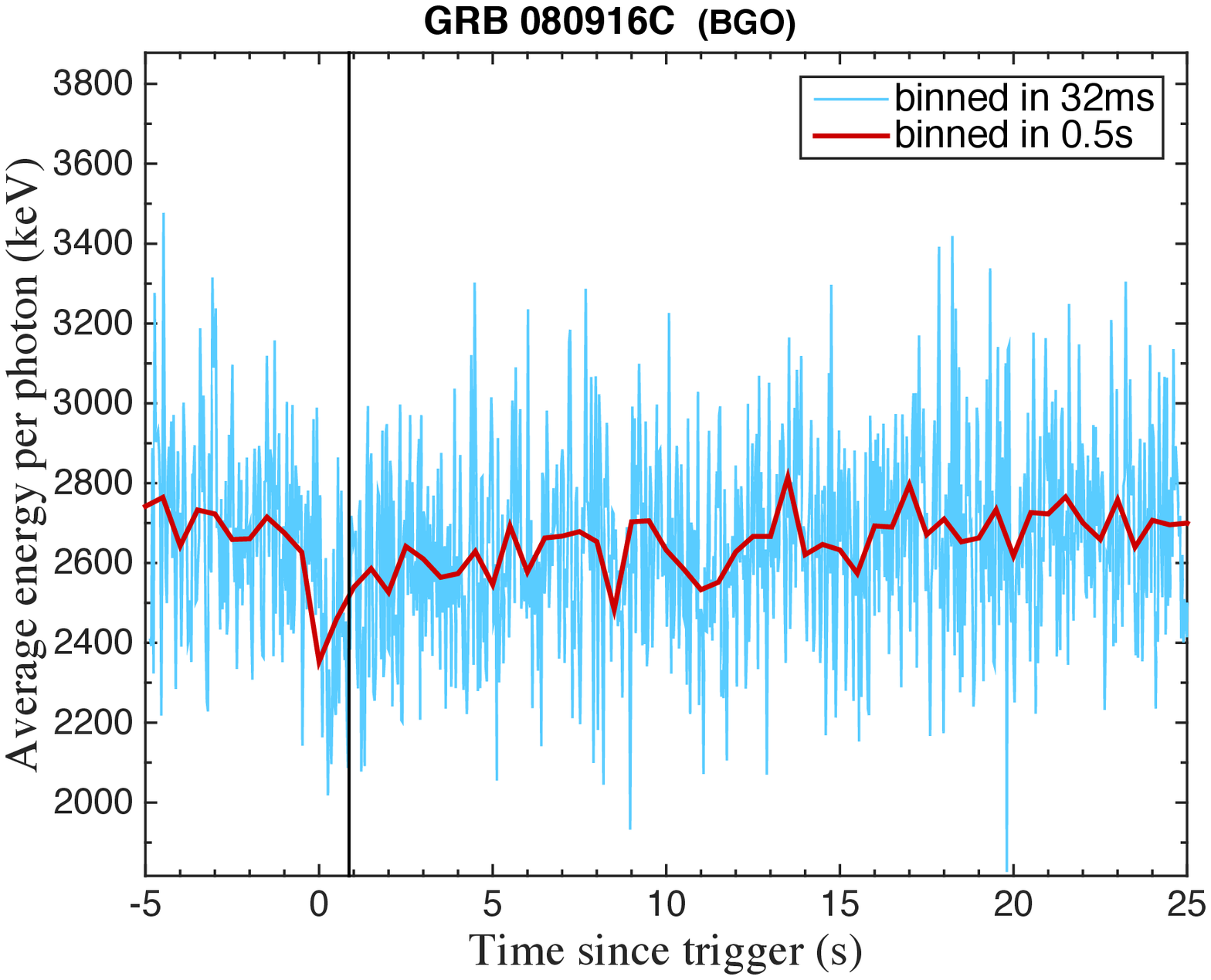}
		\includegraphics[width=0.90\linewidth]{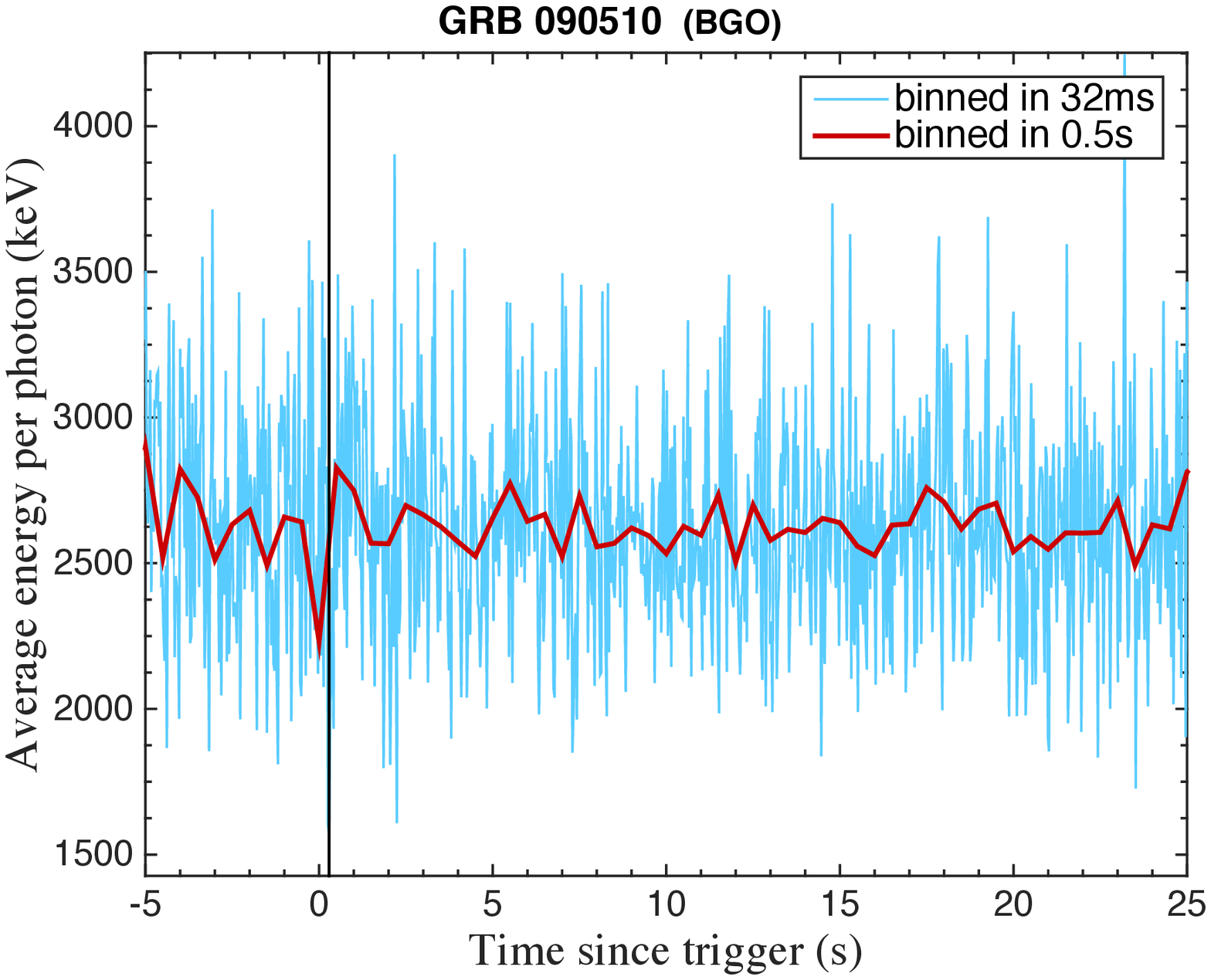}
		\includegraphics[width=0.90\linewidth]{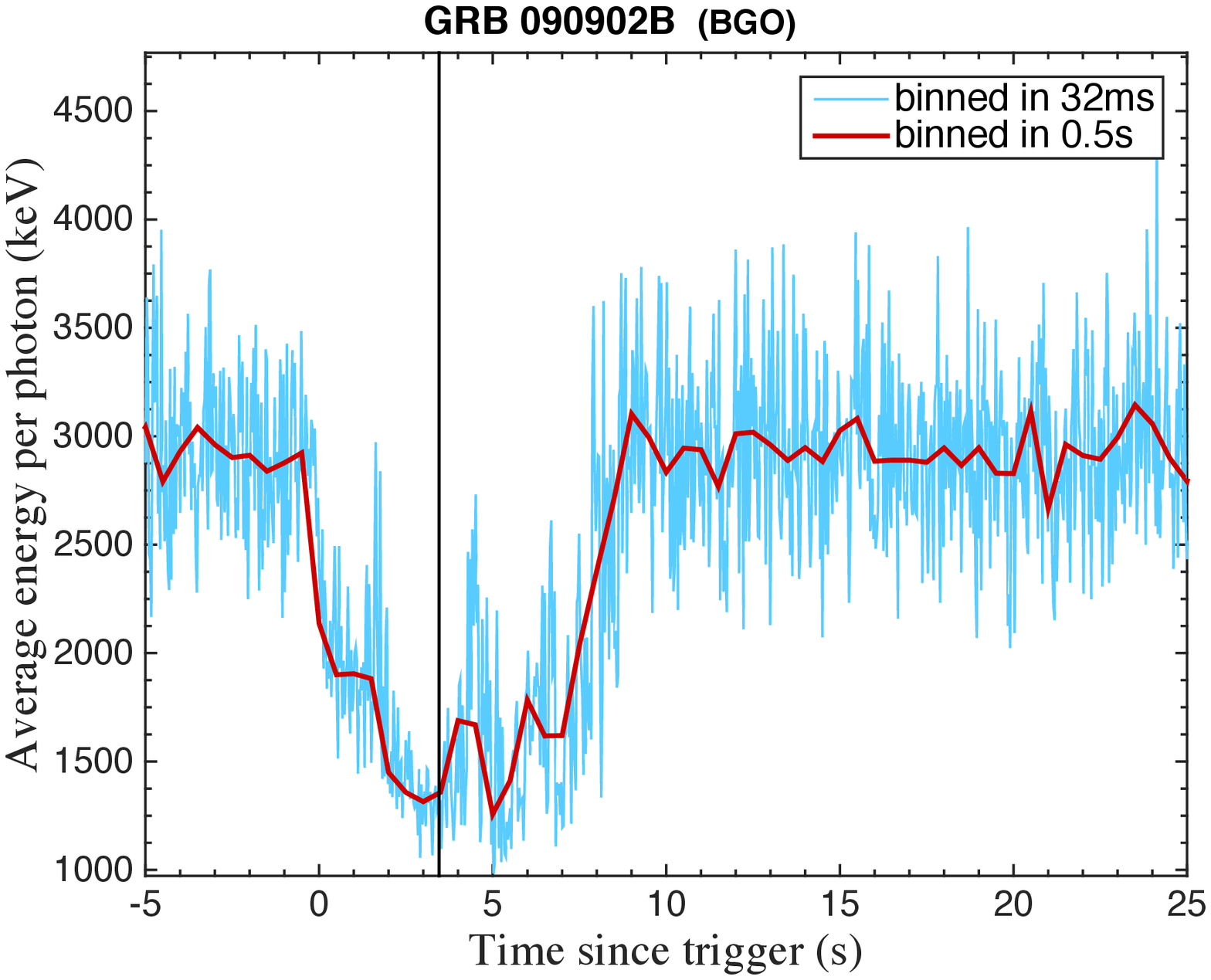}
		\includegraphics[width=0.90\linewidth]{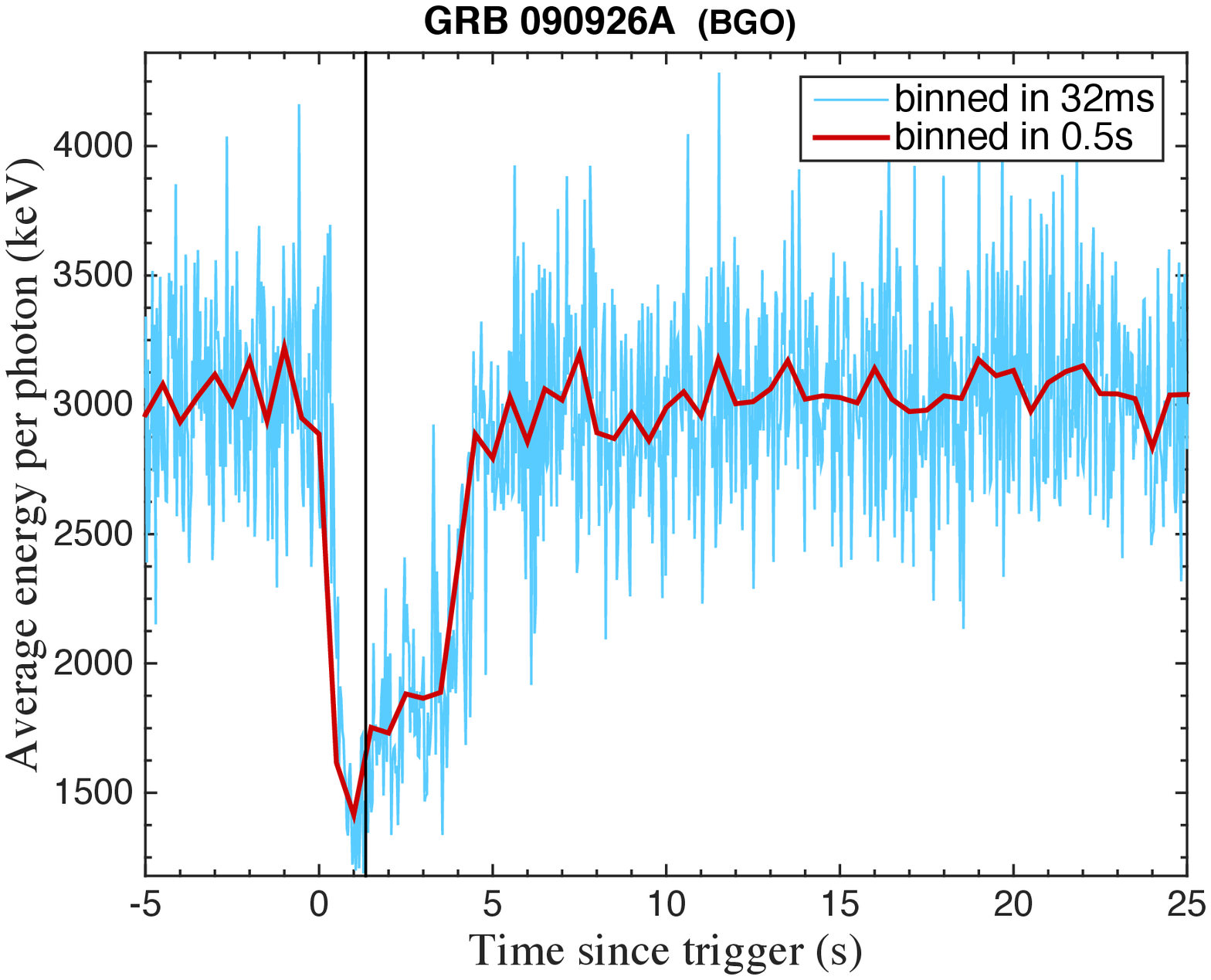}
	\end{minipage}
	\caption[Curve Set II-a]{Curves in Set-BGO (part 1): Light curves and average energy curves of GBM BGO data.}\label{bgo-1}
\end{figure*}

\begin{figure*}[htbp]
	\centering
	\begin{minipage}[t]{0.44\linewidth}
		\centering
		\includegraphics[width=0.90\linewidth]{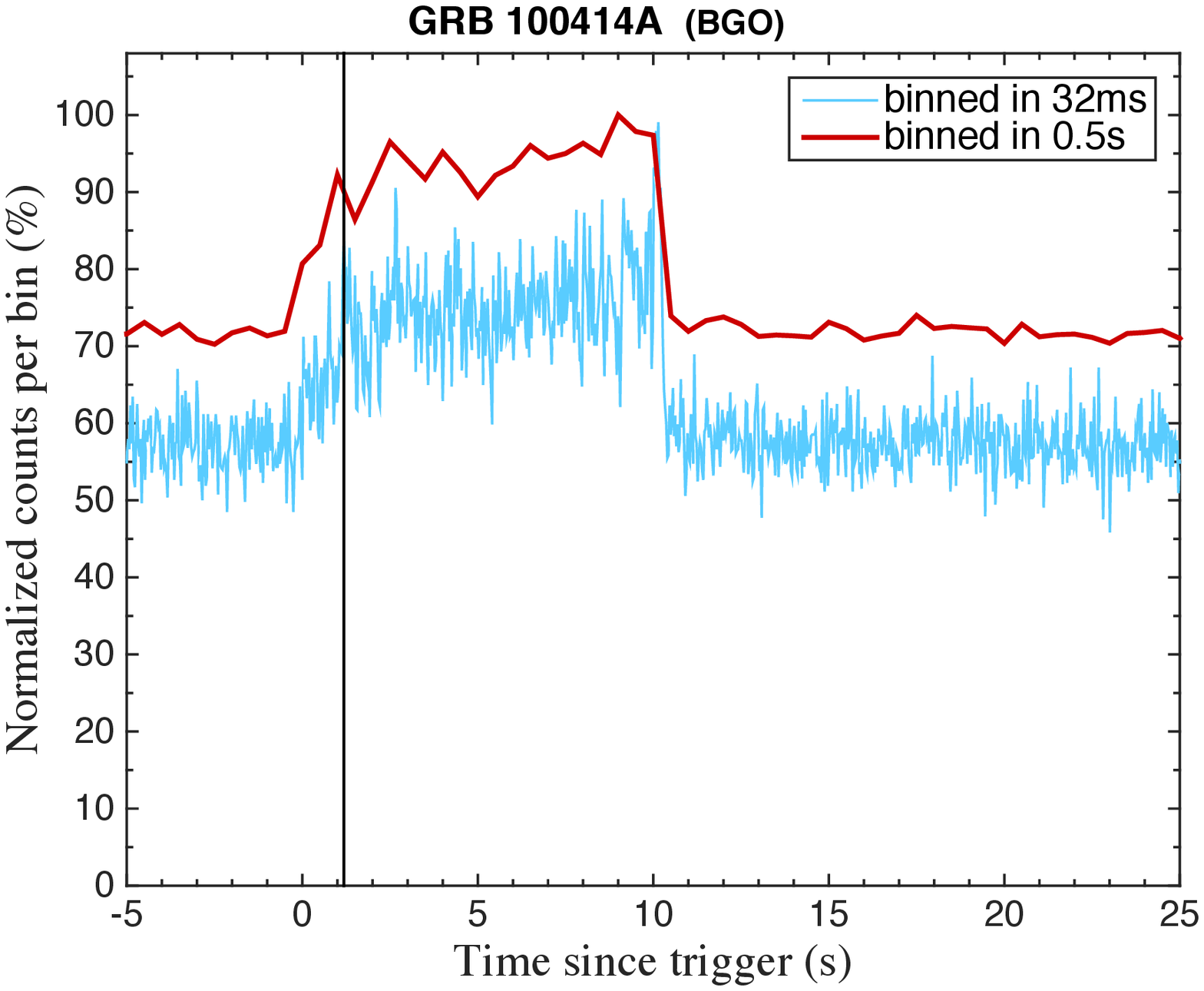}
		\includegraphics[width=0.90\linewidth]{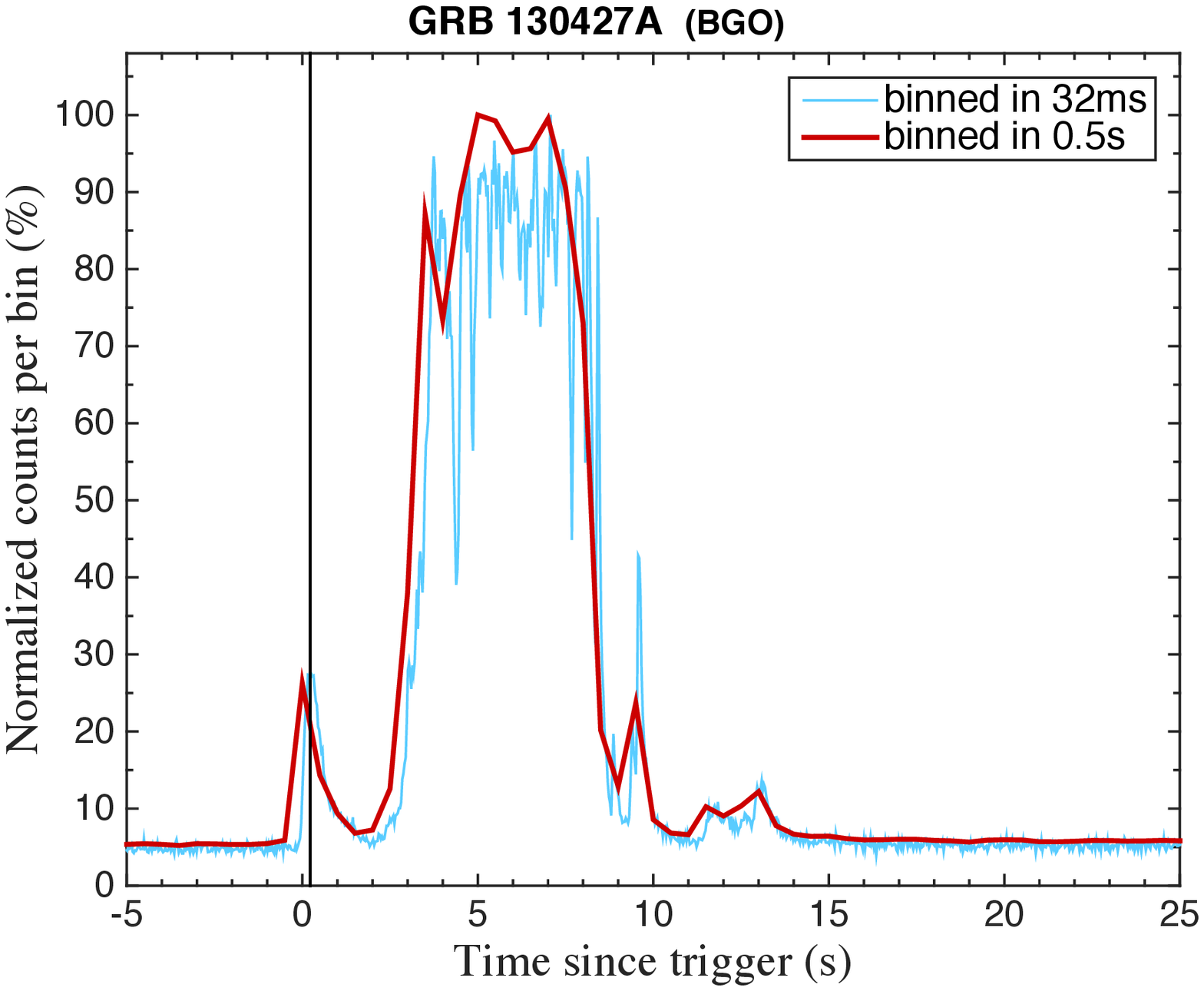}
		\includegraphics[width=0.90\linewidth]{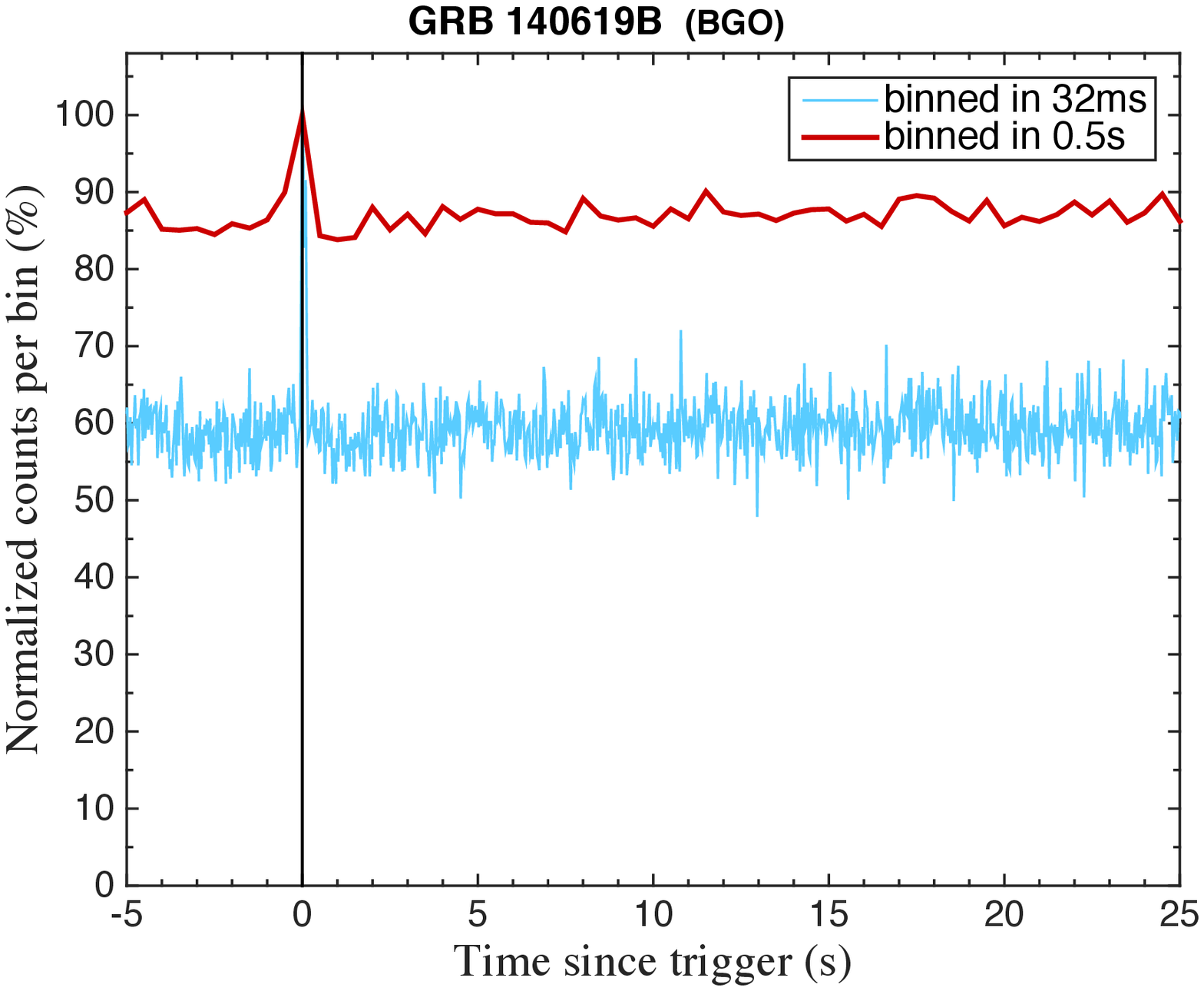}
		\includegraphics[width=0.90\linewidth]{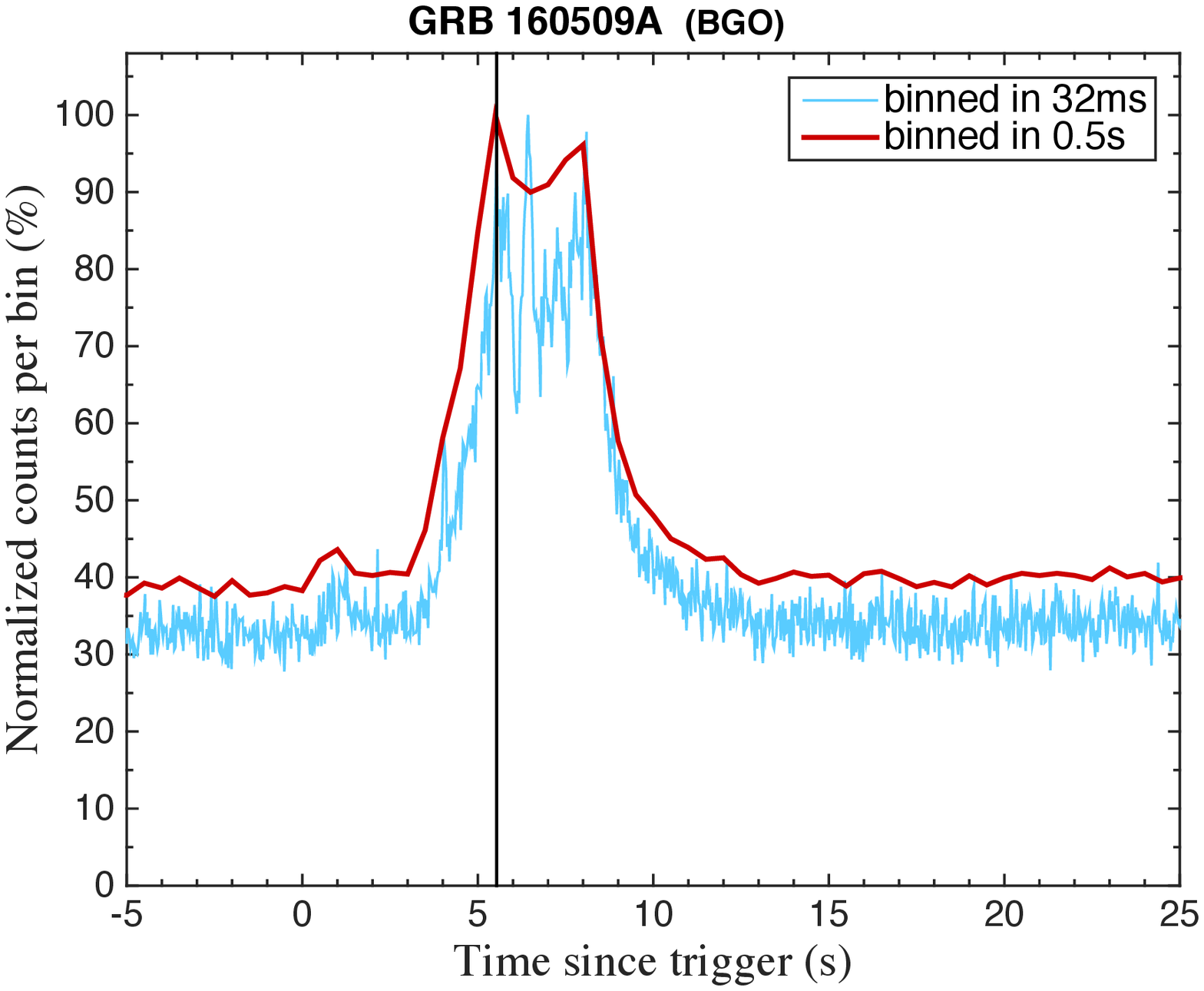}
	\end{minipage}
	\begin{minipage}[t]{0.44\linewidth}
		\centering
		\includegraphics[width=0.90\linewidth]{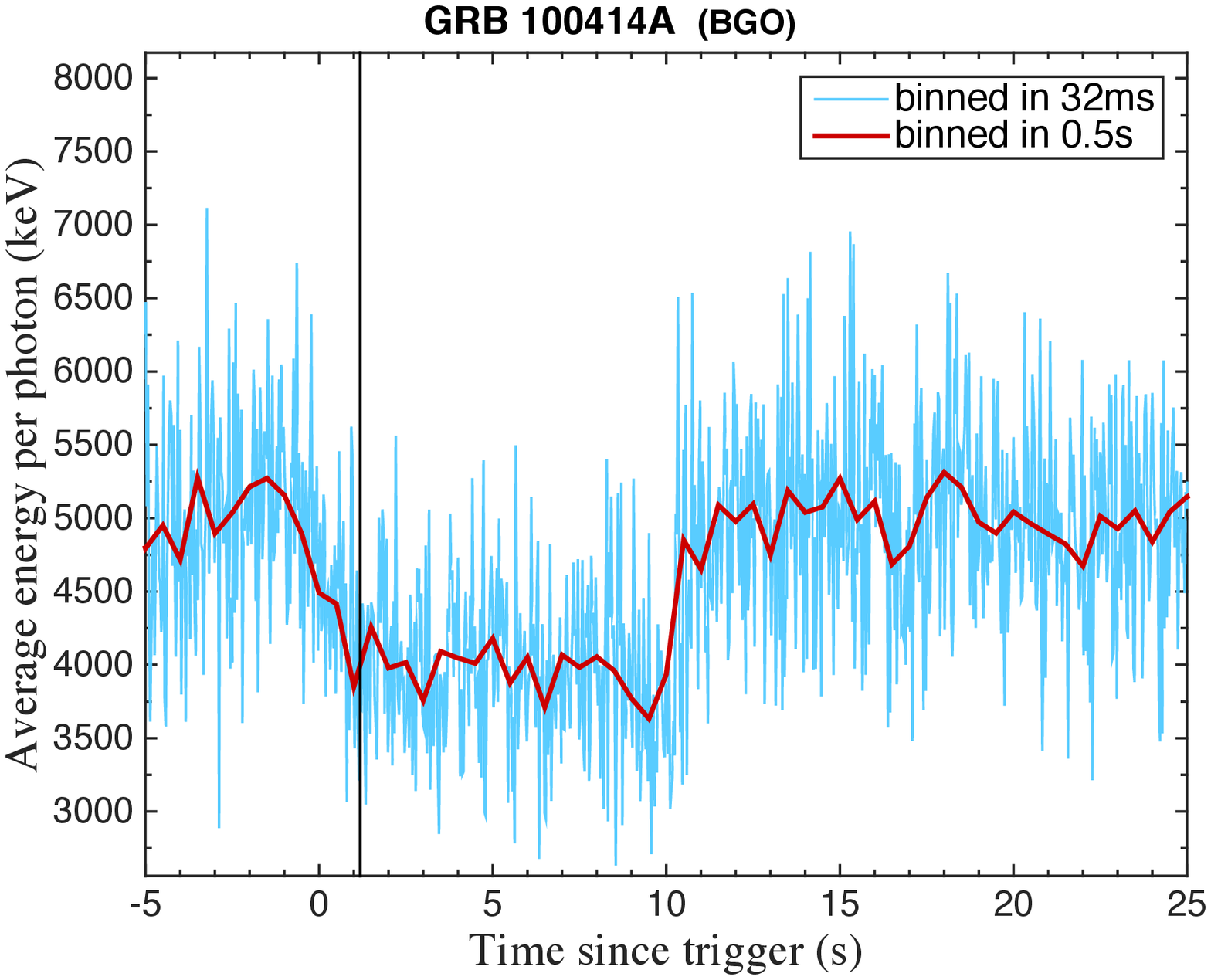}
		\includegraphics[width=0.90\linewidth]{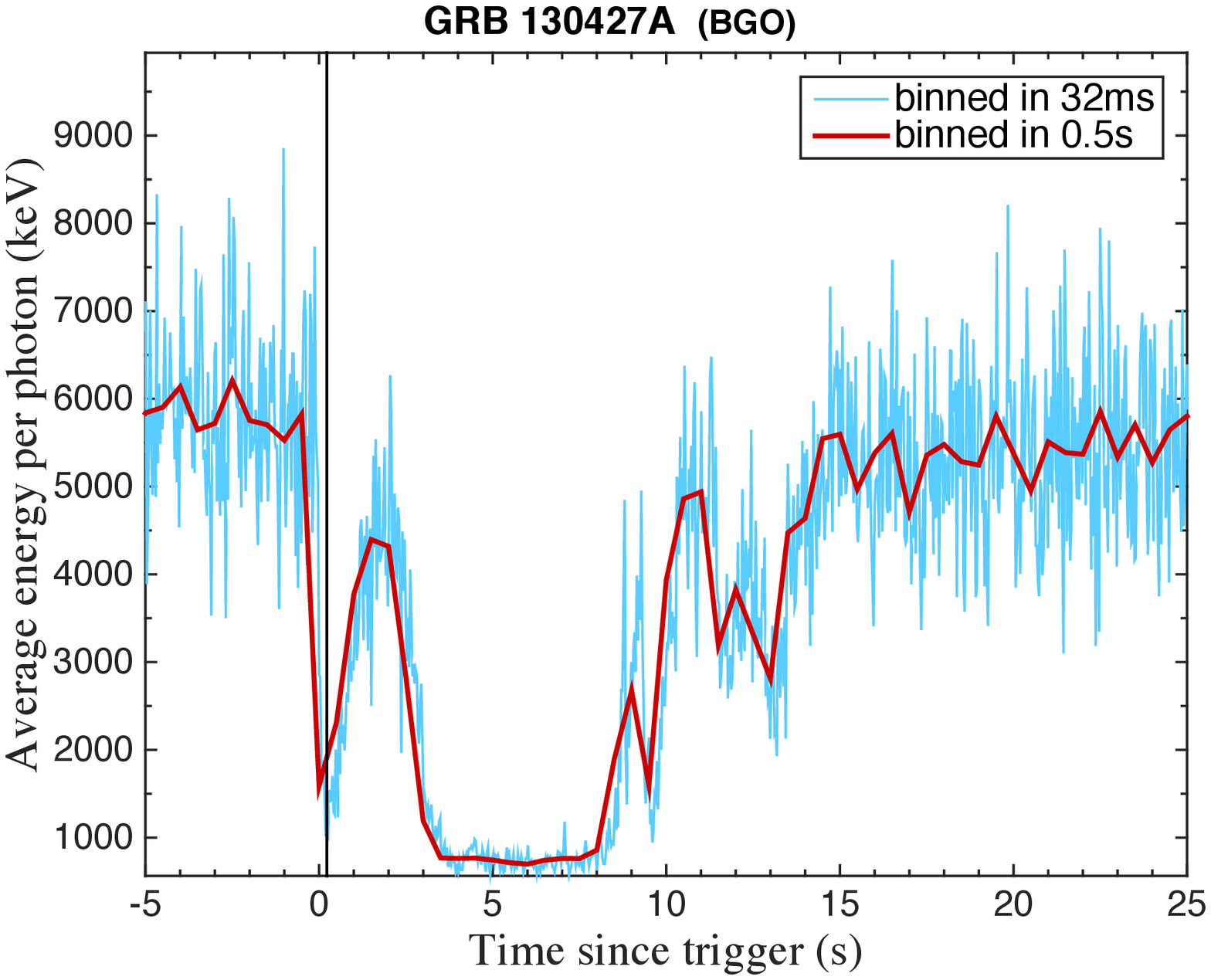}
		\includegraphics[width=0.90\linewidth]{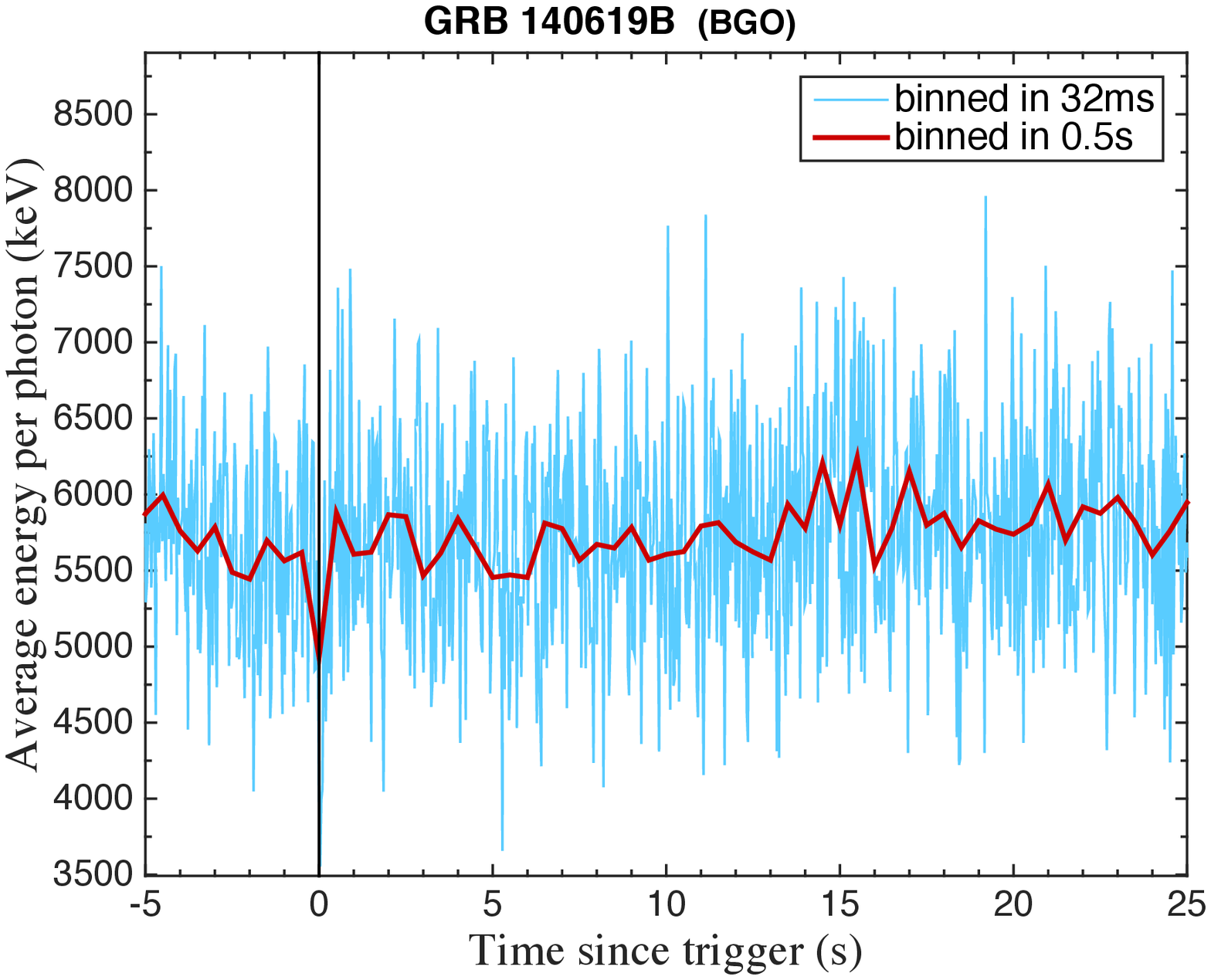}
		\includegraphics[width=0.90\linewidth]{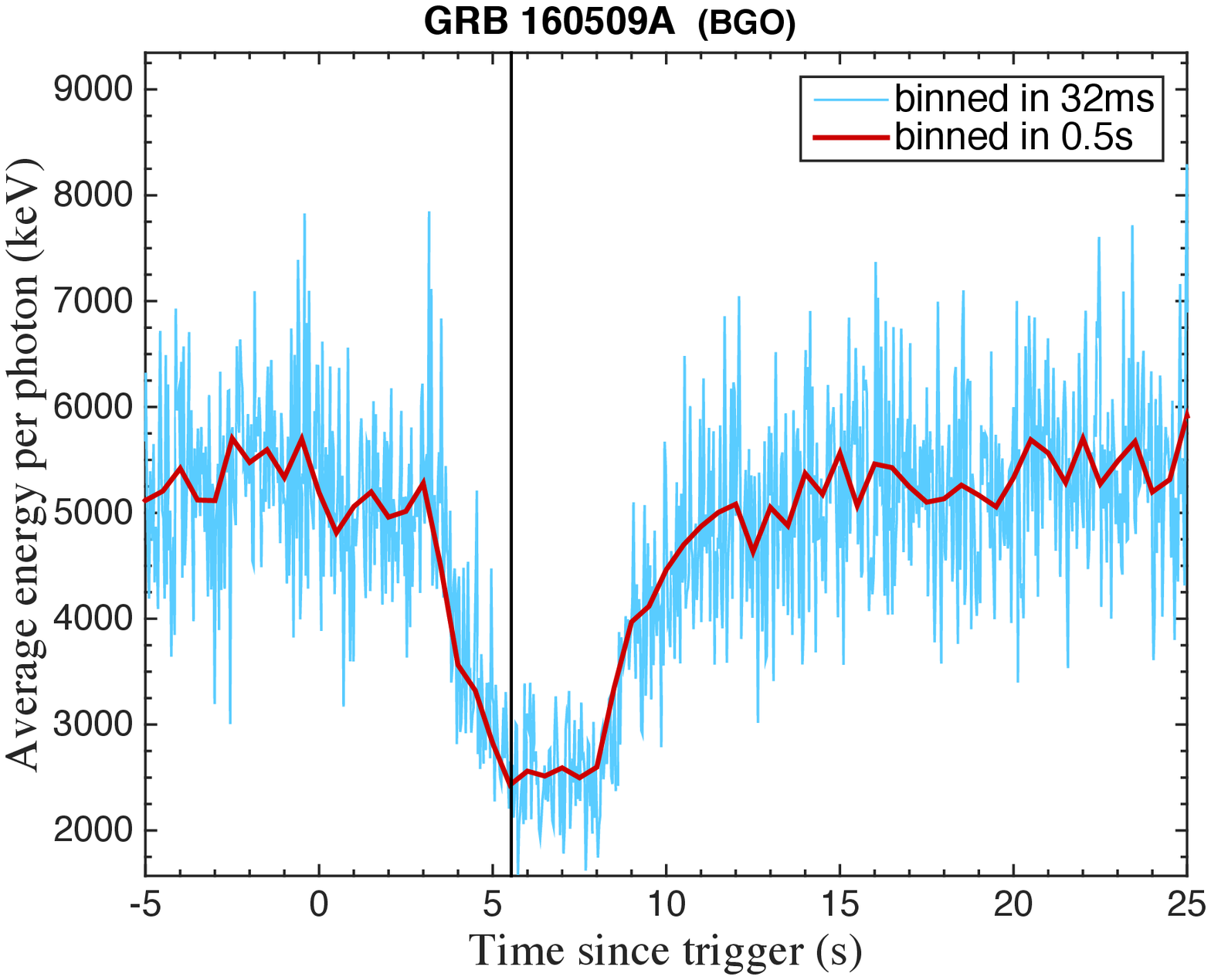}
	\end{minipage}
	\caption[Curve Set II-b]{Curves in Set-BGO (part 2): Light curves and average energy curves of GBM BGO data.}\label{bgo-2}
\end{figure*}

\begin{figure*}[htbp]
	\centering
	\begin{minipage}[t]{0.44\linewidth}
		\centering
		\includegraphics[width=0.90\linewidth]{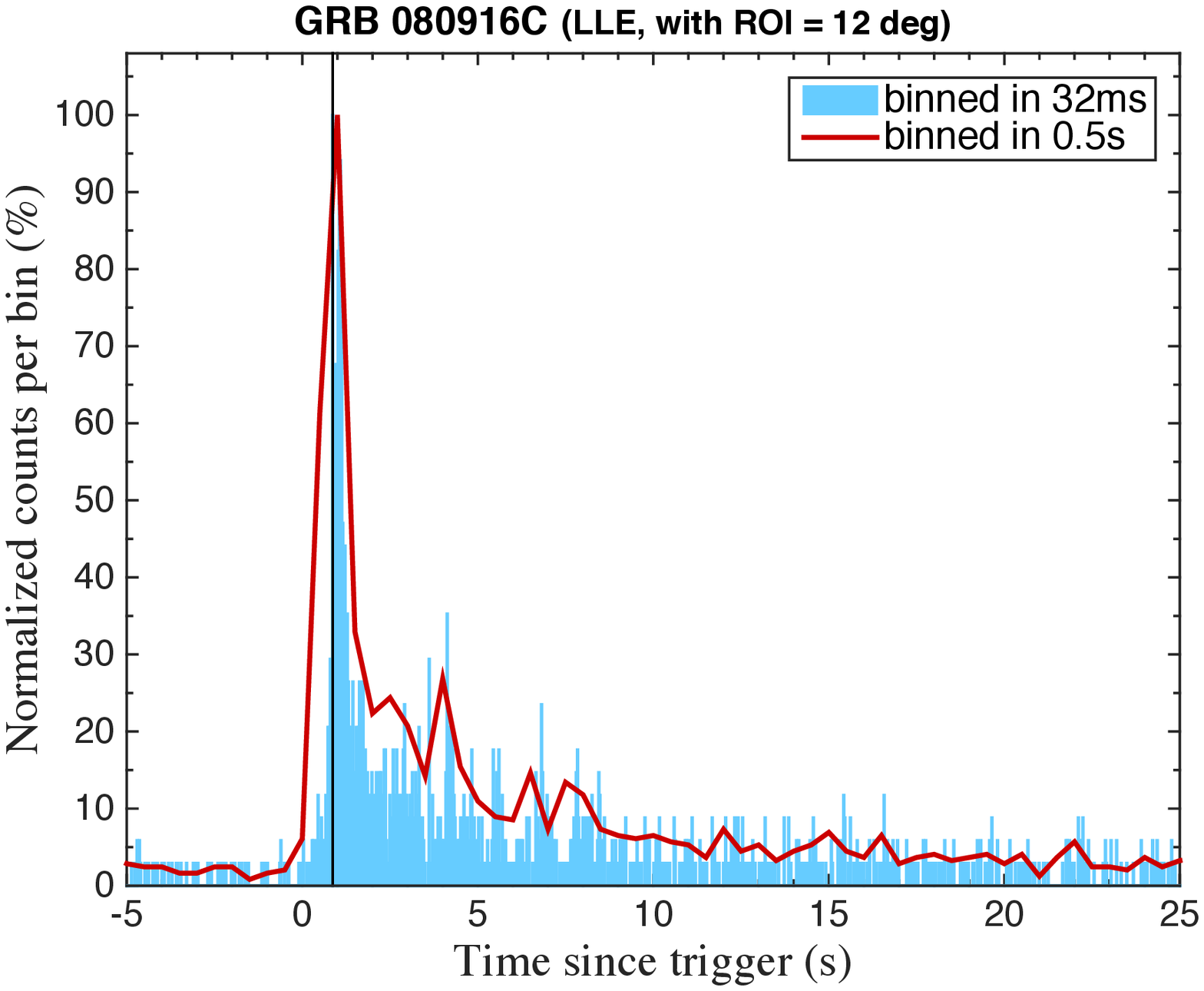}
		\includegraphics[width=0.90\linewidth]{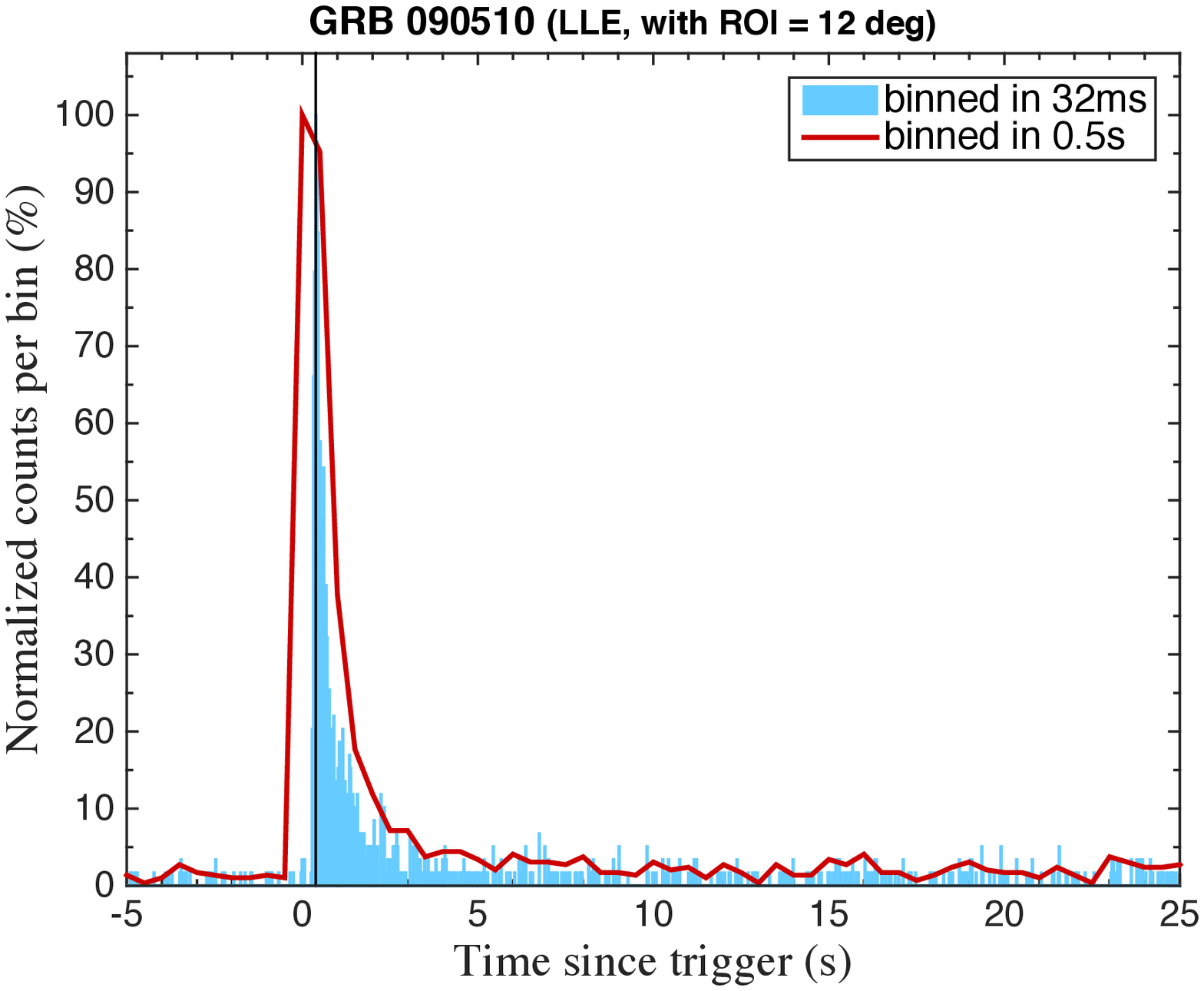}
		\includegraphics[width=0.90\linewidth]{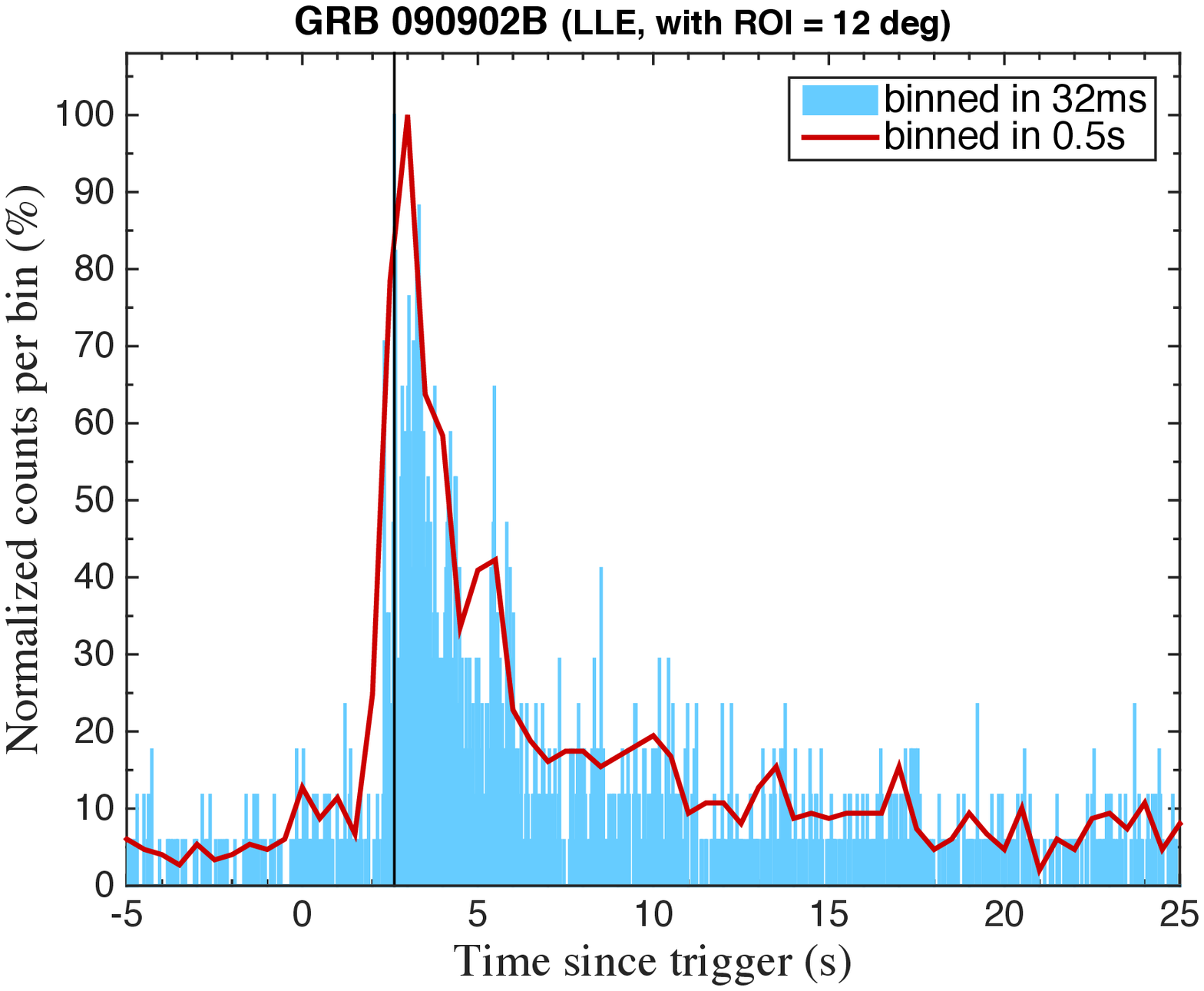}
		\includegraphics[width=0.90\linewidth]{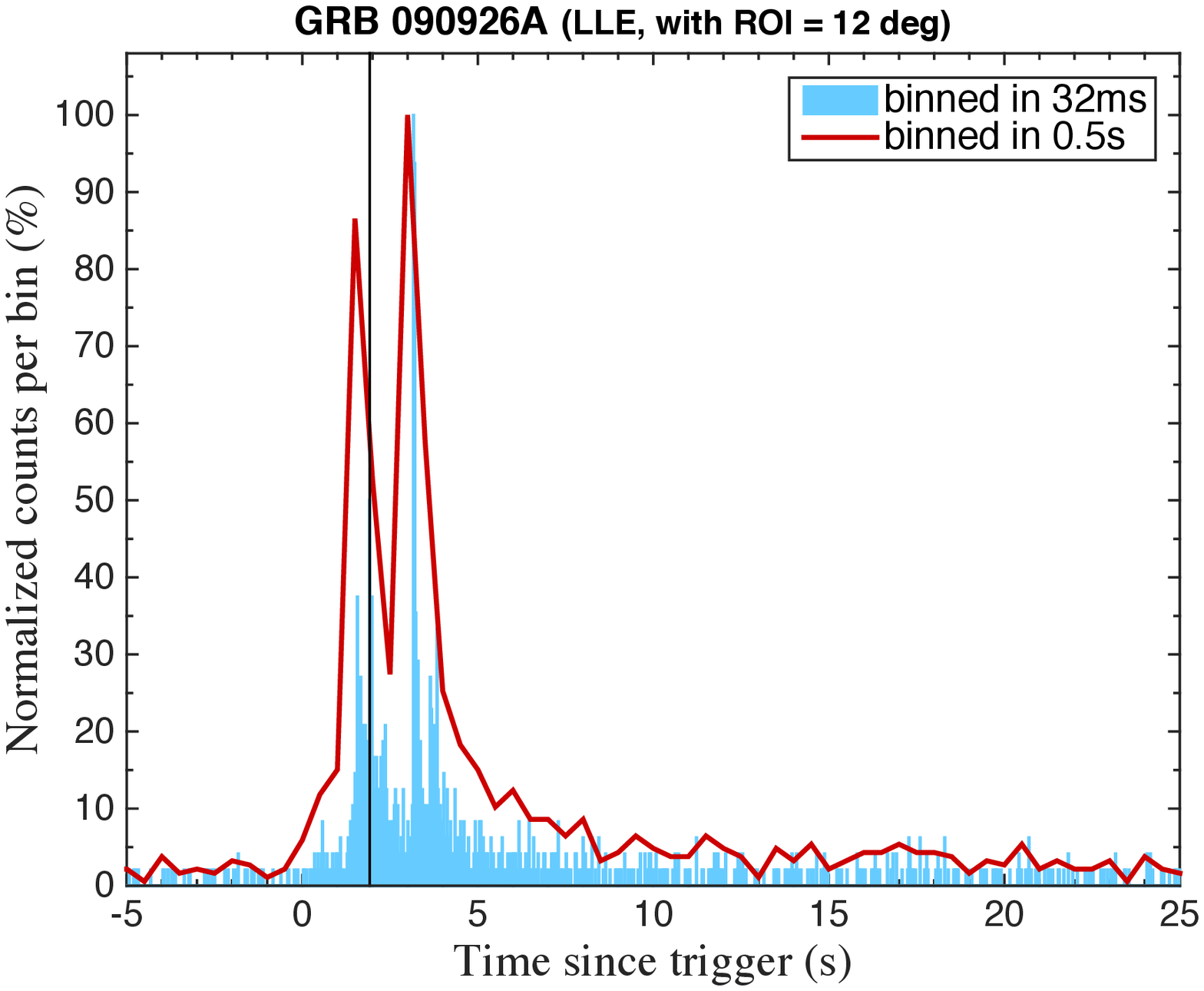}
	\end{minipage}
	\begin{minipage}[t]{0.44\linewidth}
		\centering
		\includegraphics[width=0.90\linewidth]{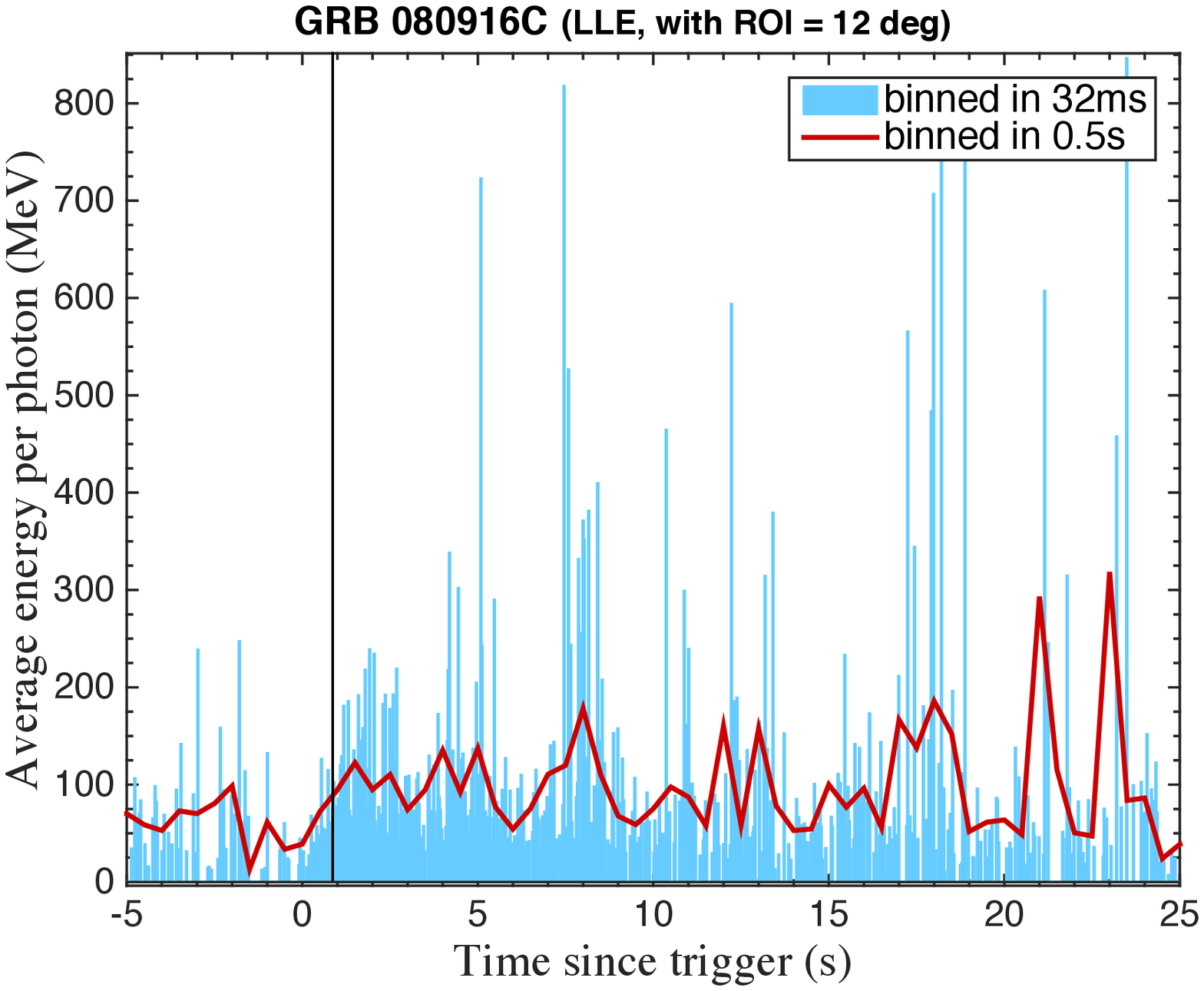}
		\includegraphics[width=0.90\linewidth]{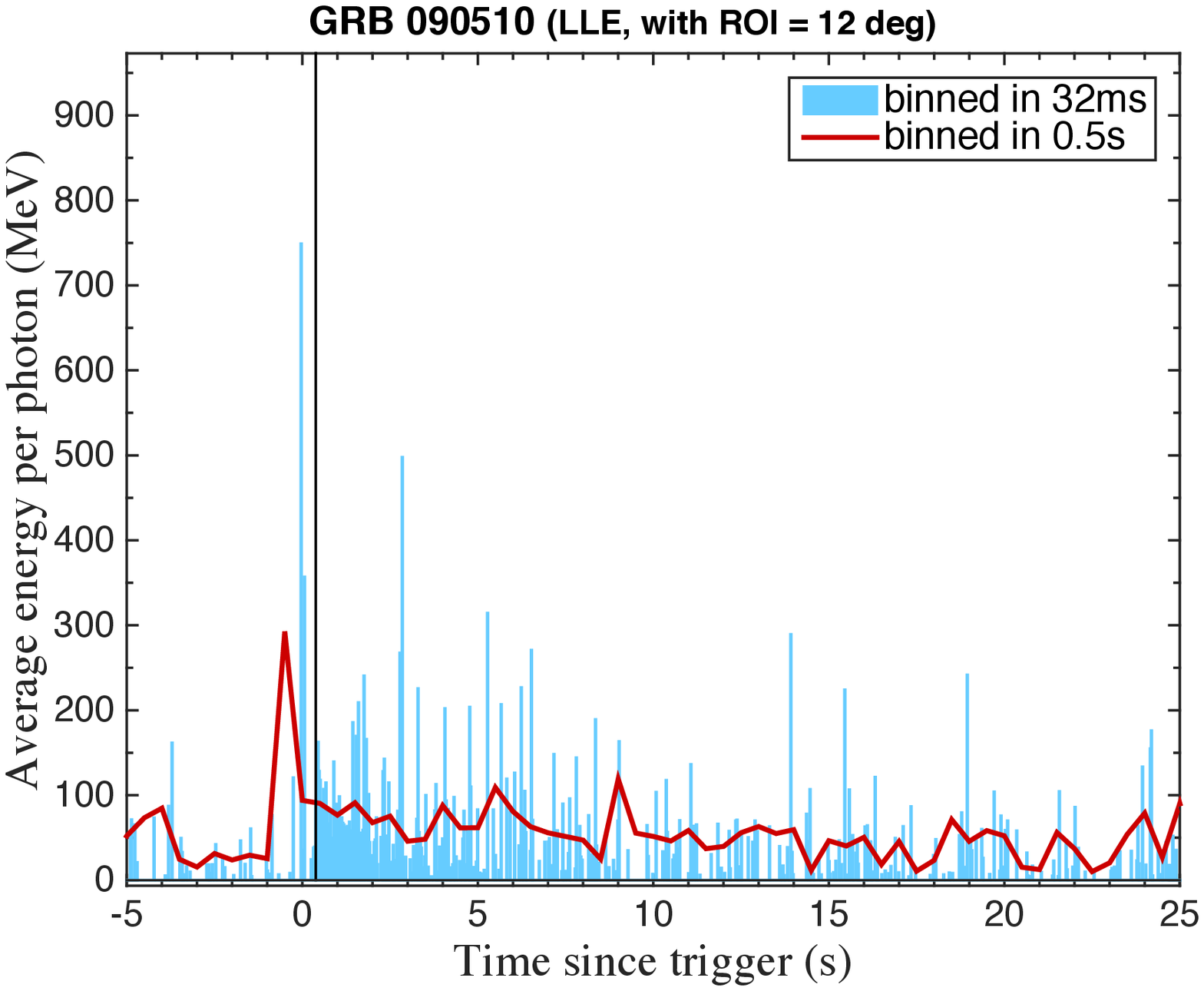}
		\includegraphics[width=0.90\linewidth]{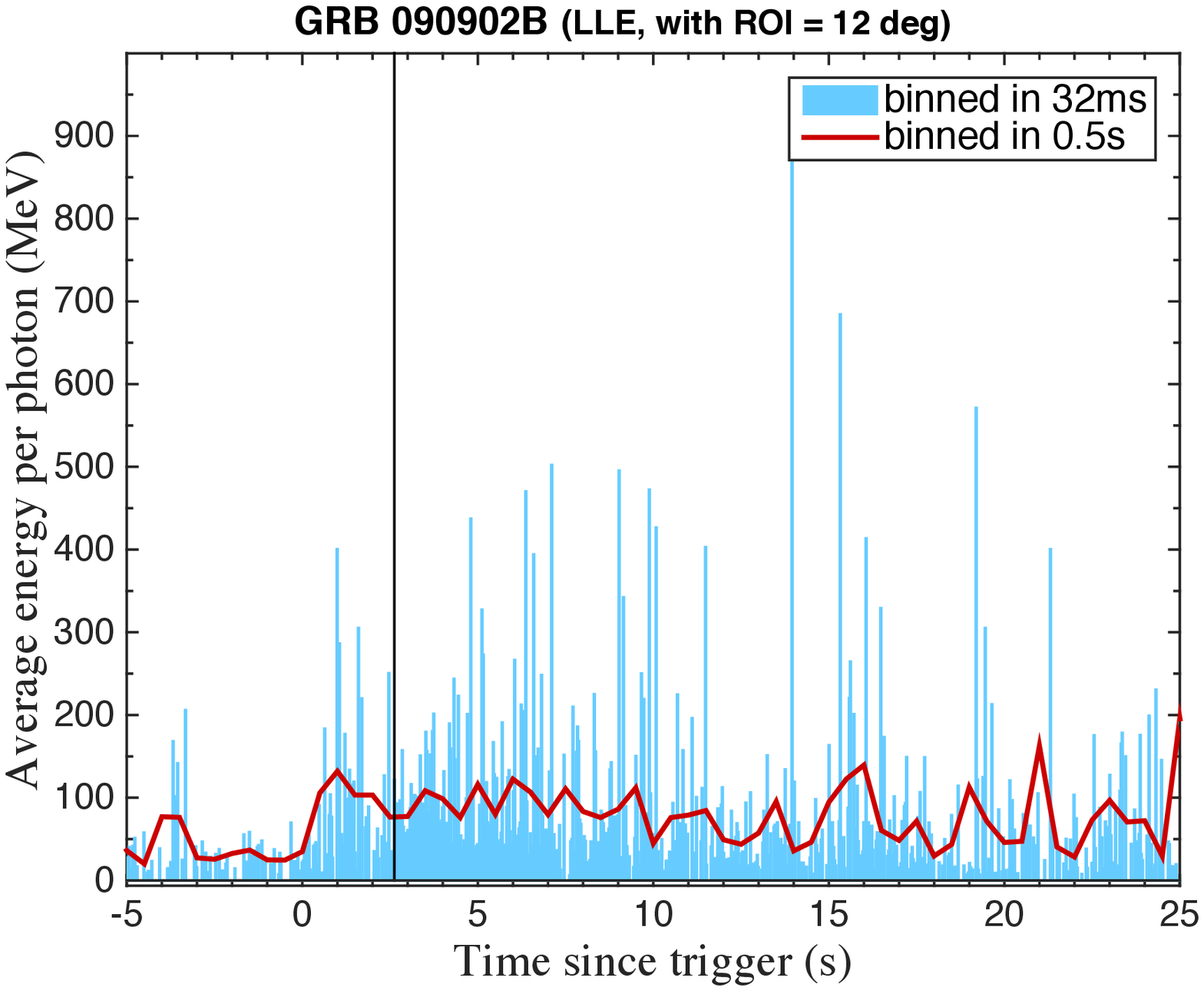}
		\includegraphics[width=0.90\linewidth]{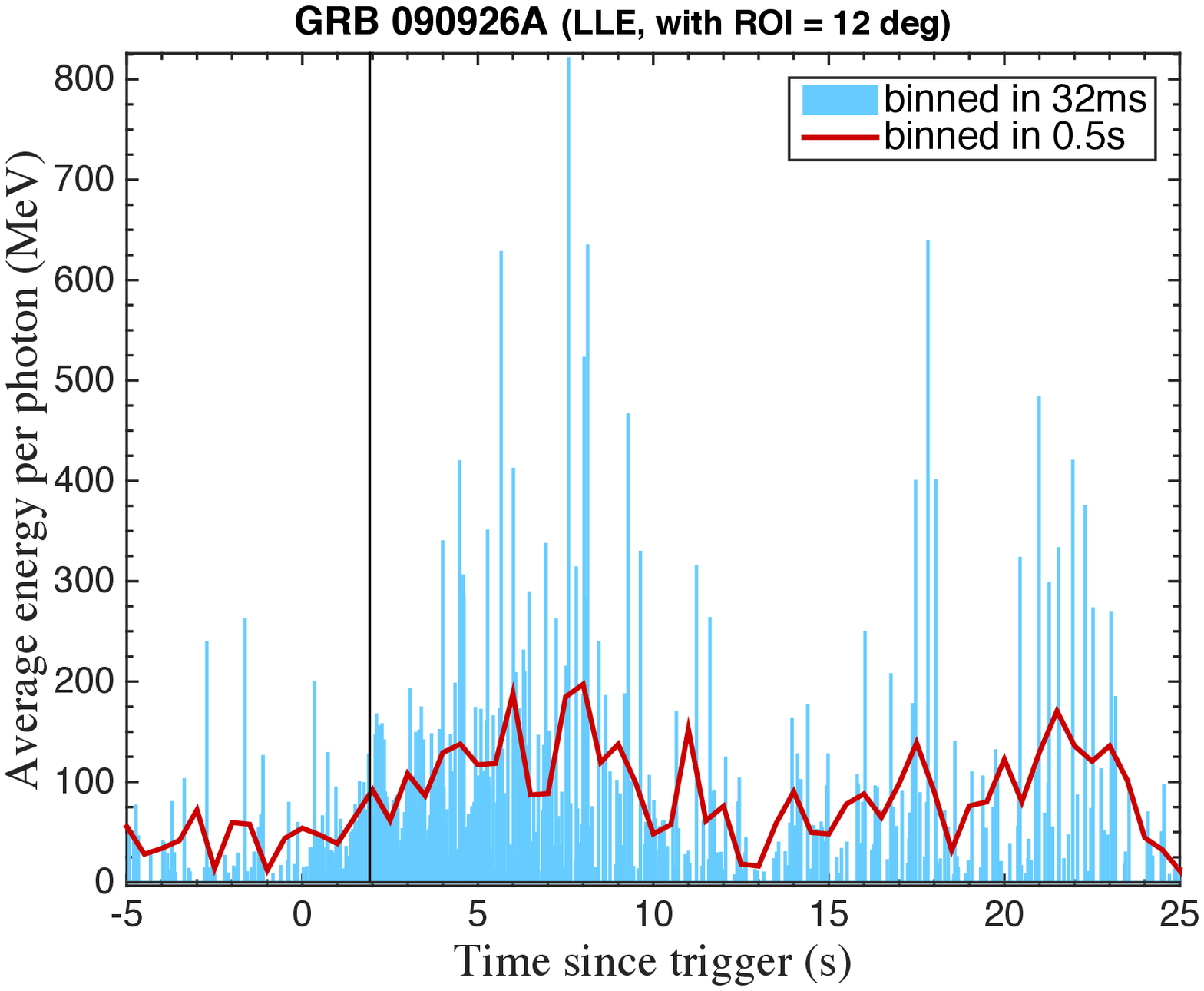}
	\end{minipage}
	\caption[Curve Set III-a]{Curves in Set-LLE (part 1): Light curves and average energy curves of LLE data.}\label{lle-1}
\end{figure*}

\begin{figure*}[htbp]
	\centering
	\begin{minipage}[t]{0.44\linewidth}
		\centering
		\includegraphics[width=0.90\linewidth]{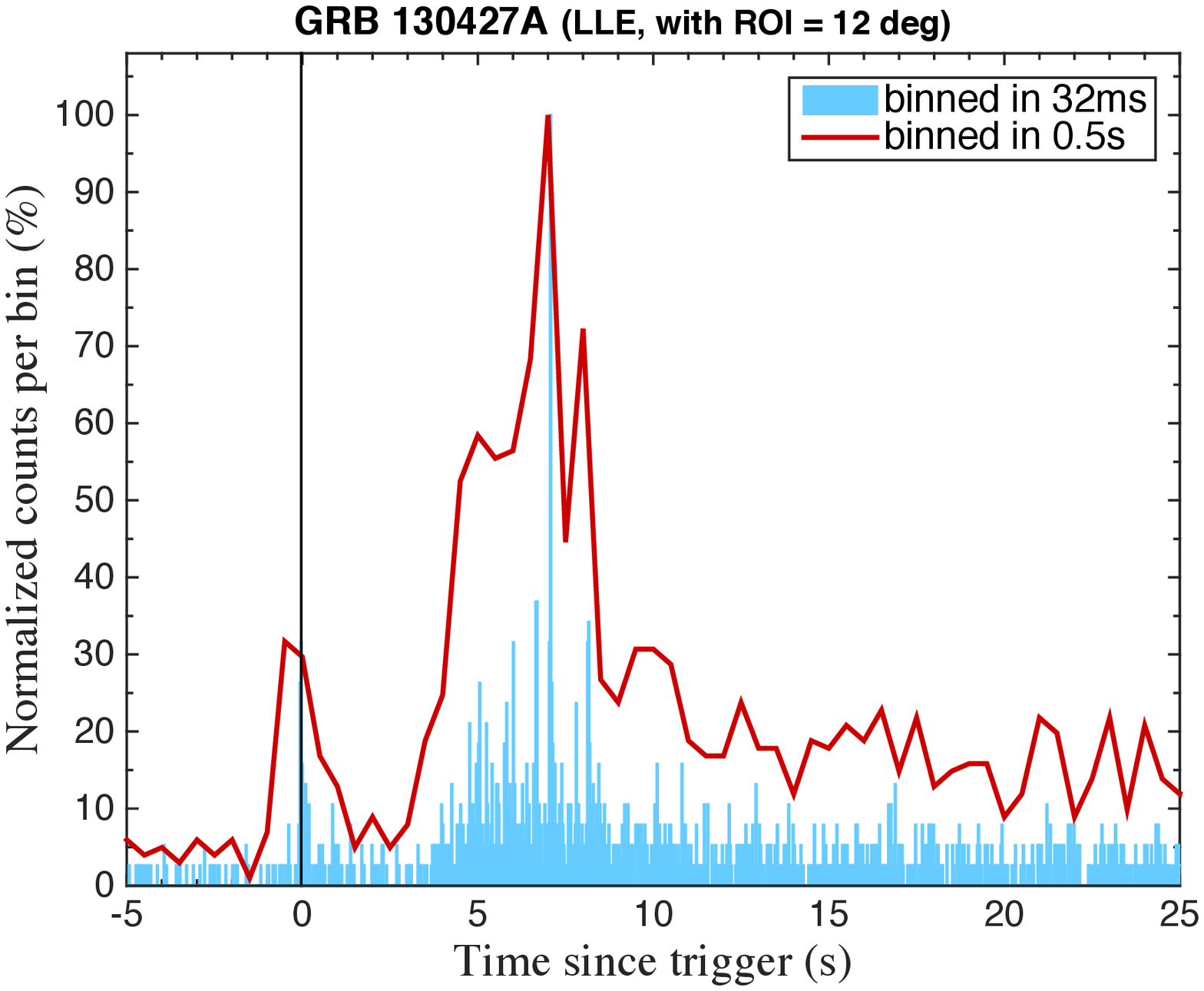}
		\includegraphics[width=0.90\linewidth]{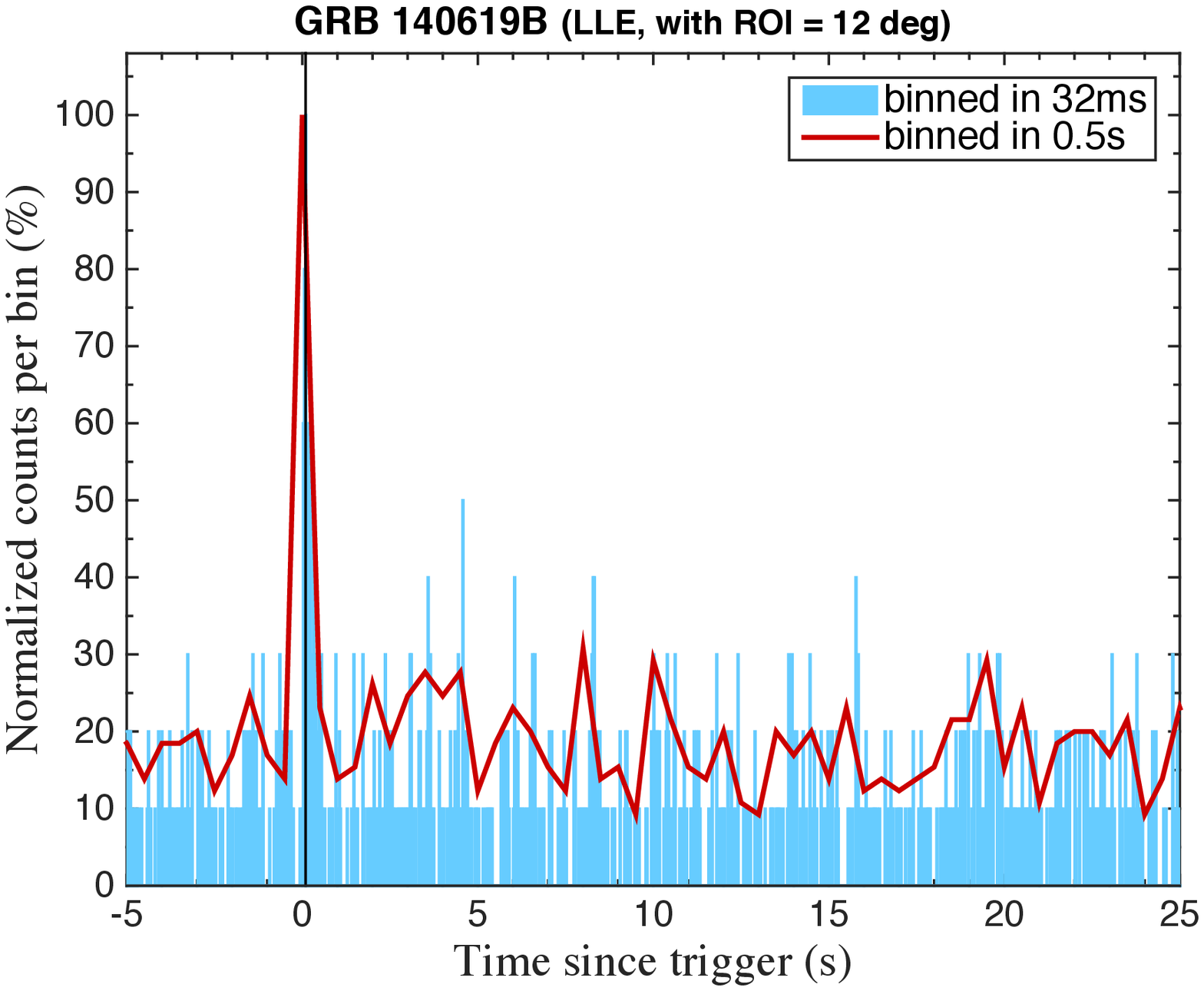}
		\includegraphics[width=0.90\linewidth]{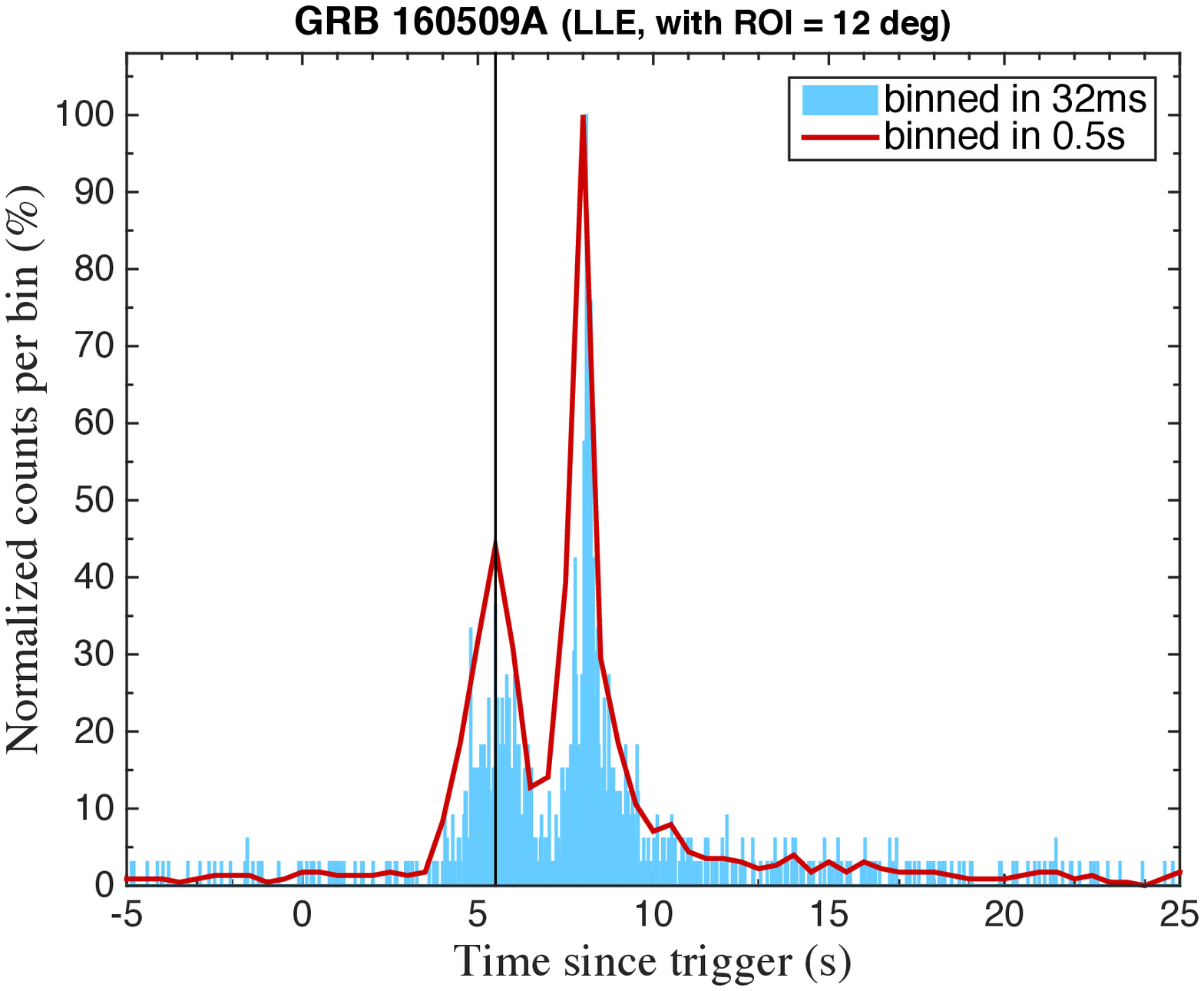}
	\end{minipage}
	\begin{minipage}[t]{0.44\linewidth}
		\centering
		\includegraphics[width=0.90\linewidth]{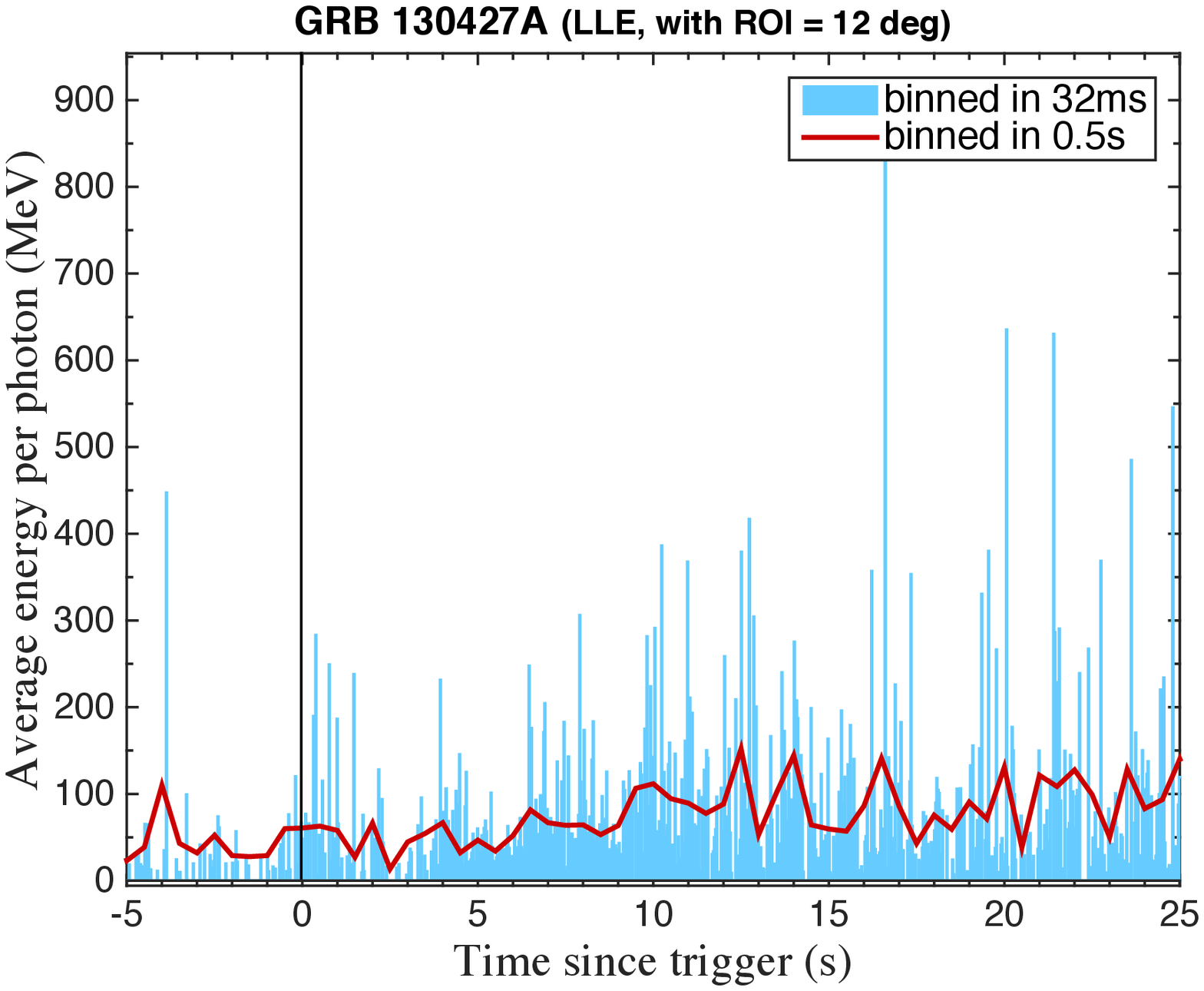}
		\includegraphics[width=0.90\linewidth]{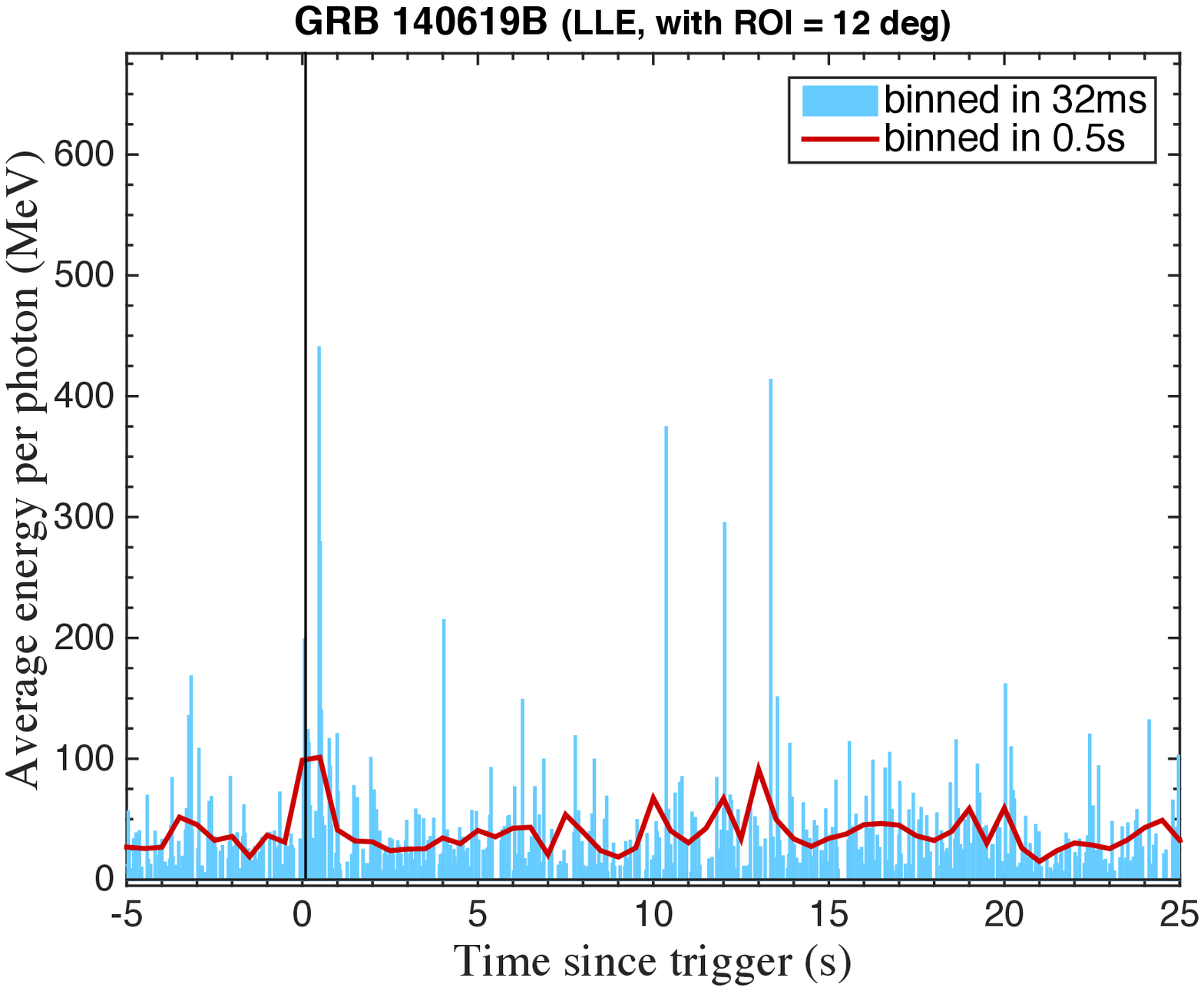}
		\includegraphics[width=0.90\linewidth]{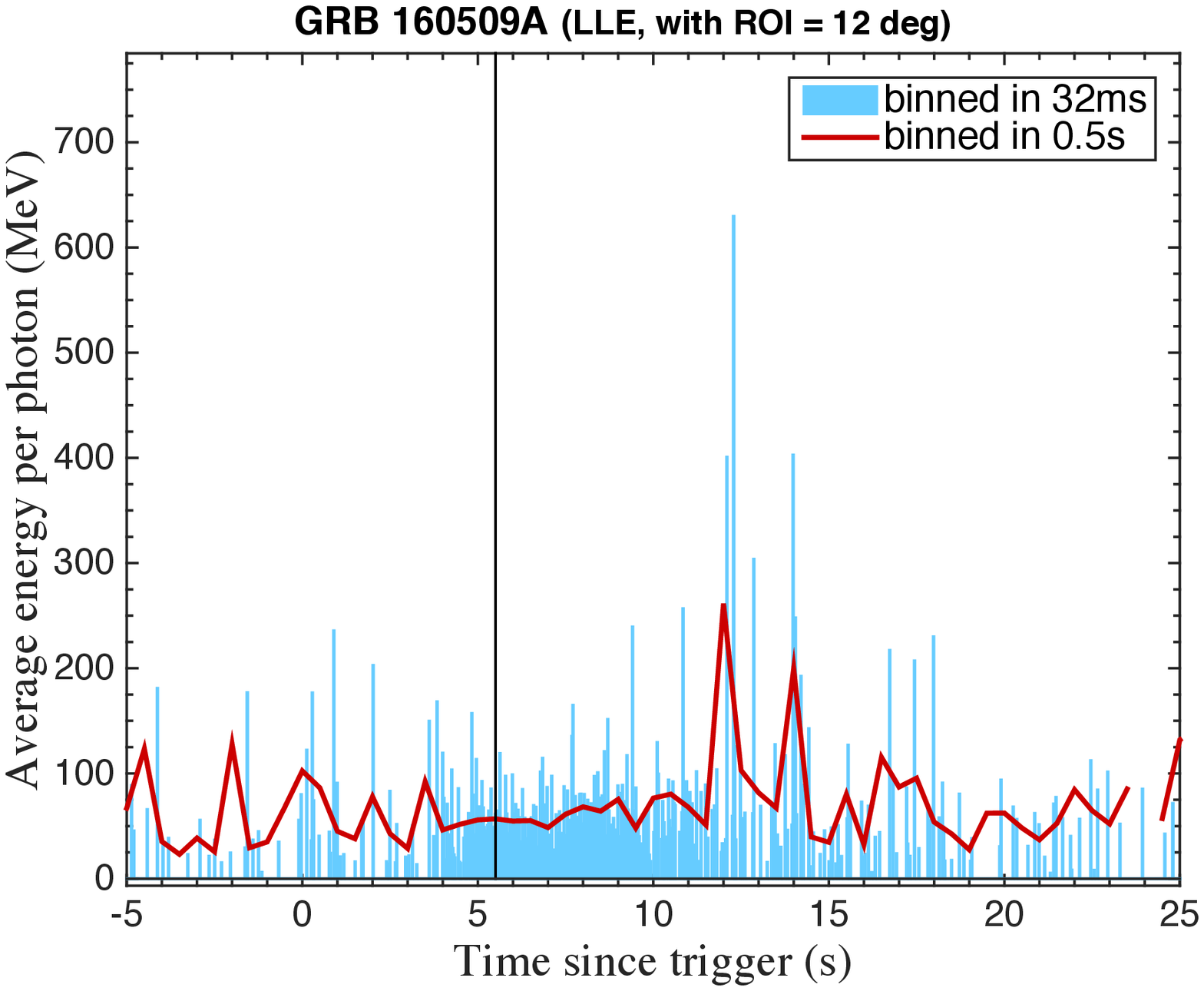}
	\end{minipage}
	\caption[Curve Set III-b]{Curves in Set-LLE (part 2): Light curves and average energy curves of LLE data.}\label{lle-2}
	
\end{figure*}

The average energy curves of Set-LLE are different from those of Set-NaI and Set-BGO. For some GRBs, it is difficult to choose characteristic times from the LLE average energy curves to match the peaks in the light curves.
One possible reason is that, for LLE data, the energy distribution of GRB photons is similar to that of background photons. For example, we plot the energy distribution histogram from the LLE data of GRB 160509A, using the events: (1) within 4-10~s~(in the source frame) since trigger and (2) before trigger (see fig.~\ref{16ED-LLE}). From the light curve, Case 1 covers an intensive GRB process and Case 2 covers the background photons. Thus Case 1 can approximately show the energy distribution of GRB photons while Case 2 reflects the energy distribution of background photons (as shown in fig.~\ref{16ED-BGO}). From the BGO energy distributions, the GRB photons tend to have lower energies than the background photons, as the middle part of the GRB energy distribution~(Case 1) is depressed compared to the background distribution. However the LLE average energy curves show that the difference of distributions between the two cases is comparably smaller. It leads to difficulty in determining a characteristic time in some LLE average energy curves.

Although the LLE average energy curves do not help much, the peaks in the corresponding light curves are sharp and clear. From figs.~\ref{nai-1}, \ref{nai-2}, \ref{bgo-1}, \ref{bgo-2}, \ref{lle-1} and \ref{lle-2}, we find the characteristic times using the Criterion 3 for curves in Set-NaI and Set-BGO and the criterion in ref.~\cite{xhw}~(first main peak in the light curve) for Set-LLE. The characteristic times, $t_{\rm low,obs}/(1+z)$, for 3 sets of curves are listed in table~\ref{tlow}. From the table, the 3 sets of $t_{\rm low,obs}/(1+z)$ are consistent with each other, as the maximum difference of $t_{\rm low,obs}/(1+z)$ for each GRB is less than 1~s~(about two 0.5~s bins). As the peaks are chosen according to the 0.5~s-binned curves, a 1~s time interval can be assumed to be related with the same energetic process. Besides, for each GRB, the uncertainty in choosing the highest point in the 0.032~s-binned curve may also lead to a difference about one or two large bins~(0.5~s) among three $t_{\rm low,obs}/(1+z)$. Therefore, the three $t_{\rm low,obs}/(1+z)$ may reflect the same intrinsic process for each GRB.
The consistence of three sets of $t_{\rm low,obs}/(1+z)$ also suggests that the mechanism around the characteristic time influences both low energy ($\sim$10~keV) and medium energy ($\sim$100~MeV) photons. Thus, the first main peak for each GRB may represent an important process with wide spectrum of energy, and this process is treated as the low energy event in sect.~\ref{method}.

\begin{figure*}[h]
	\centering
	\begin{minipage}[t]{0.33\linewidth}
		\centering
		\includegraphics[width=\linewidth]{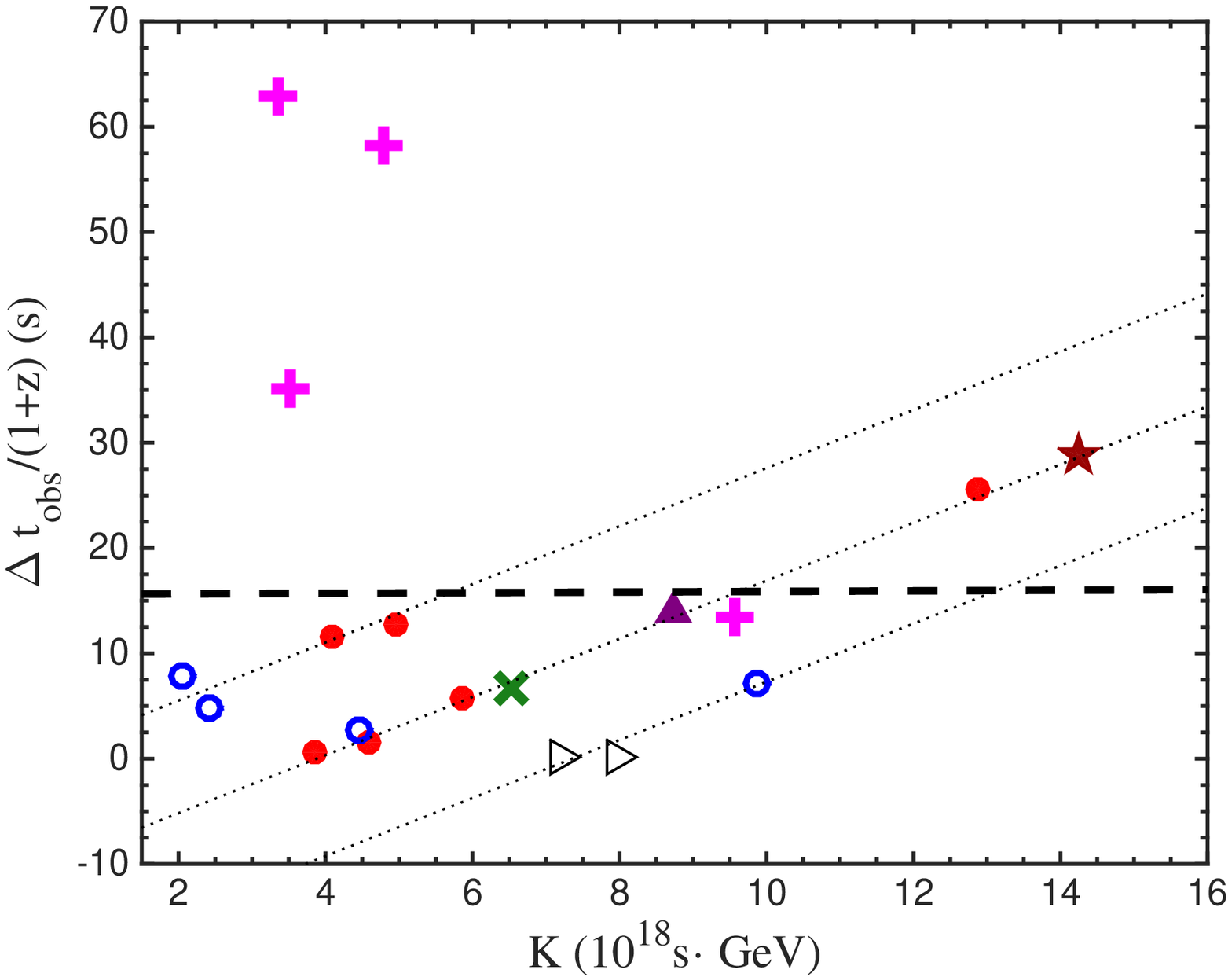}
	\end{minipage}
	\begin{minipage}[t]{0.33\linewidth}
		\centering
		\includegraphics[width=\linewidth]{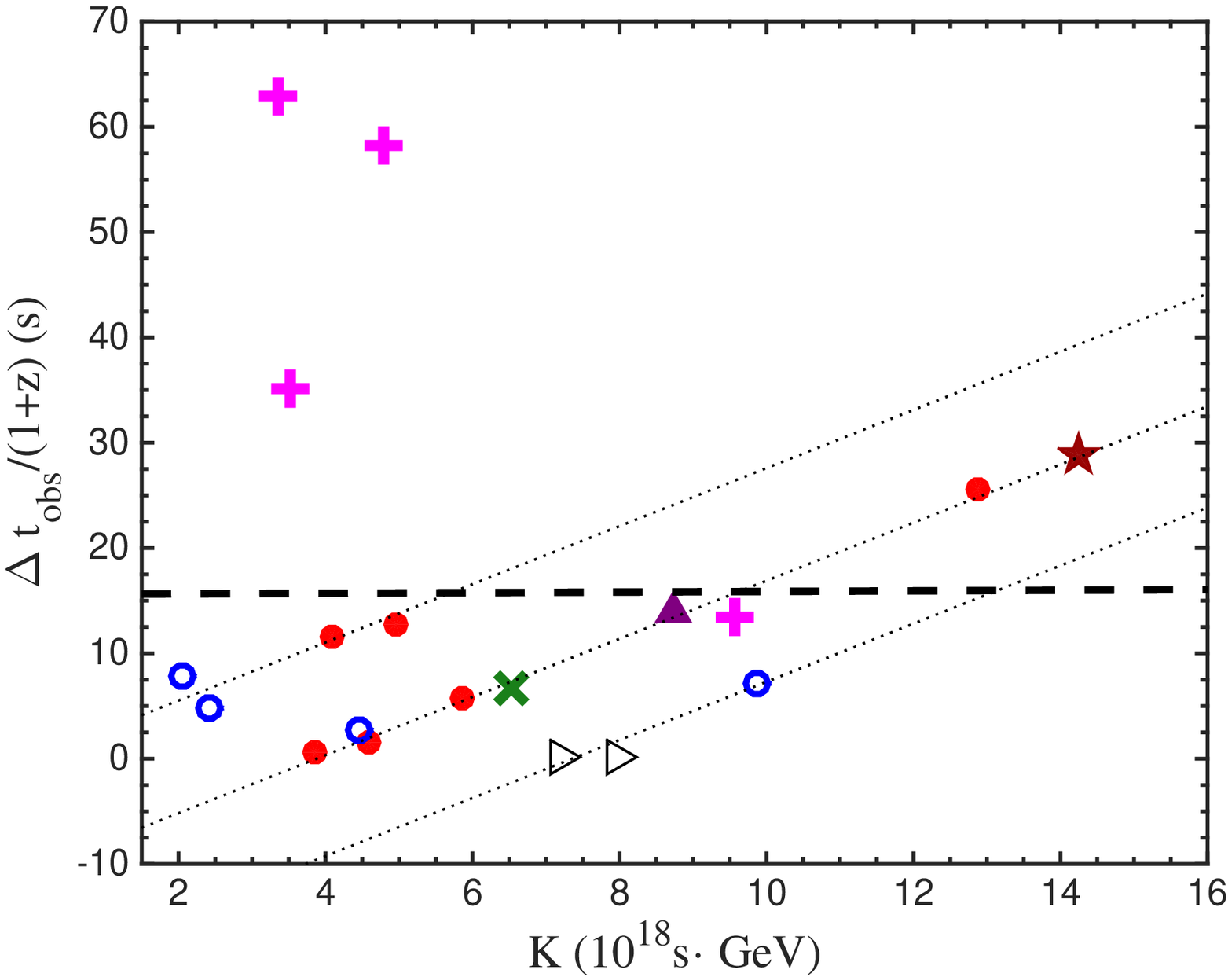}
	\end{minipage}
	\begin{minipage}[t]{0.33\linewidth}
		\centering
		\includegraphics[width=\linewidth]{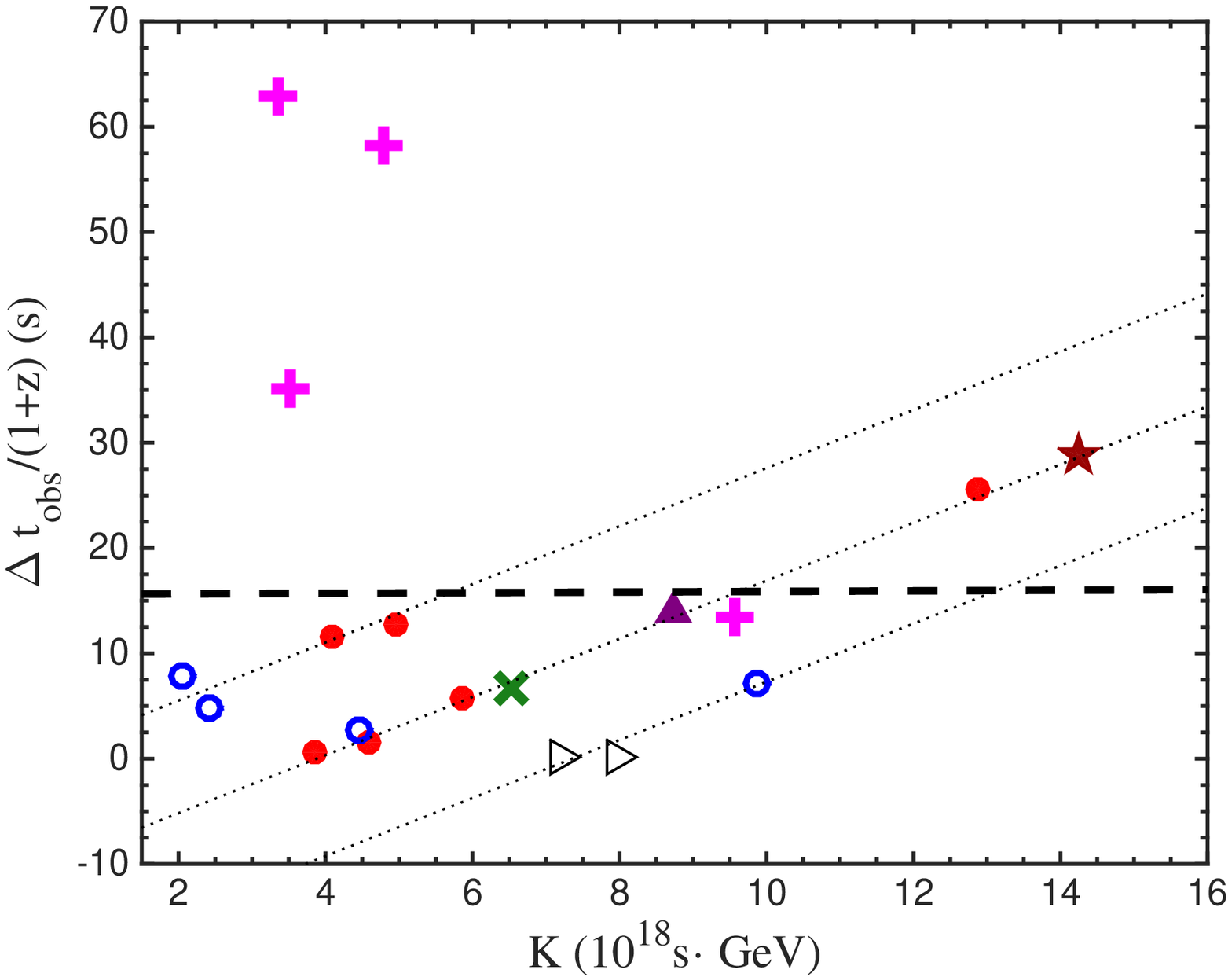}
	\end{minipage}
	\caption{The $\Delta t_{\rm obs}/(1+z)$-$K$ plots from NaI, BGO and LLE data. The dots refer to events from long bursts and the triangles refer to the events from short bursts. The middle dotted line in each plot refers to the mainline. For each plot, 9 events out of 17 fall on the mainline. As the LLE data for GRB 100414A seem not available in the website, we apply the characteristic time from NaI and BGO data~(with $t_{\rm low,obs}/(1+z)=1.184$~s) for GRB 100414A for the $\Delta t_{\rm obs}/(1+z)$-$K$ plot of LLE data. The event of GRB 100414A in the LLE plot is marked by a hollow circle.}\label{instrumentsplot}
\end{figure*}

Also the results are consistent with $t_{\rm low,obs}/(1+z)$ in former sections, and the $\Delta t_{\rm obs}/(1+z)$-$K$ plots for three sets~(as shown in fig.~\ref{instrumentsplot}) are remarkably similar to the plots in former sections. All the fittings arrive at a slope $(0\pm 3)\times 10^{-18}\rm ~GeV^{-1}$, showing consistency with figs.~\ref{allphotonsdiag}, \ref{fig:energy}, \ref{fig:aveenergy}, and \ref{fig:c3}.

\subsection{Estimation of Uncertainties }\label{uncertainty}
According to the method in sect.~\ref{method}, the uncertainty of the final $E_{\rm LV}$ comes from the uncertainties of the times and energies of high and low energy events, the uncertainty of redshifts and the error from the linear fit.
In former sections, we fit the middle 9 events with linear function and obtain the mainline. The uncertainty there only contains the error caused by the fitting. With this method, we obviously overestimate the accuracy of $E_{\rm LV}$.
Now we try to estimate the uncertainty in consideration of all uncertainties caused by the method. As the accurate value of $E_{\rm LV}$ is now less important than its order of magnitude, we simply do a rough estimation.

To estimate the uncertainty of the slope of the mainline, we need to obtain the uncertainties of $K$ and $\Delta t_{\rm obs}/(1+z)$ for each point in the $\Delta t_{\rm obs}/(1+z)$-$K$ plot.
According to sect.~\ref{method}, the error of $K$ for each high energy event is decided by the error of energy and the error of redshfit.
For high energy photons, the energy resolution is a function of incidence angle and energy. For simplicity, we take energy resolution as $\Delta E/E\simeq10\%$~(see fig.~18 in ref.~\cite{fermi-1}). As that the error of redshift is absent for some GRB, and that for most GRB on the mainline, the uncertainty of the redshift is less than 5\% as shown in table~\ref{tab:phinfo}, we ignore the uncertainty of redshift and simply apply a 10\% error in $K$ for each events.

\begin{figure}[H]
	\centering
	\includegraphics[width=0.90\linewidth]{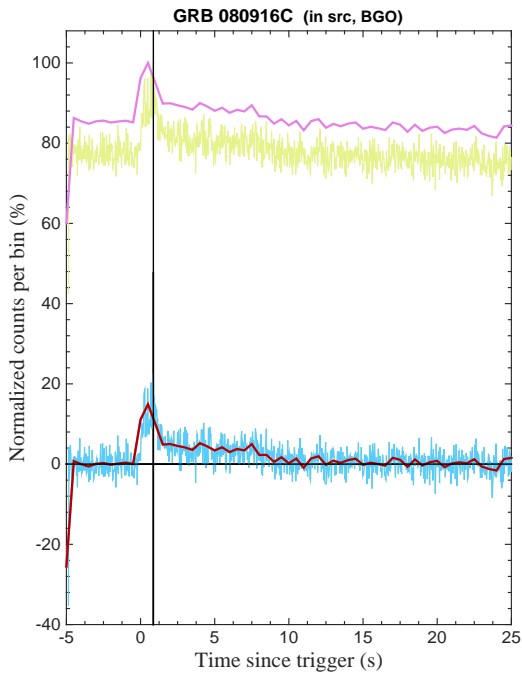}
	\caption{BGO light curves for GRB 080916C with and without background subtraction. The upper two curves refer to the light curves without background subtraction~(also shown in fig.~\ref{bgo-1}). After subtracting the background we obtain the lower two curves. The time $t_{\rm low,obs}/(1+z)$ determined from the light curves with and without background subtraction are the same, marked by the vertical line.} \label{bgsubtract}
\end{figure}

The error in the $\Delta t_{\rm obs}/(1+z)=t_{\rm high,obs}/(1+z)-t_{\rm low,obs}/(1+z)$ mainly comes from the determination of $t_{\rm low,obs}/(1+z)$ because the time resolution of high energy events~\cite{fermi-1} is negligible compared to the uncertainty of $t_{\rm low,obs}/(1+z)$.
These $t_{\rm low,obs}/(1+z)$ are read mainly from GBM data and the background photons are non-negligible as shown in the light curves before. However, the background does not influence the determination of low energy characteristic time, because that the background counts for all discussed GRBs are quite stable and that we choose a peak relative to the background. For example, for BGO light curve of GRB 080916C whose background photons occupy a large part, we do a background subtraction using the method in the commonly used tool, GTBurst~\cite{gtburst},
i.e., fitting the background before and after GRB with a polynomial function, and plot its light curve in fig.~\ref{bgsubtract}.
The low energy photon characteristic time determined from this background subtracted curve is also $t_{\rm low,obs}/(1+z)=0.864$~s, which is consistent with table~\ref{tlow}. As the background subtraction does not influence the determination of first main peak, we do not apply background subtraction in all other figures.

%\begin{figure}[H]
%	\centering
%	\includegraphics[width=0.90\linewidth]{BGO-080916-s.eps}
%	\caption{BGO light curve for GRB 080916C with background subtraction. The $y$-axis is zoomed in order that the $y$~coordinate of the highest point equals to 1.} \label{bgsubtract}
%\end{figure}

As a rough estimation, the uncertainty of $t_{\rm low,obs}/(1+z)$ is chosen as the maximum difference of $t_{\rm low,obs}/(1+z)$ under different criteria, i.e., about 1~s.
The fluctuations of the background may lead to uncertainty in selecting the highest point around the first main peak, but the peak in 0.5~s-binned curve is clear so the uncertainty caused by the background fluctuation should be $\sim$0.5~s. The difference of $t_{\rm low,obs}/(1+z)$ for a certain GRB actually contains the uncertainty of choosing the highest point around the peak.
%The process inside the source does not happen in an instant.
%We treat the different $t_{\rm low,src}$ for a certain GRB as measurements on the characteristic time of low energy photons and use the extreme difference as an estimation for uncertainty.
%so we use a 1 s uncertainty for $t_{\rm low,src}$ for simplicity.
%%Thus, we apply an 10\% uncertainty in $K$ and 1~s uncertainty in  $\Delta t_{\rm obs}/(1+z)$. As the sets of $t_{\rm low,src}$ in different sections are very similar, we just choose the data in fig.~\ref{fig:c3} to estimate the uncertainty of $E_{\rm LV}$. We use orthogonal distance regression to fit the 9 events around the mainline with both $x$ and $y$ uncertainties (see fig.~\ref{odr}). We obtain the slope and its standard error as $(2.65 \pm 0.11)\times 10^{-18}\rm GeV^{-1}$. Thus $E_{\rm LV}=(3.8\pm 0.2)\times 10^{\rm 17}$~GeV. %This shows that the uncertainty of the $E_{\rm LV}$ has a order of $10^{16}$GeV.
%%Moreover, if we fit the points
%%on all the 3 lines by a single straight line without the events of
%%130427A(2), 130427A(3), and 130427A(4), then we arrive at $E_{\rm LV}=5.8_{-2.1}^{+7.9}\times 10^{\rm 17}$~GeV, which is compatible with the previous
%%results but with larger error.

These uncertainties do not significantly change the fittings, from which we arrive at the lower bound on $E_{\rm LV}$ with $|E_{\rm LV}|\ge 3\times 10^{17}$~GeV.

%and therefore it is implausible to confirm that light speed variation exists.
%\begin{figure}
%	\centering
%	\includegraphics[width=0.90\linewidth]{odrfit.eps}
%	\caption{Fitting the events on the mainline with $x$ and $y$ uncertainties.}\label{odr}
%\end{figure}

\section{Conclusion}\label{conclusionsec}

We aim to check whether the speed of light changes with the photon energy by using the gamma-ray burst data from the FERMI telescope~(FGST).
%A linear form modification on the speed of light is assumed in this paper, $v(E)=c(1-E/E_{\rm LV})$.
%We examine the work in refs.~\cite{xhw,xhw160509} by performing the analysis in the observer reference frame and in the source reference frame. We find that the results are not sensitive to technical details, such as the bins and the energy bands for low energy photons.
We review the work in refs.~\cite{xhw,xhw160509}
and also suggest two other criteria on determining the characteristic time $t_{\rm low}$ for low energy photons.
The first is to choose the first main peak of received energy curve. It results in nearly the same $\Delta t_{\rm obs}/(1+z)$-$K$ plot as from the light curve.
The second is to choose the first significant peak (or dip) of the average energy curve, which reflects the change of energy distribution. We obtain a slightly different $\Delta t_{\rm obs}/(1+z)$-$K$ plot, but the results are still consistent with those in refs.~\cite{xhw,xhw160509}.
Finally, we offer a new criterion that uses the average energy curve in addition to the light curve to determine the characteristic time for low energy photons.
We apply the new criteria to three different sets of
GBM NaI, GBM BGO, and LAT LLE data of low energy photons from FERMI detectors. It is remarkable that we obtain consistent results for the characteristic time of three different sets of low energy photons. The regularity that several high energy photon events from different GRBs fall on a same line~\cite{xhw,xhw160509} still persists under different situations.
With this more comprehensive analysis,
we arrive at consistent results of the lower bound on $E_{\rm LV}$ with $|E_{\rm LV}|\ge 3\times 10^{17}$~GeV.
%that imply little evidence of light speed variation.
%we arrive at the results to support the regularity of high energy photon events from different GRBs~\cite{xhw,xhw160509}. Such regularity indicates a linear form modification of %light speed $v(E)=c(1-E/E_{\rm LV})$, where $E_{\rm LV}\simeq(3.8\pm 0.2)\times10^{17}$~GeV from the data.
A recent work studied the LV energy scale from statistical estimators\cite{ellis18}, with lower bound
exceeding either $8.4\times 10^{17}$~GeV or $2.4\times 10^{17}$~GeV, which is consistent with our results.

\section{Acknowledgments}
This work is supported by National Natural Science Foundation of China (Grant No.~11475006). It is also supported by Hui-Chun Chin and Tsung-Dao Lee Chinese Undergraduate Research Endowment~(CURE).

\end{document}